\documentclass{aa}
\usepackage{graphicx}
\usepackage{txfonts}  
\newcommand{\ba}[4]{\mbox{$#1 ^2{\rm #2}^{#3}_{\rm #4}$}}
\newcommand{\eu}[5]{\mbox{$#1\,^#2{\rm #3}^{#4}_{\rm #5}$}}
\newcommand{\iso}[2]{\mbox{$^{#1}{\rm #2}$}}
\newcommand{\Teff}{T_{\rm eff}}

\newcommand{\Vmic}{V_{\rm mic}}
\newcommand{\Vmac}{V_{\rm mac}}
\newcommand{\eps}[1]{\log\varepsilon_{\rm #1}}
\newcommand{\kms}{km\,s$^{-1}$}
\newcommand{\kH}{$S_{\rm H}$}
\begin{document}

\title{A non-LTE study of
neutral and singly-ionized calcium in
late-type stars
\thanks{Based on observations collected at the European Southern Observatory,
Chile, 67.D-0086A, and the German Spanish Astronomical Center,
Calar Alto, Spain}}

\author{L. Mashonkina \inst{1,2}
\and A.J. Korn \inst{3}
\and N. Przybilla \inst{4}}

\offprints{L. Mashonkina}

\institute{
Institut f\"ur Astronomie und Astrophysik der Universit\"at
M\"unchen, Scheinerstr. 1, 81679 M\"unchen, Germany\\
\email{lyuda@usm.lmu.de}
\and
Institute of Astronomy, Russian Academy of Science, Pyatnitskaya 48,
119017 Moscow, Russia\\
\email{lima@inasan.ru}
\and
Department of Astronomy and Space Physics, Uppsala University,
Box 515, SE - 75120 Uppsala, Sweden
\and
Dr. Remeis-Sternwarte Bamberg, Sternwartstrasse 7,
D-96049 Bamberg, Germany}

\date{Received  / Accepted }


\abstract{
{\it Aims.}
Non-local thermodynamical equilibrium (NLTE) line formation for
neutral and singly-ionized calcium  is considered
through a range of spectral types when the Ca abundance varies from
the solar value down to [Ca/H] = $-$5. We evaluate the influence of
departures from LTE on Ca abundance determinations
and inspect the possibility of using \ion{Ca}{i} /
\ion{Ca}{ii} line-strength ratios as indicators of surface
gravity for extremely metal-poor stars.

\noindent
{\it Methods.}
A comprehensive model atom for \ion{Ca}{i} and \ion{Ca}{ii} is presented. Accurate radiative and electron collisional atomic data are incorporated. The role of
inelastic collisions with hydrogen atoms in the statistical equilibrium of \ion{Ca}{i/ii} is estimated empirically
 from inspection of their different influence on
the \ion{Ca}{i} and \ion{Ca}{ii} lines in selected stars with well
determined stellar parameters and high-quality observed spectra.

\noindent
{\it Results.}
The dependence of NLTE effects on the atmospheric parameters is discussed. Departures from LTE significantly affect
the profiles of \ion{Ca}{i} lines over the whole range of stellar parameters considered. However, at [Ca/H] $\ge -2$, NLTE abundance correction of individual lines may be small in absolute value due to the different influence of NLTE effects on
line wings and the line core. At lower Ca abundances,
NLTE leads to systematically depleted total absorption in the line and positive abundance corrections, exceeding +0.5\,dex for \ion{Ca}{i}
$\lambda\,4226$ at [Ca/H] = $-$4.9. In contrast, NLTE effects strengthen the \ion{Ca}{ii} lines and lead to negative abundance corrections. NLTE
corrections are small, $\le 0.02$\,dex, for the \ion{Ca}{ii} resonance lines
over the whole range of stellar parameters considered. For the IR lines of multiplet
$3d$ - $4p$, they  grow
in absolute value with decreasing Ca abundance exceeding 0.4\,dex in metal-poor stars
with [Fe/H] $\le-3$. As a test and first application of the \ion{Ca}{i/ii} model atom,
Ca abundances are determined on the basis of plane-parallel LTE model atmospheres for the Sun, Procyon (F IV-V), and seven metal-poor stars, using high S/N and high-resolution spectra at visual and near-IR wavelengths. Lines of \ion{Ca}{i} and \ion{Ca}{ii} give consistent abundances for all objects (except Procyon) when collisions with hydrogen atoms are taken into account. The derived absolute solar Ca abundance (from \ion{Ca}{i} and \ion{Ca}{ii} lines) is $\eps{Ca,\odot}$ = 6.38\,$\pm$\,0.06. For Procyon, the mean Ca abundance from \ion{Ca}{i} lines is markedly subsolar, [Ca/H] = --0.14$\pm$0.03.
All metal-poor stars within our sample show an overabundance of calcium relative to iron with [Ca/Fe] abundance ratios of 0.26 to 0.46 typical for the halo population. The $W$(\ion{Ca}{i}4226) / $W$(\ion{Ca}{ii}8498) equivalent width ratio is predicted to be sensitive to
surface gravity for extremely metal-poor stars, while this is not the case for the ratio involving the \ion{Ca}{ii} resonance line(s).
}

\authorrunning{L. Mashonkina, A.J. Korn, \& N. Przybilla}
\titlerunning{A non-LTE study of 
Ca\,{\sc i/ii}}

\maketitle

\section{Introduction}

Calcium is one of the best observable chemical elements in
late-type stars. The resonance lines of \ion{Ca}{i} at
$\lambda\,4226\,\AA$ and \ion{Ca}{ii} at $\lambda\,3933$ and
3968~\AA\, lie in the visual spectral range and can be measured
even in extremely metal-poor stars with metallicity [Fe/H] $< -5$
(Christlieb et al. \cite{chri02}; Frebel et al. \cite{frebel05}).
In such stars, Ca is the only chemical element which is visible in
two ionization stages, and the \ion{Ca}{i} and \ion{Ca}{ii} lines
can be potent tools in the derivation of accurate values for
fundamental stellar parameters and for the Ca abundance itself. Calcium is
an important chemical element to study the history of
$\alpha-$process nucleosynthesis in the Galaxy. The subordinate
lines of neutral Ca located in the relatively uncrowded
yellow-to-red spectral regions are suitable for spectroscopic
analysis over a wide range of Ca abundance from
super-solar values down to [Ca/H] = $-$4 (Cayrel et al. \cite{cayrel04}).
The subordinate lines of ionized Ca at $\lambda\,8498$, 8542,
and 8662~\AA\, are among the strongest features in the
near-infrared spectra of late-type stars with metallicity down to
[Fe/H] = --3 (Mallik \cite{mallik97}).
The \ion{Ca}{ii} triplet lies at the focus of cool star research
with the advent of large spectroscopic surveys of the Galaxy like the
Radial Velocity Experiment (RAVE, Steinmetz et al.~\cite{steinmetzetal06})
and the upcoming ESA Gaia satellite mission (Perryman et al.~2001). These broad lines are
also powerful abundance and metallicity indicators out to large distances
via medium-resolution spectroscopy, favorably located near the flux maximum
of red giants. Applications range from quantitative analyses of individual
stars in globular clusters (e.g. Armandroff \& da Costa~\cite{ardaco91};
Rutledge et al.~\cite{rutledgeetal97}) to stars and stellar populations
in nearby galaxies (e.g. Tolstoy et
al.~\cite{tolstoyetal01}; Ibata et al.~\cite{ibataetal05}). Moreover,
the \ion{Ca}{ii} triplet is useful for the analysis of unresolved stellar
systems like early-type galaxies (e.g. Saglia et al.~2002).
However, as an increasing number of high-quality (echelle) spectra of stars
and unresolved stellar systems becomes available, an equally high level in
the theoretical modelling is required.

Previous investigations of the statistical equilibrium (SE) of
neutral Ca in the Sun and Procyon (Watanabe \& Steenbock
\cite{ca1}) and in the models with effective
temperatures $\Teff$ = 4500\,K -- 6200\,K, surface gravities $\log
g$ = 4.5 -- 1.0 and metal abundance, [Fe/H] = 0 and $-$1 (Drake
\cite{Drake}) found rather small departures from LTE.
Drake (\cite{Drake}) concluded that
``overionization effects (for \ion{Ca}{i}) are not expected to
increase in severity in metal-poor stars".
 NLTE analysis of neutral Ca in the sample of 252 dwarf and
subgiant stars was performed by Idiart \& Thevenin
(\cite{th-mg}). Previous NLTE investigations for \ion{Ca}{ii}
were concerned only with the resonance lines in the Sun (we cite
only the first paper, Shine \& Linsky \cite{ca74}) and the IR
lines of multiplet $3d - 4p$ ($\lambda\,8498$, 8542 and 8662\,\AA)
in moderately metal-poor stars by J\o rgensen et al. (\cite{ca92})
for [Fe/H] $\ge -1$ and by Andretta et al.
(\cite{andretta05}) for [Fe/H] $\ge -2$.

The need for a new analysis
is motivated by two points. Firstly, a fairly extensive set of
accurate atomic data on photoionization cross-sections and
oscillator strengths was recently calculated in the Opacity
Project (OP; see Seaton et al. (\cite{OP}) for a general review).
We include in the model atom 66 terms of \ion{Ca}{i}, contrary to
16 terms in the works of Watanabe \& Steenbock (\cite{ca1}) and
Drake (\cite{Drake}), and 36 terms of \ion{Ca}{ii},
contrary to 3 terms considered by J\o rgensen et al. (\cite{ca92}) and Andretta et al. (\cite{andretta05}).
This allows to
investigate the detailed line formation of the high excitation lines of
\ion{Ca}{i} and \ion{Ca}{ii}. All our results are based on line-profile
analysis and take advantage of the advanced theory of
collisional broadening by atomic hydrogen treated recently by
Anstee \& O'Mara (\cite{omara_sp}), Barklem \& O'Mara
(\cite{omara_pd}, \cite{omara_ion}) and Barklem et al.
(\cite{omara_df}). Hereafter, these four important papers are
referred to collectively as $A\&O'M$. The second point is connected
with an extension of the metallicity range observed in stars down
to [Fe/H] = $-$5.45 (Aoki et al. \cite{frebel06}). It is highly
desirable to have an understanding of the likely influence of any
departures from LTE over this whole range. In stars
differing significantly in their metal abundance, different
subsets of Ca lines are used to calculate the
\ion{Ca}{ii}/\ion{Ca}{i} ionization equilibrium and Ca abundance.
We investigate here NLTE line formation of our extended list of
\ion{Ca}{i} and \ion{Ca}{ii} lines in the metallicity range
between [Fe/H] = 0 and $-$4.34. An exception is the resonance lines
of \ion{Ca}{i} and \ion{Ca}{ii} and the \ion{Ca}{ii} lines of
multiplet $3d - 4p$ in solar metallicity stars: their cores
and inner wings are most probably influenced by the chromospheric
temperature rise and a non-thermal and depth-dependent chromospheric
velocity field neither of which are part of the homogeneous photospheric models used in this study.

For the statistical equilibrium of atoms in cool stars, an important
issue debated for decades, from Gehren (\cite{gehren75}) to Belyaev \&
Barklem (\cite{belyaev03}), is the role of inelastic collisions with
hydrogen atoms. One of our aims is to empirically constrain the
efficiency of this type of collisions in the SE
of \ion{Ca}{i/ii} from inspection of their different influence on
the \ion{Ca}{i} and \ion{Ca}{ii} lines in the Sun, Procyon and
seven metal-poor ($-1.35 \ge$ [Fe/H] $\ge -2.43$) stars with well
determined stellar parameters and high quality observed spectra.

The paper is organized as follows. The model atom of
\ion{Ca}{i/ii} and atomic data are described in Sect.\,\ref{NLTE}.
Departures from LTE for \ion{Ca}{i} and \ion{Ca}{ii}, NLTE trends
with metallicity, effective temperature and surface gravity, and
errors of NLTE results caused by the uncertainties of atomic data
are discussed in Sect.\,\ref{effect}. In Sect.\,\ref{sun}, the solar
Ca spectrum is studied to provide the basis for further
differential analysis of stellar spectra. We determine the
absolute solar Ca abundance using four different sets of oscillator
strengths based on laboratory measurements, OP calculations, and
collected from the NIST (http://physics.nist.gov/PhysRefData) and VALD
(Kupka et al. \cite{vald}) databases and, thus, estimate the
accuracy of the available atomic data. Observations and stellar
parameters of our sample of stars are described in
Sect.\,\ref{obs}. For the selected stars, Sect.\,\ref{stars1}
investigates whether or not Ca abundances derived from the
\ion{Ca}{i} and \ion{Ca}{ii} lines agree. NLTE formation
of the \ion{Ca}{i} resonance line in the atmospheres of metal-poor
stars is tested in Sect.\,\ref{stars2}. The predictions are made
in Sect.\,\ref{extrim} for the \ion{Ca}{i}/\ion{Ca}{ii} line-strength
ratios as indicators of surface gravity for extremely
metal-poor stars. Our recommendations and conclusions are given in
Sect.\,\ref{conclusion}.

\section{The method of NLTE calculations for \ion{Ca}{i/ii}} \label{NLTE}

\subsection{The model atom} \label{model}

{\it Energy levels.} Calcium is almost completely ionized
throughout the atmosphere of stars with effective temperature
between 5000\,K and 6500\,K, with a small fraction (no more than
several parts in a thousand) of \ion{Ca}{i} in line formation
layers. Minority species are particularly sensitive to NLTE
effects because any small change of the ionization rates largely
changes their populations. In order to provide close
collisional coupling of \ion{Ca}{i} to the continuum electron
reservoir and consequently establish a realistic ionization balance between the
atomic and singly-ionized species, the atomic model for calcium has
to be fairly complete. Energy levels up to 0.17/0.67 eV below the
ionization threshold are included explicitly in our \ion{Ca}{i/ii}
model atom. Only the ground state of \ion{Ca}{iii} is considered.
The \ion{Ca}{i} levels belong to singlet and triplet terms of the
$4snl$ ($n \le 9$ and $l \le 4$), $3dnl$ ($nl$ = $4p$ and
$3d$), and $4p^2$ electronic configurations. Triplet fine
structure is neglected except for the \eu{4p}{3}{P}{\circ}{} and
\eu{3d}{3}{D}{}{} splitting. Singlet and triplet $ng$, $8f$, $9d$,
and $9f$ levels are combined into a single level due to their small
energy differences. The final model atom includes 63 levels of
\ion{Ca}{i}. The \ion{Ca}{ii} levels belong to doublet terms of
the $nl$ ($n = 4 - 9$ and $l \le 5$), $10s$, and $10p$ electronic
configurations. Fine structure splitting sub-levels are included
explicitly for the terms \ba{3d}{D}{}{}, \ba{4p}{P}{\circ}{}, and
\ba{4d}{D}{}{}. $ng$ and $nh$ ($n =$ 7, 8, and 9) levels are
combined into a single level. The final model atom contains 37
levels of \ion{Ca}{ii}.

\begin{figure}
\resizebox{88mm}{!}{\includegraphics{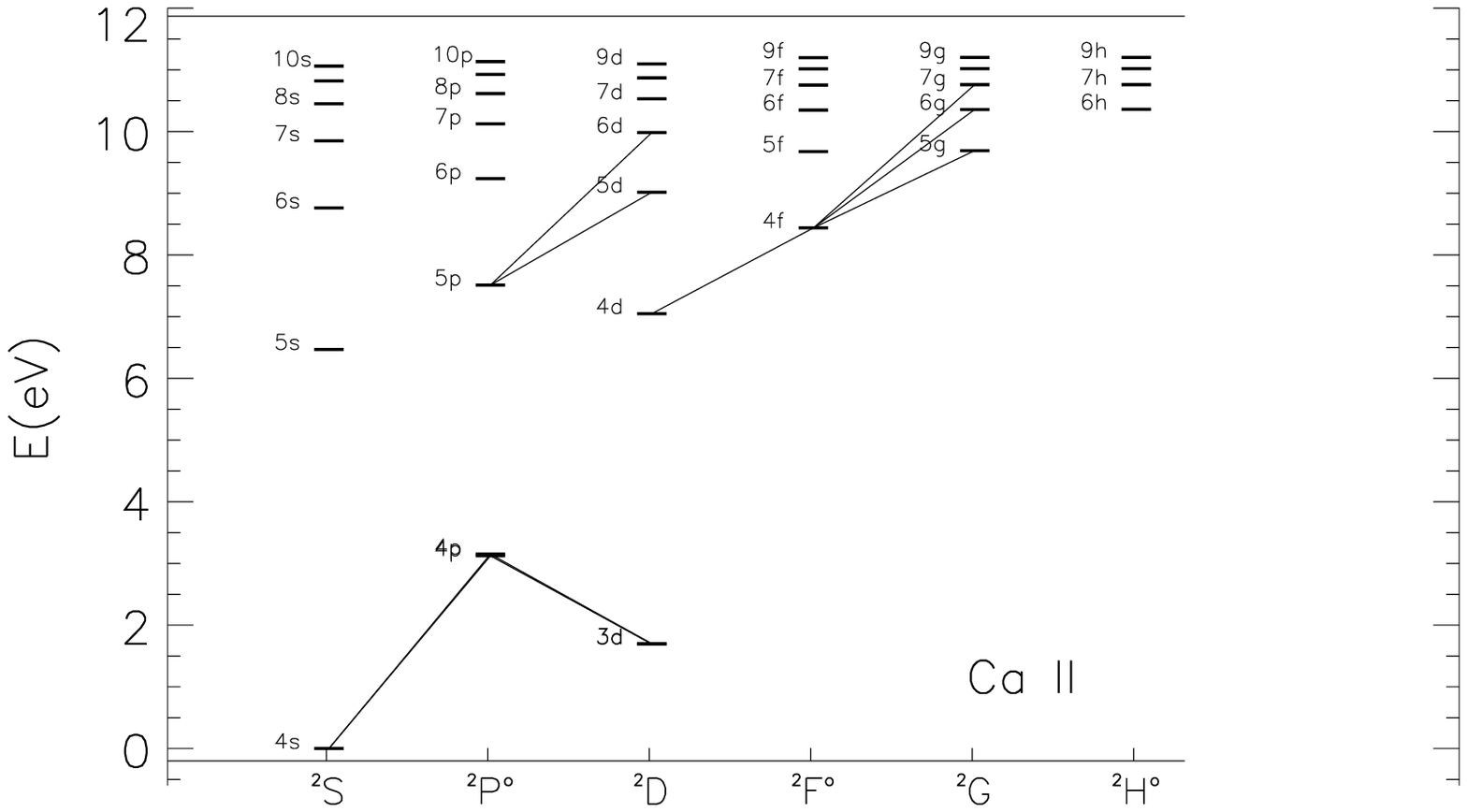}}

\vspace{-10mm} \resizebox{88mm}{!}{\includegraphics{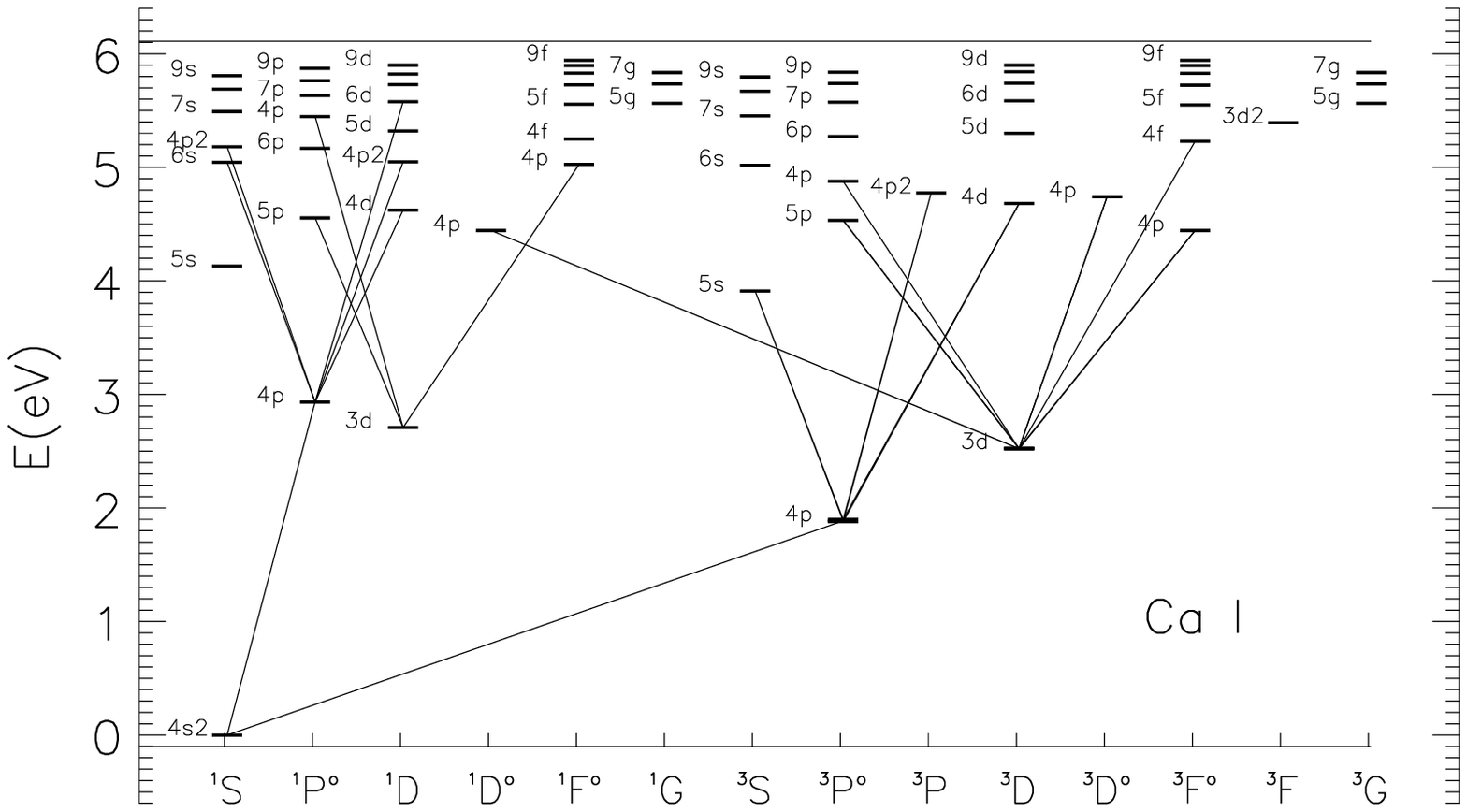}}
\vspace{-10mm}

\vspace{-3mm} \caption[]{The \ion{Ca}{i/ii} model atom. The
\ion{Ca}{i} and
 \ion{Ca}{ii} spectral lines used in Ca abundance analysis arise
 in the transitions shown as continuous lines.
}
\label{atom}
\end{figure}

The energy levels were taken from the NIST atomic spectra database
(Sugar \& Corliss \cite{ca_nist})  for all the \ion{Ca}{i}
electronic configurations with $n \le 5$ and \ion{Ca}{ii}
electronic configurations with $n \le 6$. For the remainder,
Opacity-Project data from the TOPBASE database (Cunto \& Mendoza
\cite{topbase}) are used. The corresponding term diagram is shown
in Fig. \ref{atom}.

{\it Radiative transitions.} Our NLTE calculations rely heavily on
the OP radiative data. They cover the whole range of allowed
transitions in our model atom. Currently available OP calculations
do not include the effects of spin-orbit interaction, and only
average-multiplet oscillator strengths are reported. Where
required, we decompose the $LS$ multiplet averages into their
$LSJ$ fine structure components in correspondence with their
relative line strengths. Additional radiative coupling between the
different spin systems of \ion{Ca}{i} is provided by the
transition $4s^2$ - \eu{4p}{3}{P}{\circ}{1} (Drozdowski et al.
\cite{ca6572}) and by all intercombination transitions listed by
Smith \& Raggett (\cite{ca_fij}). The radiative transfer
equation is solved, in total, for 421 line transitions in
\ion{Ca}{i} and 213 transitions in \ion{Ca}{ii}.

We compared OP $f_{ij}$ for \ion{Ca}{i} with the experimental
values from Shabanova (\cite{ca_np}), Smith \& Gallagher
(\cite{ca4226}), Smith \& O'Neil (\cite{ca_75}), Smith \& Raggett
(\cite{ca_fij}) and Smith (\cite{ca_88}), in total for 24
multiplets with $\log gf$ ranging between 0.73 and $-$2.25. The
difference $\Delta \log gf (laboratory - {\rm OP})$ shows no
correlation with the multiplet strength, and the mean value equals
$-0.07 \pm 0.28$. For \ion{Ca}{ii}, measured oscillator strengths
are available only for the multiplets $4s - 4p$ and $3d - 4p$
(Theodosiou \cite{ca3933}). They agree within 12\%\ with the
corresponding OP values. OP data for the remaining \ion{Ca}{i} and
\ion{Ca}{ii} transitions were compared with calculations of Kurucz
(\cite{cdrom18}). For the 151 strongest ($\log gf$(OP) $> -2.5$)
\ion{Ca}{i} multiplets, the mean difference $\Delta \log gf$
(Kurucz -- OP) = $-$0.03$\pm$0.93. Kurucz calculations predict, on
average, lower transition probabilities compared to OP data for
the 58 strongest ($\log gf$(OP) $> -2.5$) \ion{Ca}{ii} multiplets,
and the mean difference $\Delta \log gf$ (Kurucz - OP) =
$-0.17\pm$0.86. With respect to systematical errors OP data for
\ion{Ca}{i} and \ion{Ca}{ii} transitions are expected to be
accurate within ~15\%. Additional arguments for the validity of
this statement come from the analysis of solar Ca spectrum (see
Sect.\,\ref{sun}).

Photoionization from all energy levels is treated utilizing OP
cross-sections as available through the TOPBASE database.
Cross-sections for the combined $ngh$ levels in \ion{Ca}{ii} are
assumed to be equal to that for the corresponding $ng$ levels.

{\it Collisional transitions.} In our calculations we take into
account inelastic collisions with electrons and hydrogen atoms
leading to both excitation and ionization.
Drawin's (\cite{D68}) formula as described by Steenbock \&
Holweger (\cite{hyd}) is widely used to calculate hydrogenic
collisions, and it suggests that their influence is comparable
to electron impact. Recently it was shown both experimentally
(Belyaev et al. \cite{belyaev99}, for Na~I) and theoretically
(Belyaev \& Barklem \cite{belyaev03}, for Li~I) that Drawin's
formula overestimates the collision rate of the resonance
transitions by several orders of magnitude. However, for SE
calculations, estimates are required for transitions between all
states which might affect the population of the states of
interest. In this study, we constrain the efficiency of
hydrogenic collisions empirically. It is represented by a scaling factor
\kH\ applied to Steenbock \& Holweger's formula. The
cross-sections calculated using this formula were multiplied by
\kH\ = 0 (no hydrogenic collisions), 0.01, 0.1, and 1 in order to
agree Ca abundances derived from the \ion{Ca}{i} and \ion{Ca}{ii}
lines in the Sun and selected stars.

For electron impact excitation, detailed results are available
from the $R-$matrix calculations of Samson \& Berrington
(\cite{samson}) for the transitions from the ground state to all the
excited levels up to \eu{3d4p}{1}{P}{\circ}{} in \ion{Ca}{i} and from
calculations of Burgess et al. (\cite{burgess}) in a non-exchange
distorted wave approximation for the transitions between the
lowest seven terms (up to \ba{4f}{F}{\circ}{}) in \ion{Ca}{ii}.
Uncertainties in data obtained by the $R-$matrix method are
typically on the order of a few 10\%, and Samson \& Berrington
note excellent agreement of their results with experimentally
derived rates for 4$s^2$ - \eu{4s4p}{1}{P}{\circ}{}. For the
remaining bulk of the transitions, approximate formulae must be
used. Basically, we apply the impact parameter method (IPM) of
Seaton (\cite{Seaton}) to the allowed transitions and the Allen's
formula (\cite{Allen}) with a collision strength of 1.0 to the
optically forbidden transitions. Electronic collision rates based
on Samson \& Berrington and Burgess et al. data where they are
available, IPM data and the Allen's formula for the remaining
transitions are referred to below as the standard collisional
recipe. For test purposes (see Sect.\,\ref{error}), we treat the
allowed transitions absent in the Samson \& Berrington and Burgess
et al. papers using the van Regemorter's formula (\cite{Reg}).

 Electron impact ionization cross-sections are calculated applying
 the formula of Seaton (\cite{seat_i}) with threshold photoionization
cross-sections from the OP data.

\subsection{Model atmospheres. Programs and methodical notes}

Calcium is assumed to be a trace element because its contribution
to the continuous opacity and the reservoir of free electrons is by,
at least, one order of magnitude smaller than that from the more
abundant elements Mg, Si, and Fe. Thus, we obtain statistical
equilibrium populations for \ion{Ca}{i/ii} while keeping the
atmospheric structure fixed. All calculations are performed with
plane-parallel, homogeneous, LTE, and blanketed model atmospheres
computed with the MAFAGS code (Fuhrmann et al. \cite{Fuhr1}).

We use a revised version of the DETAIL program (Butler \& Giddings
 \cite{detail}) based on the accelerated lambda iteration
following the efficient method described by Rybicki \&
Hummer (\cite{rh91}, \cite{rh92}) in order to solve the coupled
radiative transfer and statistical equilibrium equations. All
bound-bound ($b-b$) and bound-free ($b-f$) transitions of
\ion{Ca}{i} and \ion{Ca}{ii} are explicitly taken into account in
the SE calculations. The 15 strongest $b-b$ transitions in \ion{Ca}{i}
and all $LSJ$ transitions $4s - 4p$ and $3d - 4p$ in \ion{Ca}{ii}
are treated using Voigt profile. Microturbulence is accounted for
by inclusion of an additional term in the Doppler width. The van der
Waals damping parameters based on the Anstee \& O'Mara's
(\cite{omara_sp}) theory are taken from the VALD database (Kupka
et al. \cite{vald}). The remaining $b-b$ transitions are treated
using depth-dependent Doppler profiles.

In addition to the continuous background opacity, the line opacity
introduced by both \ion{H}{i} and metal lines is taken into
account by explicitly including it in solving the radiation
transfer. The metal line list has been extracted from Kurucz'
(\cite{cdrom18}) compilation and contains about 650 000 atomic
and molecular lines between 1300\,\AA\, and 300 000\,\AA. Ca lines are
excluded from the background.

The obtained departure coefficients are then used to compute the
synthetic line profiles via the SIU program
(www.usm.uni-muenchen.de/people/reetz/siu.html). In this step of
calculations, Voigt profile functions are adopted and the same
microturbulence value $\Vmic$ as in DETAIL is applied. Oscillator
strengths and van der Waals damping constants of the \ion{Ca}{i}
and \ion{Ca}{ii} lines and their accuracy are discussed in
Sect.\,\ref{sun}.

Ca is represented in the solar system matter by several isotopes
with the isotope abundance ratio \iso{40}{Ca} : \iso{42}{Ca} :
\iso{43}{Ca} : \iso{44}{Ca} : \iso{46}{Ca} : \iso{48}{Ca} = 96.9 :
0.647 : 0.135 : 2.09 : 0.004 : 0.187 (Anders \& Grevesse
\cite{AG}). In this study, we account for isotope structure of the
\ion{Ca}{ii} $\lambda\,8498$ line\footnote{Of the three IR triplet
lines we look at \ion{Ca}{ii} $\lambda\,8498$ only, as {\em a)} it
is the weakest of the three and {\em b)} it is least perturbed by
Paschen lines.} with the isotope shifts measured by
N\"ortersh\"auser et al. (\cite{ca_is}). Wavelengths and the
product of $gf$ and solar fractional isotope abundance $\epsilon$
are given in Table\,\ref{iso8498} for each isotope component. The
oscillator strength is taken from measurements of Theodosiou
(\cite{ca3933}), $\log gf (\lambda\,8498) = -1.416$. Isotope
shifts are much smaller ($<$10\,m\AA) for the resonance lines in
\ion{Ca}{i} and \ion{Ca}{ii} (Lucas et al. \cite{ca_is_res}) and
we treat them as single lines. No data is available for the
remaining Ca lines. We neglect the hyperfine structure of Ca lines
due to very low fractional abundance of the only odd isotope
\iso{43}{Ca} (0.135\%).

\begin{table}
\caption{Atomic data for the isotopic components of \ion{Ca}{ii}
$\lambda\,8498$. The fractional isotope abundances $\epsilon$
correspond to the solar system matter}\label{iso8498}
\begin{center}
\begin{tabular}{ccc}
\hline \noalign{\smallskip}
 \ $\lambda,\AA$ & isotope  & $\log gf\epsilon$ \\
\noalign{\smallskip} \hline \noalign{\smallskip}
  8498.023 & 40 &  $-$1.43   \\
  8498.079 & 42 &  $-$3.60   \\
  8498.106 & 43 &  $-$4.29   \\
  8498.131 & 44 &  $-$3.10   \\
  8498.179 & 46 &  $-$5.81   \\
  8498.223 & 48 &  $-$4.14   \\
\noalign{\smallskip} \hline
\end{tabular}
\end{center}
\end{table}

\section{NLTE effects as a function of stellar parameters}\label{effect}

In this section, we investigate NLTE effects for \ion{Ca}{i/ii} in
the following range of stellar parameters: $\Teff$ between 5000K and 6000K,
$\log g =$ 3.0 and 4.0, and [Fe/H] between 0 and $-$3. For the
models with [Fe/H] = 0, the Ca abundance was assumed to follow
the global metallicity. Ca enhancement with [Ca/Fe] = 0.4 was
assumed for the metal-poor models. Statistical
equilibrium was computed using the standard recipe for electronic
collisions and no hydrogenic collisions were taken into account
(\kH\ = 0).

\subsection{Statistical equilibrium of \ion{Ca}{i} and NLTE effects for spectral lines}

\begin{figure}
\hbox{
\resizebox{88mm}{!}{\rotatebox{0}{\includegraphics{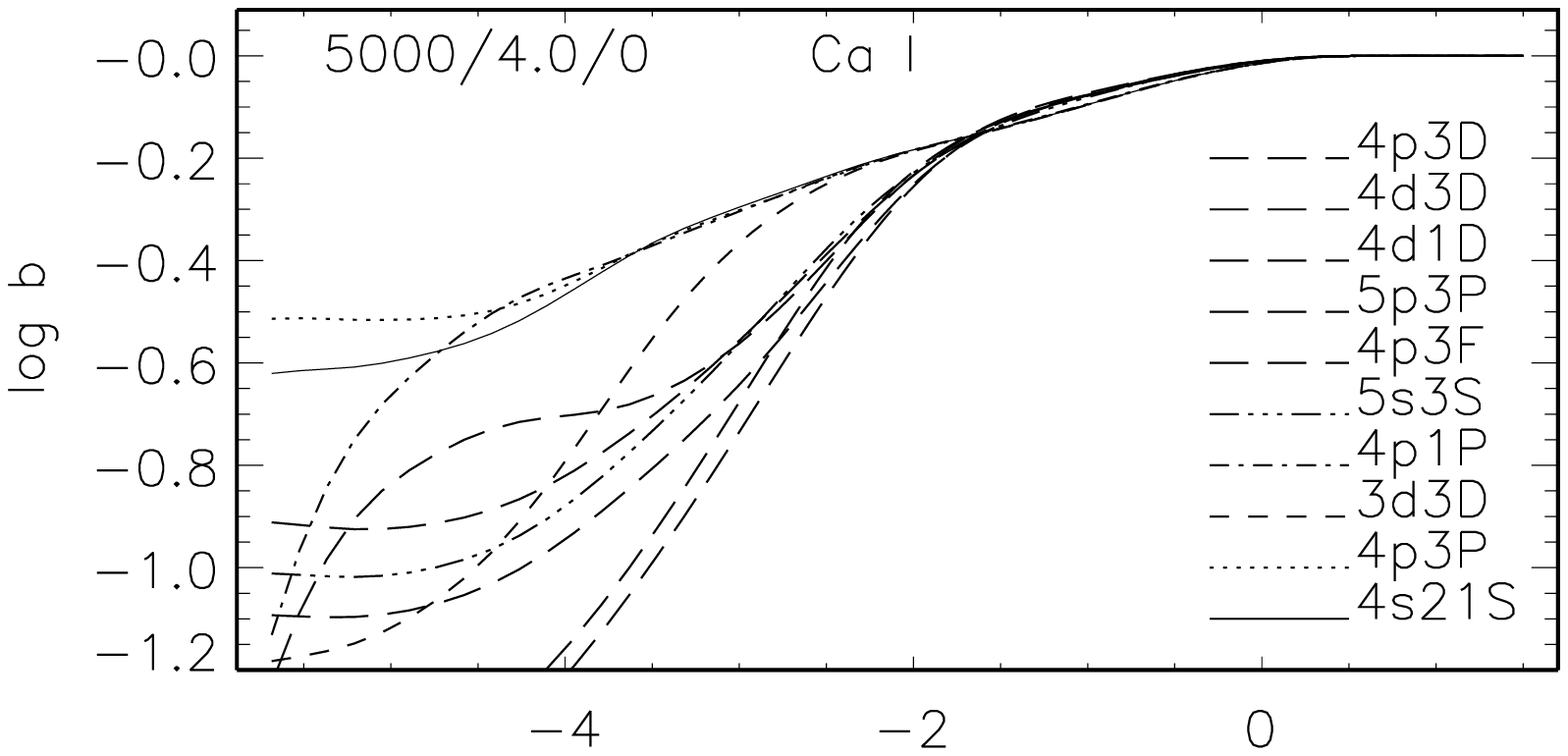}}
\hspace{-8mm} \rotatebox{0}{\includegraphics{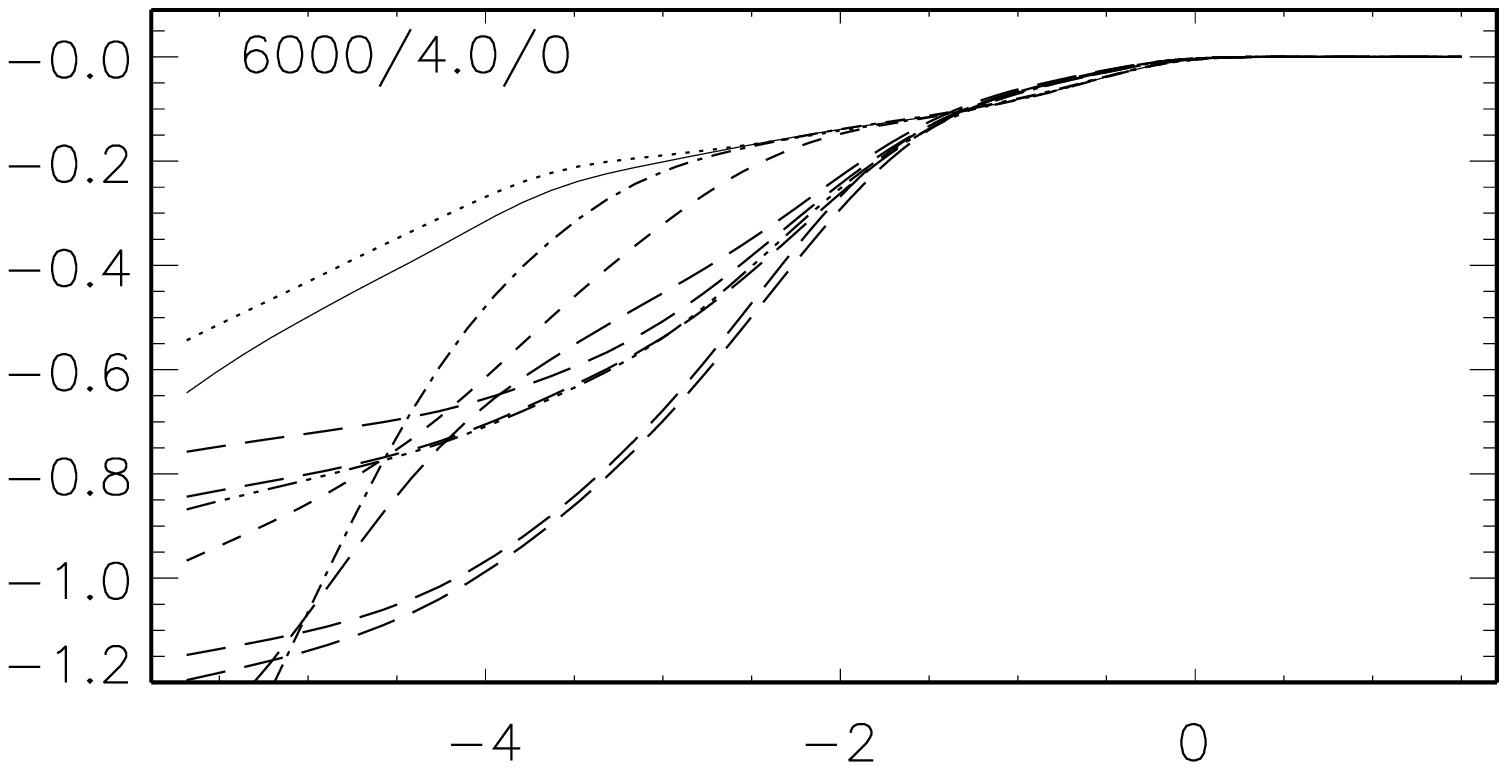}}}}
\vspace{-5mm}
\hbox{
\vspace*{-3mm}
\resizebox{88mm}{!}{\rotatebox{0}{\includegraphics{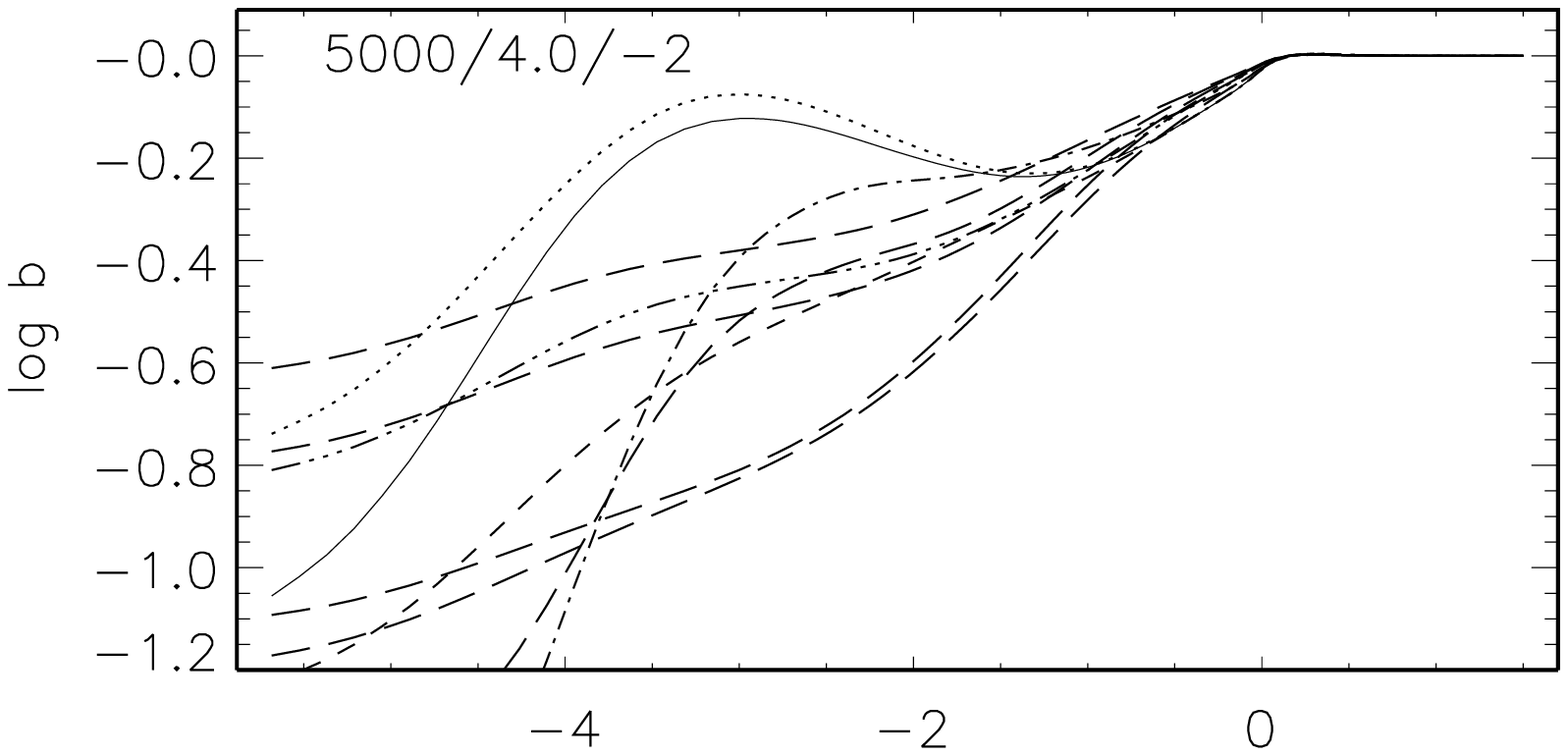}}
\hspace{-8mm} \rotatebox{0}{\includegraphics{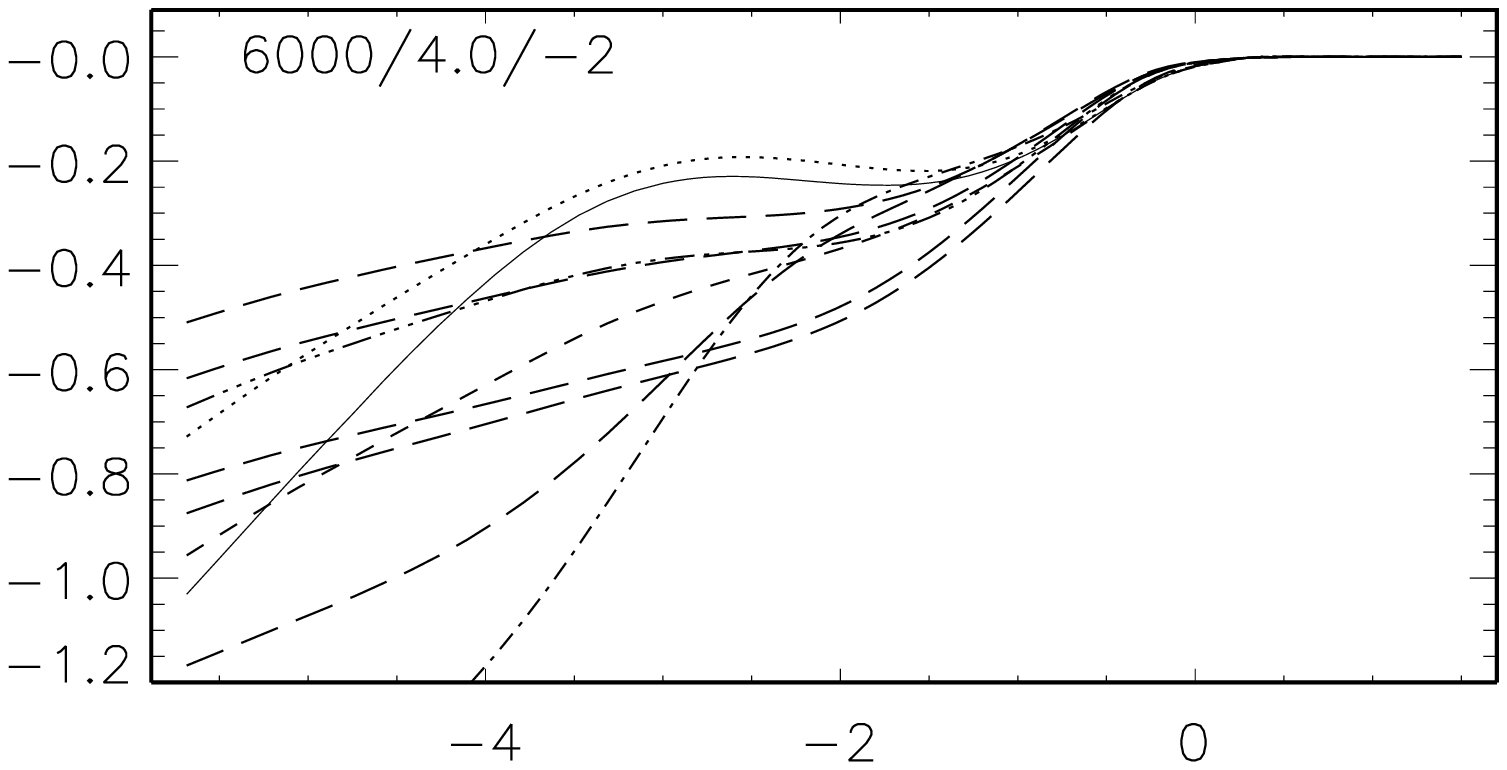}}}}
\hbox{
\vspace{-5mm}
\resizebox{88mm}{!}{\rotatebox{0}{\includegraphics{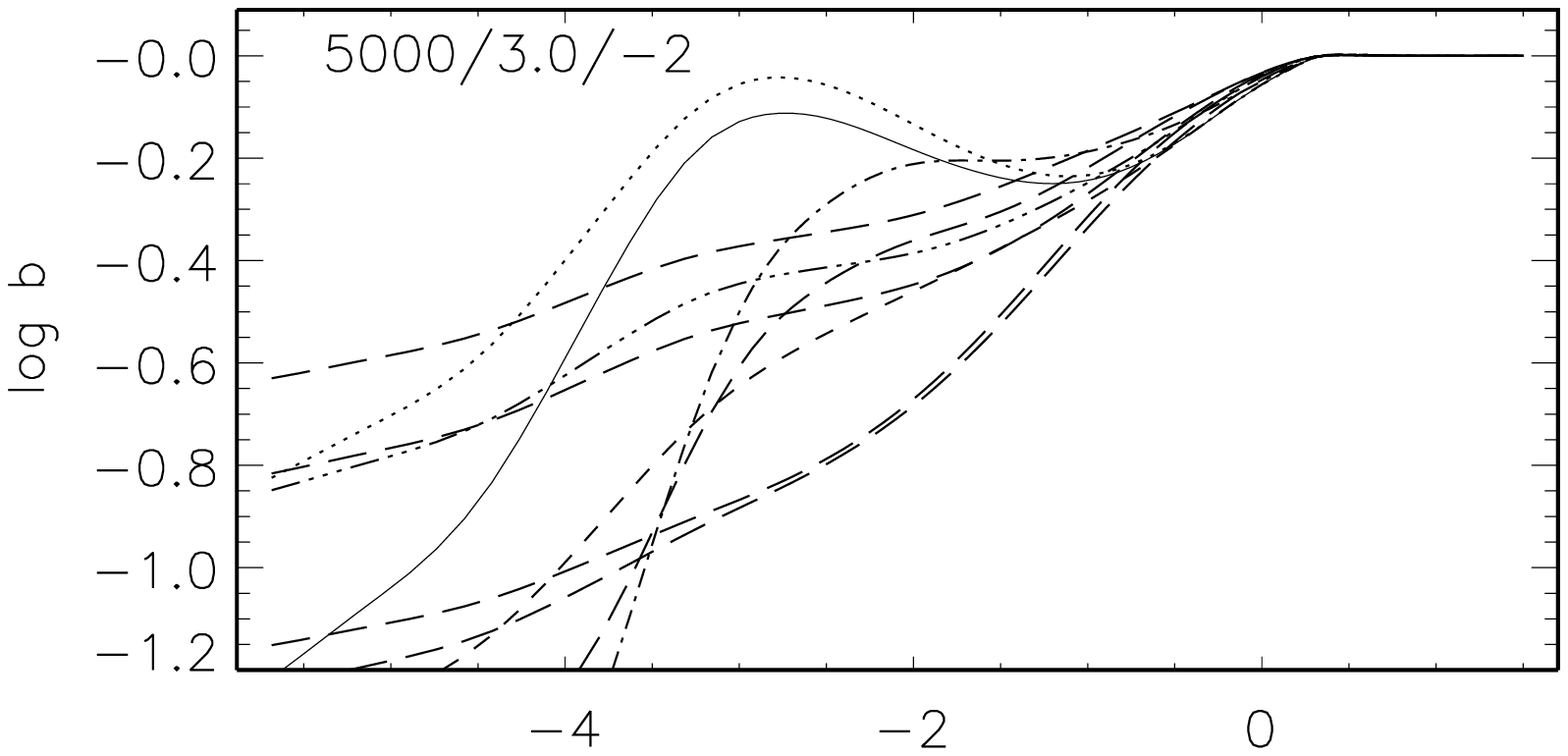}}
\hspace{-8mm} \rotatebox{0}{\includegraphics{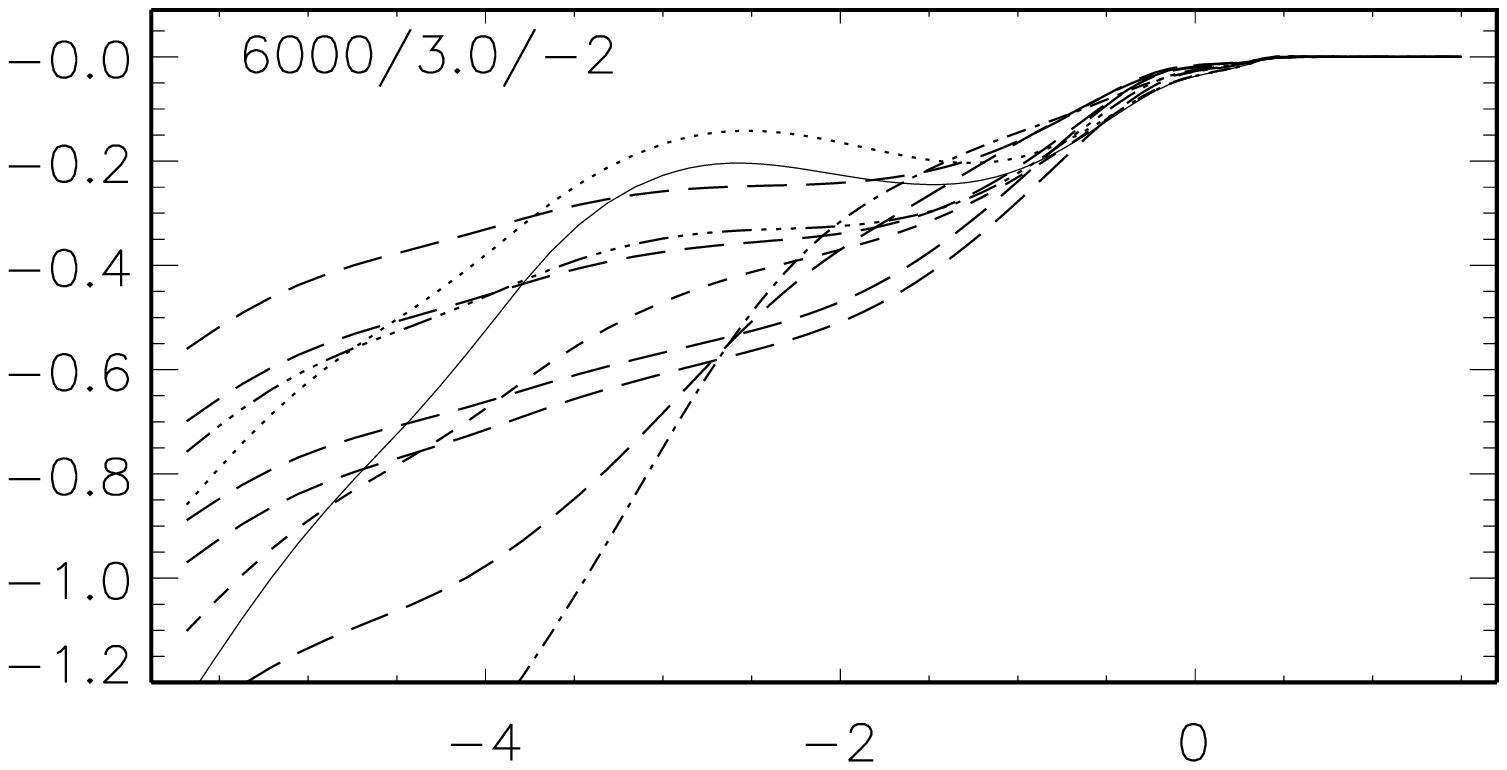}}}}
\hbox{
\vspace{-5mm}
\resizebox{88mm}{!}{\rotatebox{0}{\includegraphics{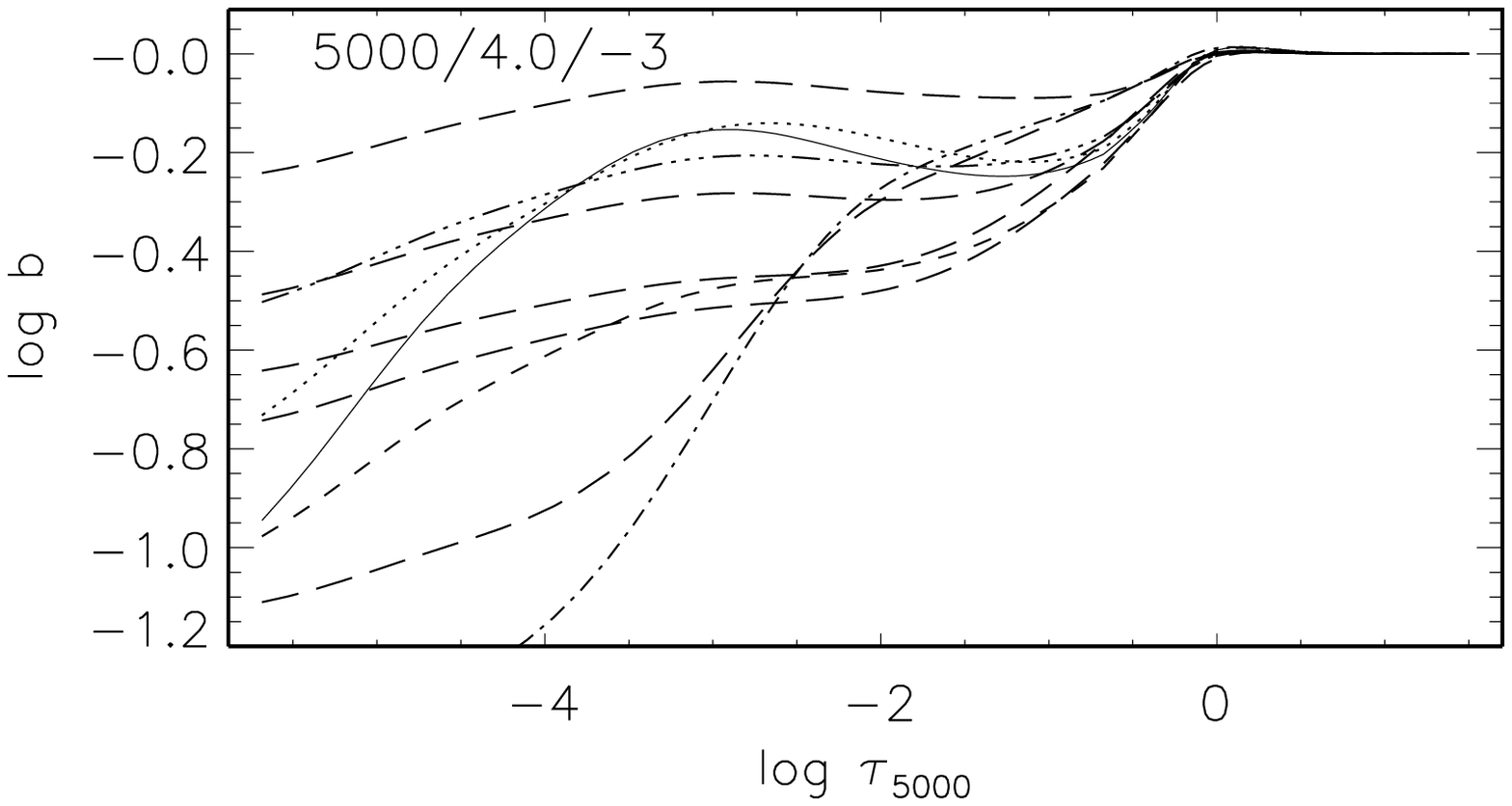}}
\hspace{-8mm} \rotatebox{0}{\includegraphics{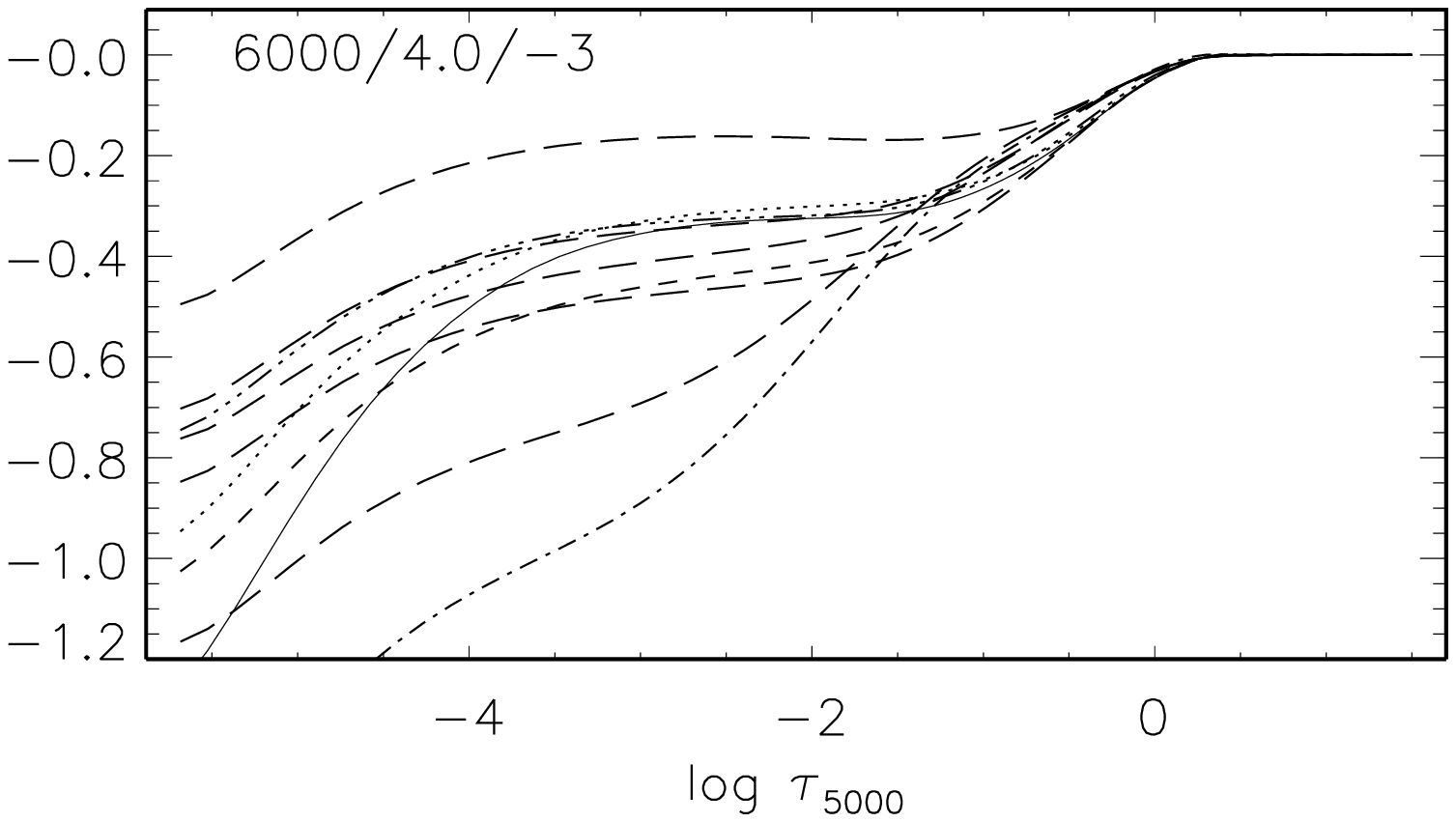}}}}

\vspace{5mm} \caption[]{Departure coefficients $\log b$ as a
function of $\log \tau_{5000}$ for selected levels of \ion{Ca}{i},
plotted for selected models of the grid. For each term of
\eu{3d}{3}{D}{}{} and \eu{4p}{3}{P}{\circ}{}, departure
coefficients of the fine structure splitting sublevels are obtained to be
close together and only the lowest sublevels, \eu{3d}{3}{D}{}{1}
and \eu{4p}{3}{P}{\circ}{0}, are shown. Stellar parameters are
quoted in each panel} \label{bfca1}
\end{figure}

Our calculations indicate a general behaviour of departure
coefficients, $b_i = n_i^{\rm NLTE}/n_i^{\rm LTE}$, independent of
effective temperature, surface gravity and metallicity. Here,
$n_i^{\rm NLTE}$ and $n_i^{\rm LTE}$ are the statistical
equilibrium and thermal (Saha-Boltzmann) number densities,
respectively. Departure coefficients of the
important levels of \ion{Ca}{i} are plotted in Fig.\,\ref{bfca1}
for selected models of our grid.
It can be seen, all levels of \ion{Ca}{i} are underpopulated
in the atmospheric layers above $\log \tau_{5000} = 0$.
Overionization is caused by superthermal radiation of
non-local origin below the thresholds of low excitation levels of
\ion{Ca}{i}. In atmospheres with solar or mildly deficient Ca
abundance, [Ca/H] $\ge -1$, the most important levels are
\eu{4p}{1}{P}{\circ}{}, \eu{3d}{3}{D}{}{}, and
\eu{4p}{3}{P}{\circ}{} with the thresholds at 3898~\AA, 3450~\AA,
and 2930~\AA, correspondingly. At the lower metallicity and Ca
abundance, depopulation processes in \ion{Ca}{i} are dominated by
enhanced ionization of the ground state due to a reduction of the
continuous absorption coefficient below the threshold of this
level at 2028\AA.

 Statistical equilibrium of \ion{Ca}{i} in the metallicity
range between 0 and --1 was discussed in detail by Drake
(\cite{Drake}). Referring to the main continuous opacity sources
in the near ultraviolet wavelength regime, due to the H$^-$ ions
and excited neutral H atoms, Drake has shown that overionization
effects decrease with increasing model temperature, decreasing
surface gravity, and decreasing Ca abundance. Our calculations for
the stellar parameters overlapping with that of Drake support
qualitatively his conclusions. However, quantitatively, in our work,
overionization effects are found to be smaller. The
explanation lies with our extended model atom that contains the
energy levels up to 0.17\,eV below the ionization threshold,
contrary to 1.06\,eV in Drake's paper. Overionization effects on
\ion{Ca}{i} are partially cancelled due to collisional coupling of
the \ion{Ca}{i} high excitation levels to the ground state of the
majority species, \ion{Ca}{ii}, that keeps thermodynamic
equilibrium population. Our model atom provides this closer
coupling and in this respect is more realistic than the one by Drake.
The smaller departures from LTE
for level populations result in weaker NLTE effects for
spectral lines and smaller differences between derived NLTE and LTE
Ca abundance, $\Delta_{\rm NLTE} = \eps{NLTE}-\eps{LTE}$. We refer
to $\Delta_{\rm NLTE}$ as the NLTE abundance correction. For
example, for atmospheric parameters $\Teff =$ 5800K, $\log g =$
4.5, [Fe/H] = 0, Drake gives nearly equal values $\Delta_{\rm
NLTE} =$ 0.11\,dex for $\lambda\,5261$ (multiplet 22),
$\lambda\,6455$ (multiplet 19), and $\lambda\,6166$ (multiplet
20). The corresponding values from our calculations are 0.04\,dex,
0.05\,dex, and 0.07\,dex.

Contrary to the models with [Fe/H] $\ge -1$, overionization
effects increase with increasing model temperature and decreasing
global metallicity, when ones goes to the lower metallicity and Ca
abundance, [Fe/H] and [Ca/H] $< -1$. The explanation lies with a
behavior of the continuous absorption coefficient below 2028\AA\
that is, in metal-poor atmospheres, mainly due to quasi-molecular
hydrogen absorption (Doyle \cite{Doyle}), the H$^-$ ion, and the
H$^+_2$ ion. The contribution of the H$^-$ ion increases with
increasing effective temperature and decreases with decreasing
metallicity, while the contribution of quasi-molecular hydrogen
absorption changes in the opposite direction. Their competition
results in effects seen in Fig.\,\ref{bfca1}.

\begin{figure}
\resizebox{88mm}{!}{\includegraphics{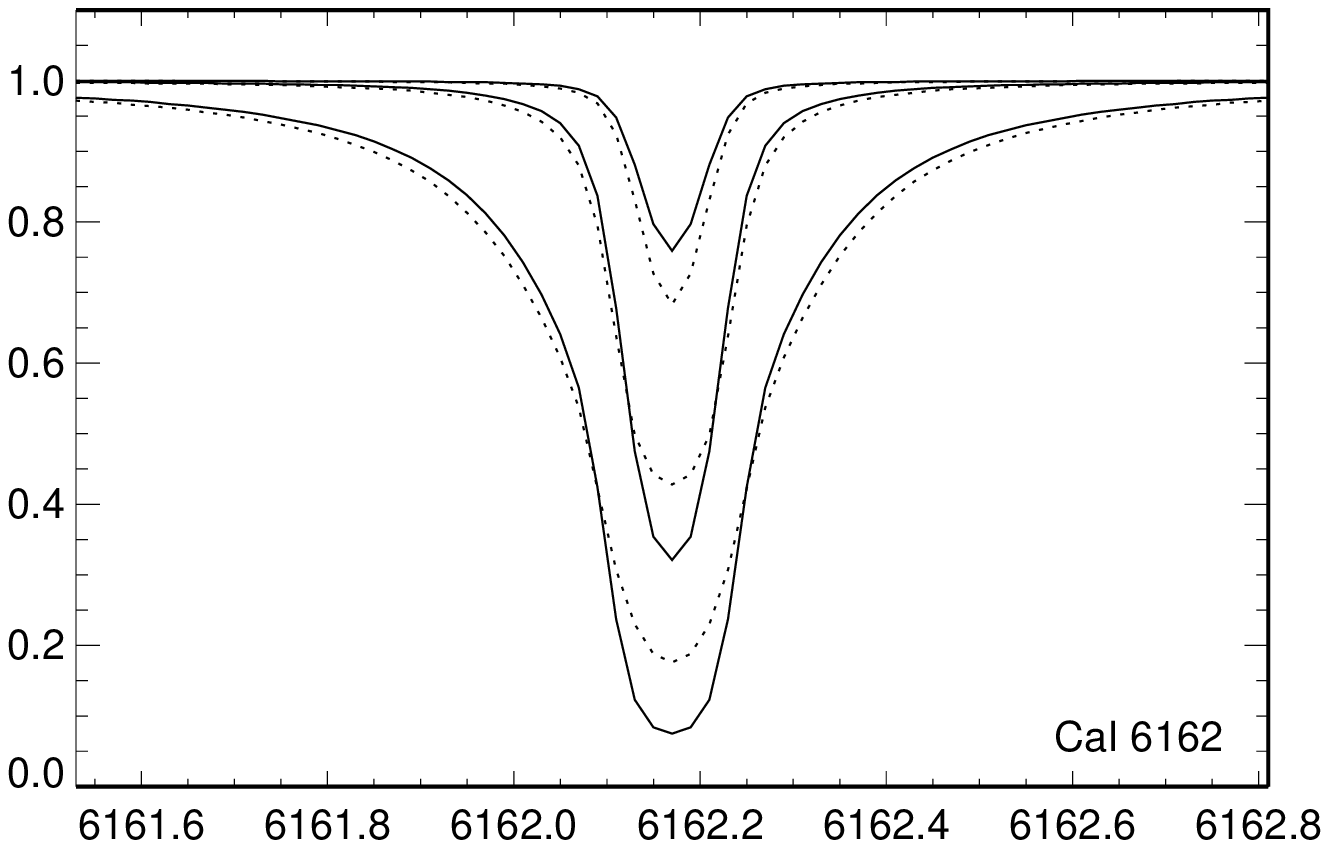}}

\vspace{-5mm}
\resizebox{88mm}{!}{\includegraphics{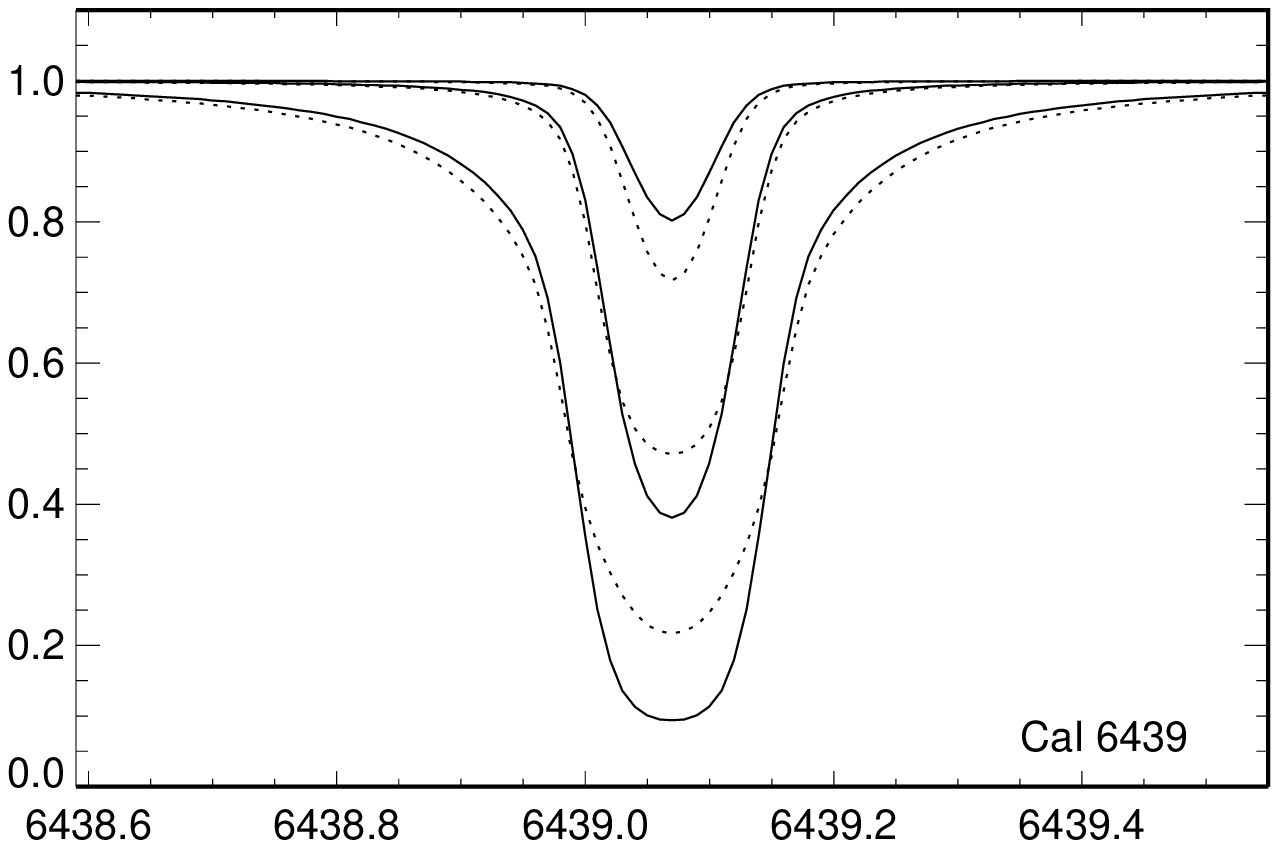}}

\vspace{-5mm}
\resizebox{88mm}{!}{\includegraphics{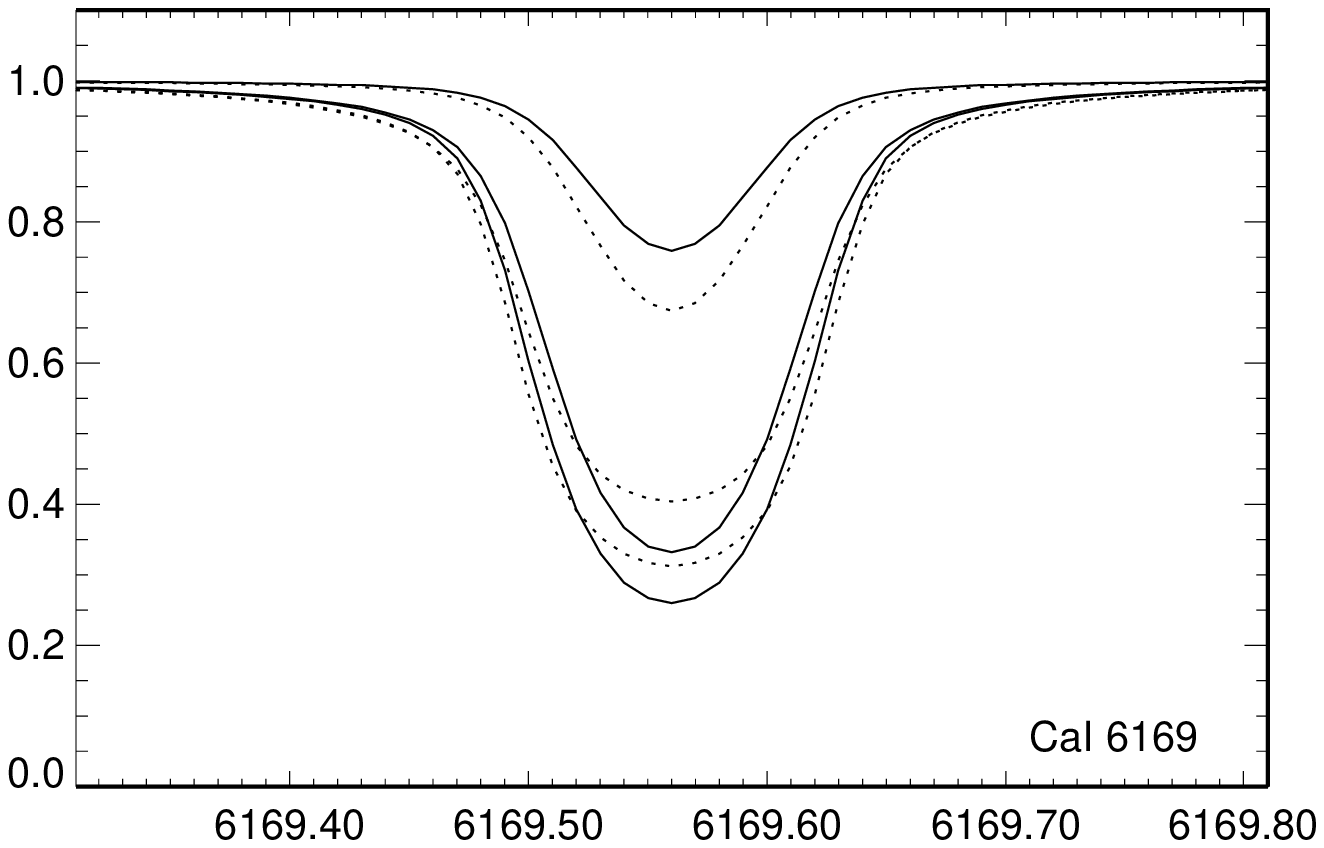}}

\vspace{-3mm}
\caption[]{NLTE (continuous line) and LTE (dotted line)
theoretical profiles of the selected \ion{Ca}{i} lines for the
models with the same $\Teff =$ 5500K and $\log g = 4.0$ but with
different metallicities. For the $\lambda\,6162$ and
$\lambda\,6439$ lines, [Fe/H] = 0, --2, and --3 from bottom to
top; for $\lambda\,6169$, [Fe/H] = 0, --1, and --2}
\label{ca1_line}
\end{figure}

NLTE effects on spectral lines are illustrated in
Fig.\,\ref{ca1_line} and Table\,\ref{corr_full}.
Fig.\,\ref{ca1_line} shows NLTE and LTE profiles of three
representative lines of neutral calcium computed for the models
with different global metal abundance. These are the low
excitation line $\lambda\,6162$ which is the strongest among all
subordinate lines, $\lambda\,6439$ with the largest oscillator
strength, $f_{ij} = 0.390$, but relatively small van der Waals
damping constant, $\log C_6 = -31.58$, which is lower by 1.28\,dex
than that of $\lambda\,6162$, and the intermediate strength line
$\lambda\,6169.5$. Table\,\ref{corr_full} presents NLTE abundance
corrections, $\Delta_{\rm NLTE}$, for 16 \ion{Ca}{i} lines. Three
other \ion{Ca}{i} lines, $\lambda\,6161$, $\lambda\,6169.0$, and
$\lambda\,6122$, have $\Delta_{\rm NLTE}$ close to those of
$\lambda\,6166$, $\lambda\,6169.5$, and $\lambda\,6162$,
respectively. NLTE and LTE line profiles and equivalent widths
were calculated with the laboratory oscillator strengths and $C_6$
values given in Table\,\ref{line_list}. Microturbulence value,
$\Vmic =$ 1\,\kms, was adopted for all the models. Departures from
LTE for spectral lines can be understood by considering the depths
of formation of various parts of the lines and by inspecting the
level departure coefficients and line source functions.
In the metallicity regime, [Fe/H] $\ge -1$, our results support the
conclusions of Drake (\cite{Drake}): for each
investigated line, NLTE effects lead to enhanced absorption in the
line core and depleted absorption in the line wings (Fig.\,\ref{ca1_line}).
This can be understood
because the line wings are formed in deep layers where overionization
depopulates all \ion{Ca}{i} levels, but the line cores are
formed at small depths, above $\log\tau_{5000} = -2$, where the upper
levels of the transitions
are underpopulated to a more extent than are the lower levels due
to photon losses in the line wings. The most prominent example is
a steep decrease of $b(\eu{4p}{1}{P}{\circ}{})$ above
$\log\tau_{5000} = -2$ in all metal-poor models
(Fig.\,\ref{bfca1}). The line source function drops below the
Planck function at these depths resulting in the enhanced
absorption in the line cores. Combined effect on the line strength is
that the NLTE abundance correction is small in most cases. Its
sign and value depends on the contributions of the line core and
wings to the overall line strength. NLTE corrections tend toward negative
values with increasing $\Teff$ and decreasing $\log g$. They are
more negative for the lines of multiplet 18, $\lambda\,6439$,
$\lambda\,6471$, $\lambda\,6493$, and $\lambda\,6499$
(Table\,\ref{corr_full}). All these
trends reflect the  behaviour of the van der Waals broadened line
wings.

Contrary to the solar metallicity models, the energy levels
become weakly coupled far inside the metal-poor atmospheres with [Fe/H] $<
-1$ due to
deficient collisions (Fig.\,\ref{bfca1}). For each model, at the
depths where the weak lines are formed, the upper levels are all
depleted to a lesser extent relative to their LTE populations than
are the lower levels. The lines are weaker relative to their LTE
strengths not only due to the general overionization but also due
to $b_u/b_l > 1$ resulting in the line source function $S_{lu}
\simeq b_u/b_l\,B_\nu > B_\nu$ and the depleted line absorption.
Here, $b_u$ and $b_l$ are the departure coefficients for the upper
and lower levels of the transition. For example, this is valid for
$\lambda\,6169.5$ in the model with $\Teff$ = 5500\,K, $\log g$ =
4.0, [Fe/H] = --2 and $\lambda\,6162$ and $\lambda\,6439$ in the
model 5500\,K / 4.0 / --3. The corresponding NLTE and LTE profiles
are shown in Fig.\,\ref{ca1_line}. If the line is strong for a given
set of stellar parameters (e.g.,
$\lambda\,6162$ and $\lambda\,6439$ for the models with [Fe/H] =
--2), NLTE effects on line profiles are similar to that for the
models with [Fe/H] $\ge -1$.

\begin{table*}
\caption{NLTE abundance corrections (dex) for the \ion{Ca}{i} and
\ion{Ca}{ii} lines depending on effective temperature, surface
gravity, and metallicity. Everywhere $\Vmic$ = 1\,\kms. No
hydrogenic collisions were taken into account (\kH\ = 0) in SE
calculations. Blanks denote theoretical NLTE equivalent widths
below 5 m\AA} \label{corr_full} \tabcolsep1.2mm \tiny
\begin{center}
\begin{tabular}{l|rrrrrrrrrrrrrrrr|rrr}
\hline\noalign{\smallskip}
 $\Teff$/$\log g$/ & \multicolumn{16}{|c|}{\ion{Ca}{i}} & \multicolumn{3}{c}{\ion{Ca}{ii}} \\
 \ \ [Fe/H] &    4226 &  4425 &  5261 &  5349 &  5512 &  5588 &  5590 &  5857 &  5867 &  6162 &  6166 &  6169 &  6439 &  6471 &  6493 &  6499 &  8498 &  8248 &  8927 \\
\hline\noalign{\smallskip}
5780/4.44/ 0 &    0.07 &  0.04 &  0.04 &  0.01 &  0.03 &  0.01 &  0.03 &  0.00 &  0.06 &  0.01 &  0.07 &  0.02 & $-$0.04 & $-$0.04 & $-$0.04 & $-$0.01 & $-$0.02 & $-$0.11 & $-$0.19 \\
5000/3/ 0 &    0.11 &  0.08 &  0.11 &  0.04 &  0.00 &  0.07 &  0.12 &  0.00 &  0.04 &  0.07 &  0.12 &  0.05 &  0.04 &  0.00 &  0.05 &  0.04 & $-$0.02 & $-$0.12 & $-$0.13 \\
5000/3/$-$2 &    0.24 &  0.17 &  0.35 &  0.35 &  0.10 &  0.07 &  0.26 &  0.09 &  0.08 &  0.00 &  0.30 &  0.24 & $-$0.04 &  0.14 &  0.17 &  0.20 & $-$0.08 &      & $-$0.11 \\
5000/3/$-$3 &    0.24 &  0.32 &      &      &      &  0.32 &      &  0.20 &      &  0.17 &      &      &  0.27 &      &  0.30 &      & $-$0.13 &      &      \\
5000/4/ 0 &    0.09 &  0.08 &  0.08 &  0.05 &  0.03 &  0.07 &  0.08 &  0.04 &  0.08 &  0.07 &  0.09 &  0.06 &  0.06 &  0.01 &  0.05 &  0.03 & $-$0.02 & $-$0.07 & $-$0.14 \\
5000/4/$-$1 &    0.14 &  0.12 &  0.13 &  0.07 &  0.06 &  0.08 &  0.11 &  0.03 &  0.11 &  0.07 &  0.16 &  0.08 &  0.01 &  0.00 &  0.02 &  0.04 & $-$0.03 & $-$0.05 & $-$0.13 \\
5000/4/$-$2 &    0.21 &  0.18 &  0.29 &  0.29 &  0.09 &  0.10 &  0.23 &  0.10 &  0.09 &  0.06 &  0.26 &  0.20 &  0.00 &  0.12 &  0.14 &  0.16 & $-$0.06 &      & $-$0.08 \\
5000/4/$-$3 &    0.23 &  0.25 &      &      &      &  0.27 &      &  0.12 &      &  0.16 &      &      &  0.22 &      &  0.24 &      & $-$0.16 &      &      \\
5500/3/ 0 &    0.10 &  0.05 &  0.07 &  0.00 & $-$0.03 & $-$0.02 &  0.06 & $-$0.07 &  0.02 & $-$0.02 &  0.10 & $-$0.02 & $-$0.11 & $-$0.05 & $-$0.06 &  0.00 & $-$0.02 & $-$0.20 & $-$0.22 \\
5500/3/$-$2 &    0.20 &  0.19 &  0.32 &  0.27 &  0.12 &  0.10 &  0.25 &  0.13 &  0.11 & $-$0.04 &  0.28 &  0.23 & $-$0.02 &  0.14 &  0.19 &  0.18 & $-$0.11 & $-$0.02 & $-$0.16 \\
5500/3/$-$3 &    0.12 &  0.29 &      &      &      &  0.29 &      &  0.24 &      &  0.18 &      &      &  0.24 &      &  0.26 &      & $-$0.29 &      &      \\
5500/4/ 0 &    0.08 &  0.05 &  0.05 &  0.01 &  0.01 &  0.03 &  0.05 & $-$0.01 &  0.05 &  0.03 &  0.08 &  0.02 & $-$0.02 & $-$0.04 & $-$0.02 &  0.00 & $-$0.02 & $-$0.10 & $-$0.13 \\
5500/4/$-$1 &    0.13 &  0.08 &  0.14 &  0.08 &  0.06 &  0.00 &  0.09 &  0.00 &  0.10 &  0.00 &  0.15 &  0.06 & $-$0.10 & $-$0.01 & $-$0.04 &  0.04 & $-$0.03 & $-$0.09 & $-$0.21 \\
5500/4/$-$2 &    0.19 &  0.18 &  0.28 &  0.25 &  0.11 &  0.10 &  0.22 &  0.13 &  0.11 &  0.03 &  0.25 &  0.21 &  0.00 &  0.12 &  0.16 &  0.15 & $-$0.08 &      & $-$0.10 \\
5500/4/$-$3 &    0.20 &  0.27 &      &      &      &  0.29 &      &  0.19 &      &  0.18 &      &      &  0.23 &      &  0.24 &      & $-$0.21 &      &      \\
6000/3/ 0 &    0.09 & $-$0.01 &  0.05 & $-$0.01 & $-$0.04 & $-$0.14 &  0.02 & $-$0.13 &  0.01 & $-$0.17 &  0.08 & $-$0.05 & $-$0.27 & $-$0.07 & $-$0.14 & $-$0.02 & $-$0.03 & $-$0.28 & $-$0.29 \\
6000/3/$-$2 &    0.10 &  0.20 &  0.26 &  0.21 &  0.13 &  0.13 &  0.22 &  0.16 &      & $-$0.01 &      &  0.21 &  0.01 &  0.11 &  0.17 &  0.14 & $-$0.16 & $-$0.04 & $-$0.20 \\
6000/3/$-$3 &    0.02 &  0.26 &      &      &      &  0.27 &      &      &      &  0.19 &      &      &  0.22 &      &      &      & $-$0.41 &      &  0.00 \\
6000/4/ 0 &    0.07 &  0.02 &  0.04 &  0.00 &  0.01 & $-$0.05 &  0.01 & $-$0.05 &  0.05 & $-$0.05 &  0.06 & $-$0.01 & $-$0.13 & $-$0.06 & $-$0.10 & $-$0.02 & $-$0.02 & $-$0.17 & $-$0.27 \\
6000/4/$-$1 &    0.11 &  0.03 &  0.13 &  0.08 &  0.05 & $-$0.10 &  0.07 & $-$0.02 &  0.07 & $-$0.10 &  0.13 &  0.05 & $-$0.22 & $-$0.01 & $-$0.08 &  0.03 & $-$0.05 & $-$0.13 & $-$0.29 \\
6000/4/$-$2 &    0.13 &  0.18 &  0.24 &  0.20 &  0.12 &  0.12 &  0.19 &  0.15 &      &  0.03 &      &  0.19 &  0.01 &  0.10 &  0.14 &  0.13 & $-$0.11 & $-$0.01 & $-$0.12 \\
6000/4/$-$3 &    0.10 &  0.25 &      &      &      &  0.29 &      &      &      &  0.19 &      &      &  0.21 &      &      &      & $-$0.23 &      &      \\
\noalign{\smallskip}\hline
\end{tabular}
\end{center}
\end{table*} %

\subsection{Statistical equilibrium of \ion{Ca}{ii} and NLTE
effects on spectral lines}

\begin{figure}
\vspace*{-3mm} \hbox{
\resizebox{88mm}{!}{\rotatebox{0}{\includegraphics{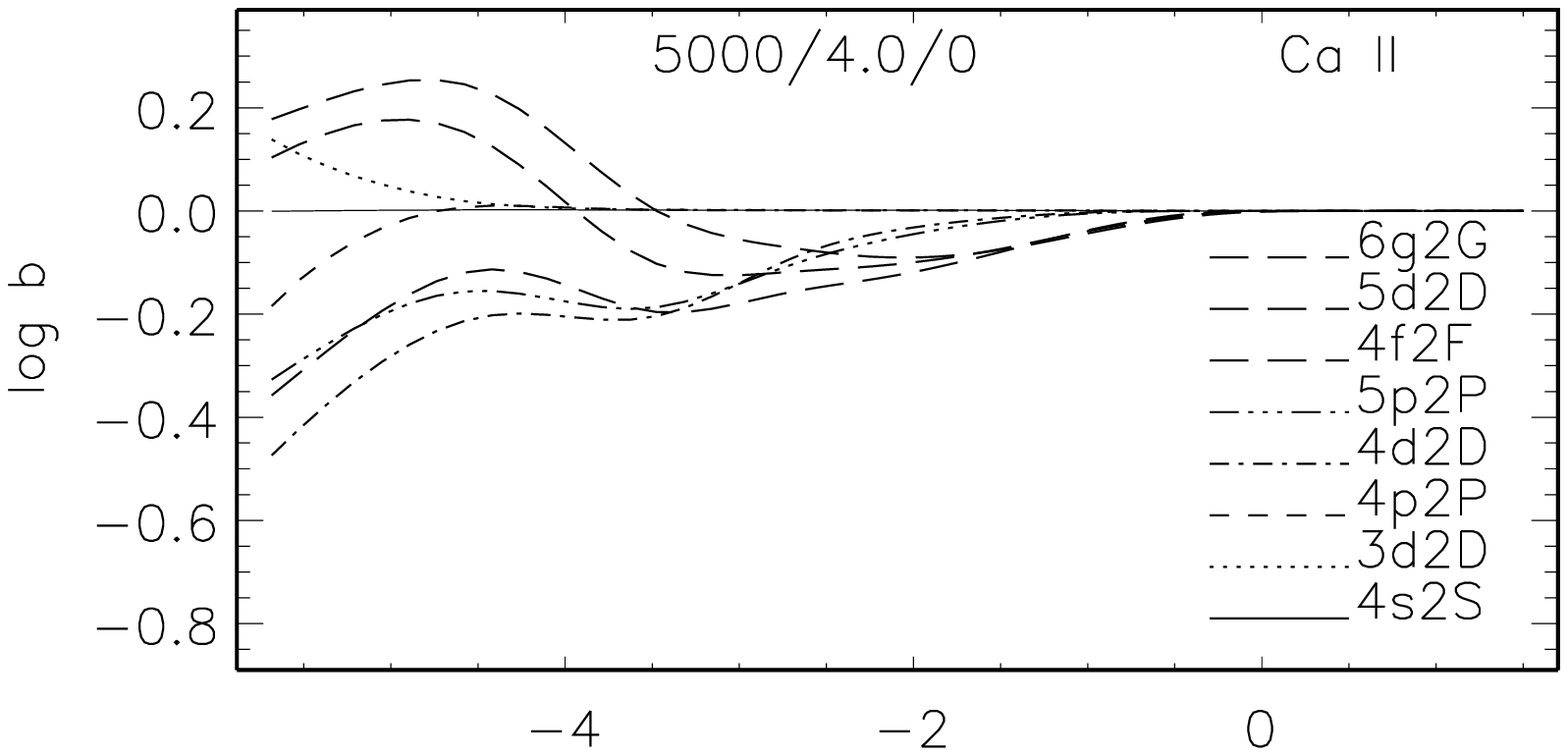}}
\hspace{-8mm} \rotatebox{0}{\includegraphics{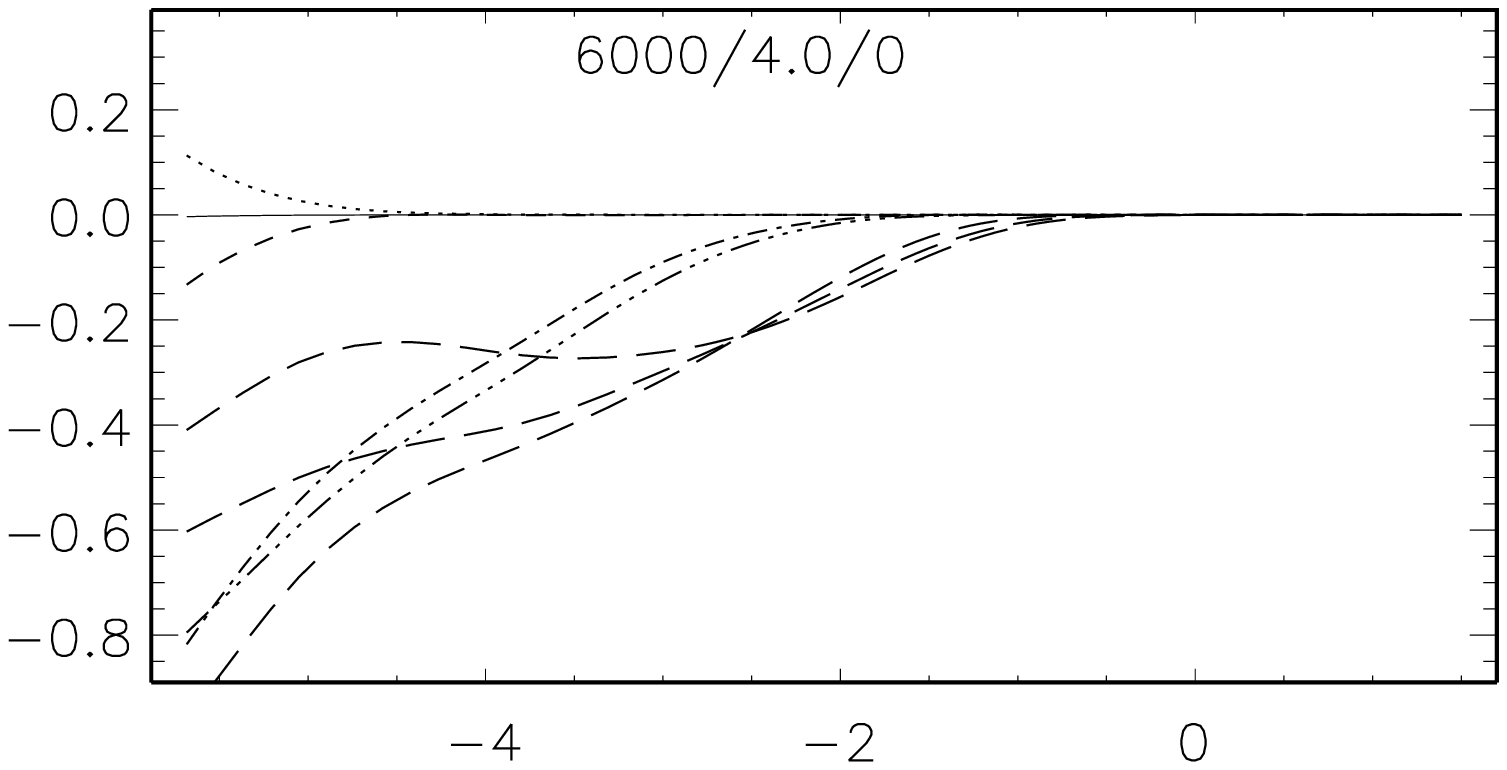}}}}
\vspace{-5mm} \hbox{
\vspace*{-3mm}
\resizebox{88mm}{!}{\rotatebox{0}{\includegraphics{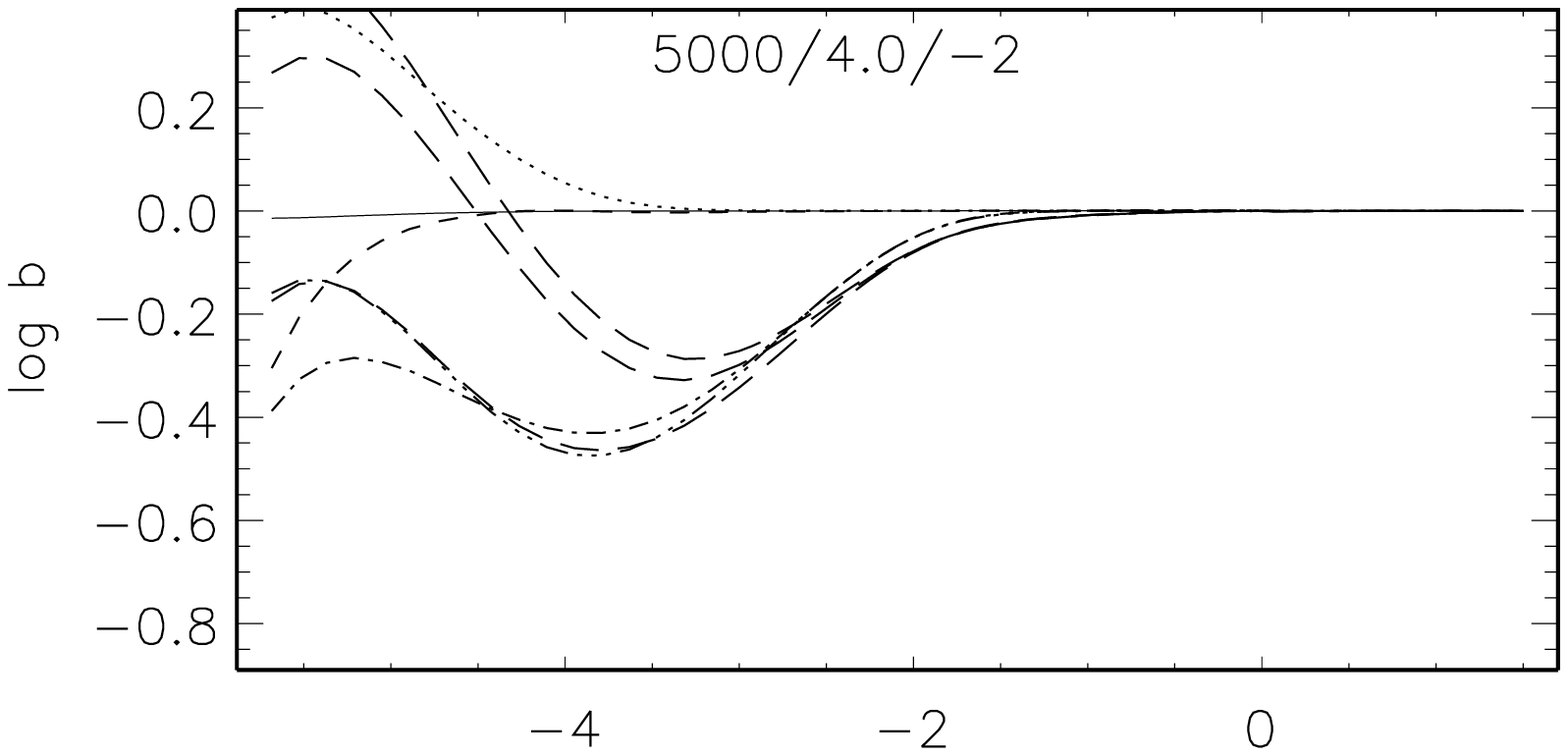}}
\hspace{-8mm} \rotatebox{0}{\includegraphics{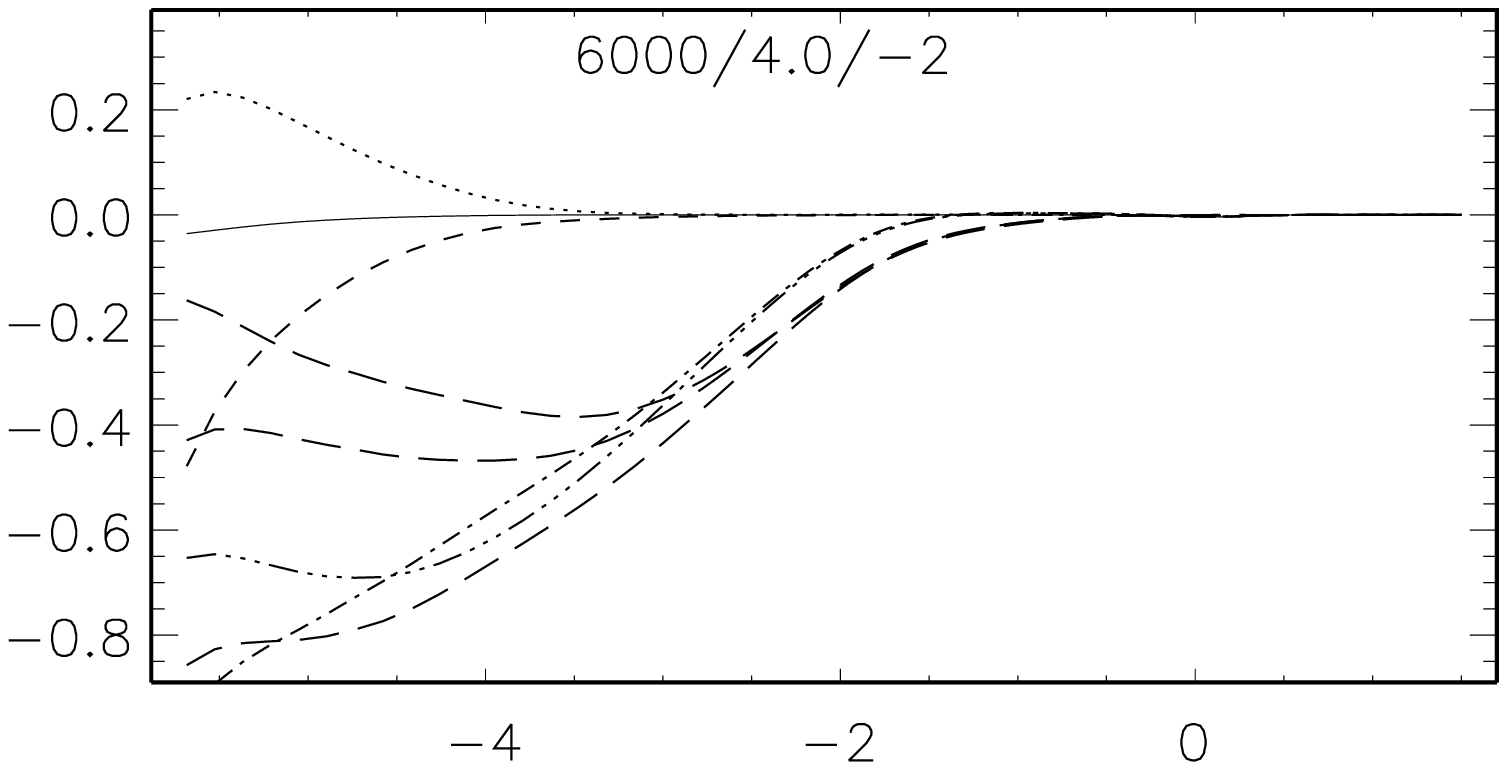}}}}
\hbox{
\vspace{-5mm}
\resizebox{88mm}{!}{\rotatebox{0}{\includegraphics{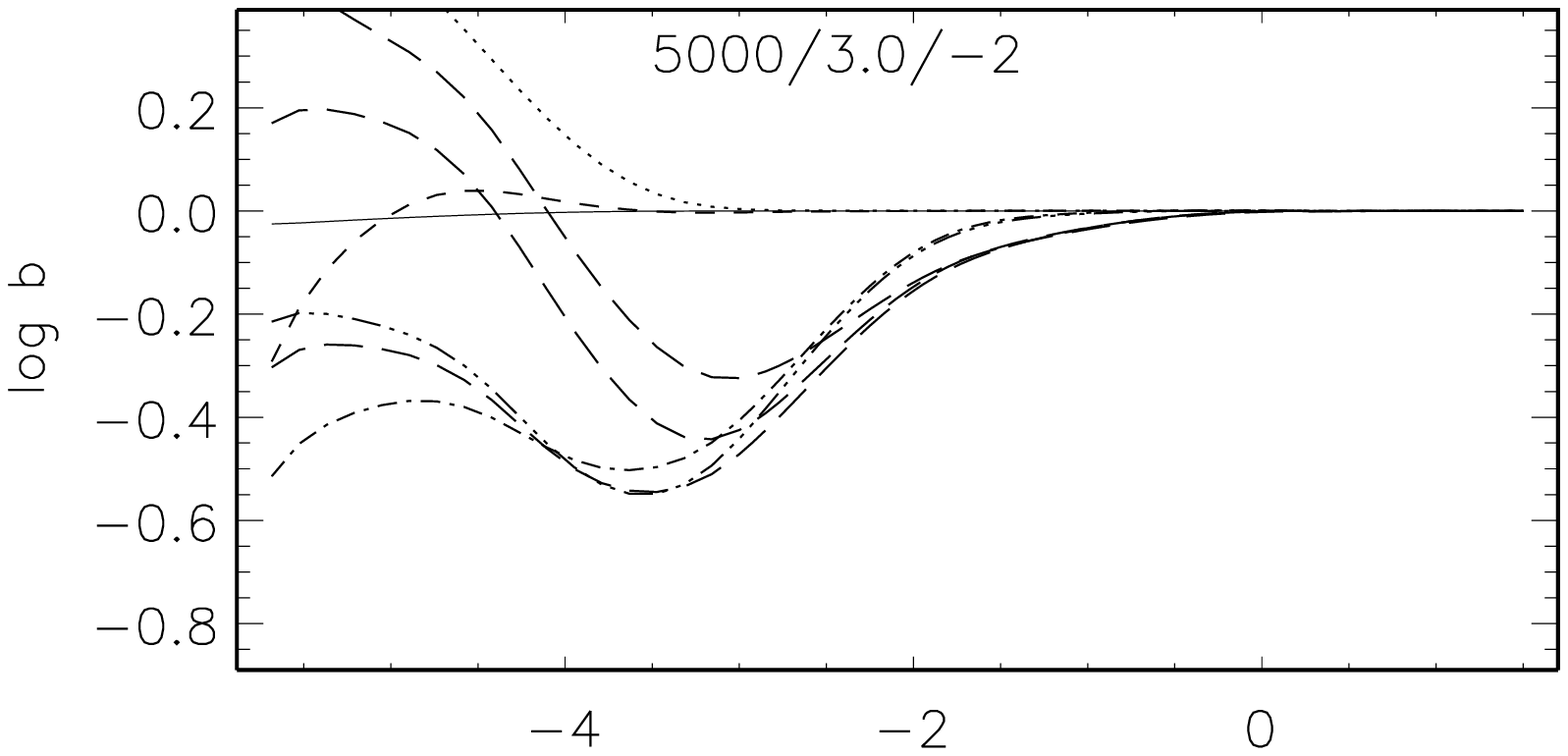}}
\hspace{-8mm} \rotatebox{0}{\includegraphics{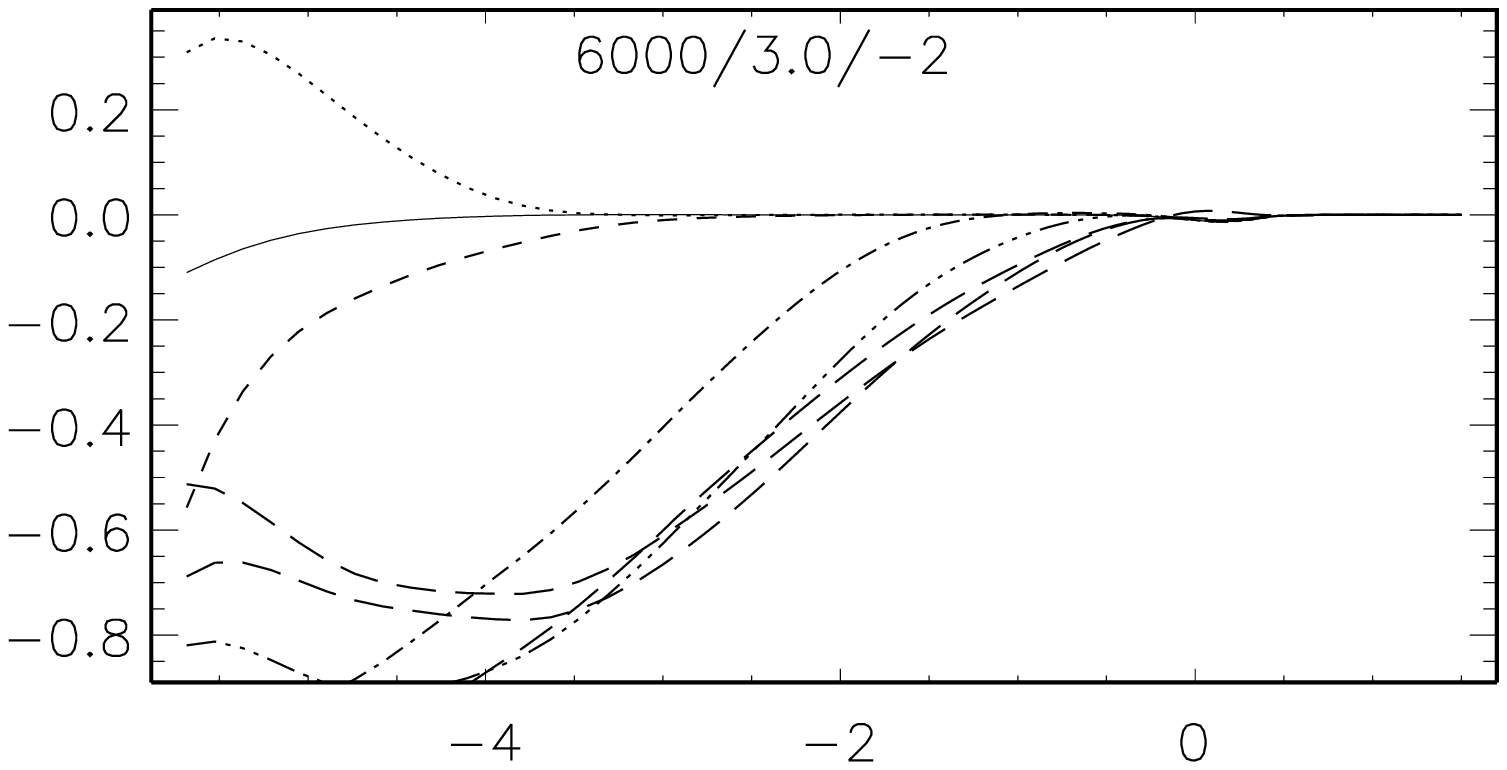}}}}
\hbox{
\vspace{-5mm}
\resizebox{88mm}{!}{\rotatebox{0}{\includegraphics{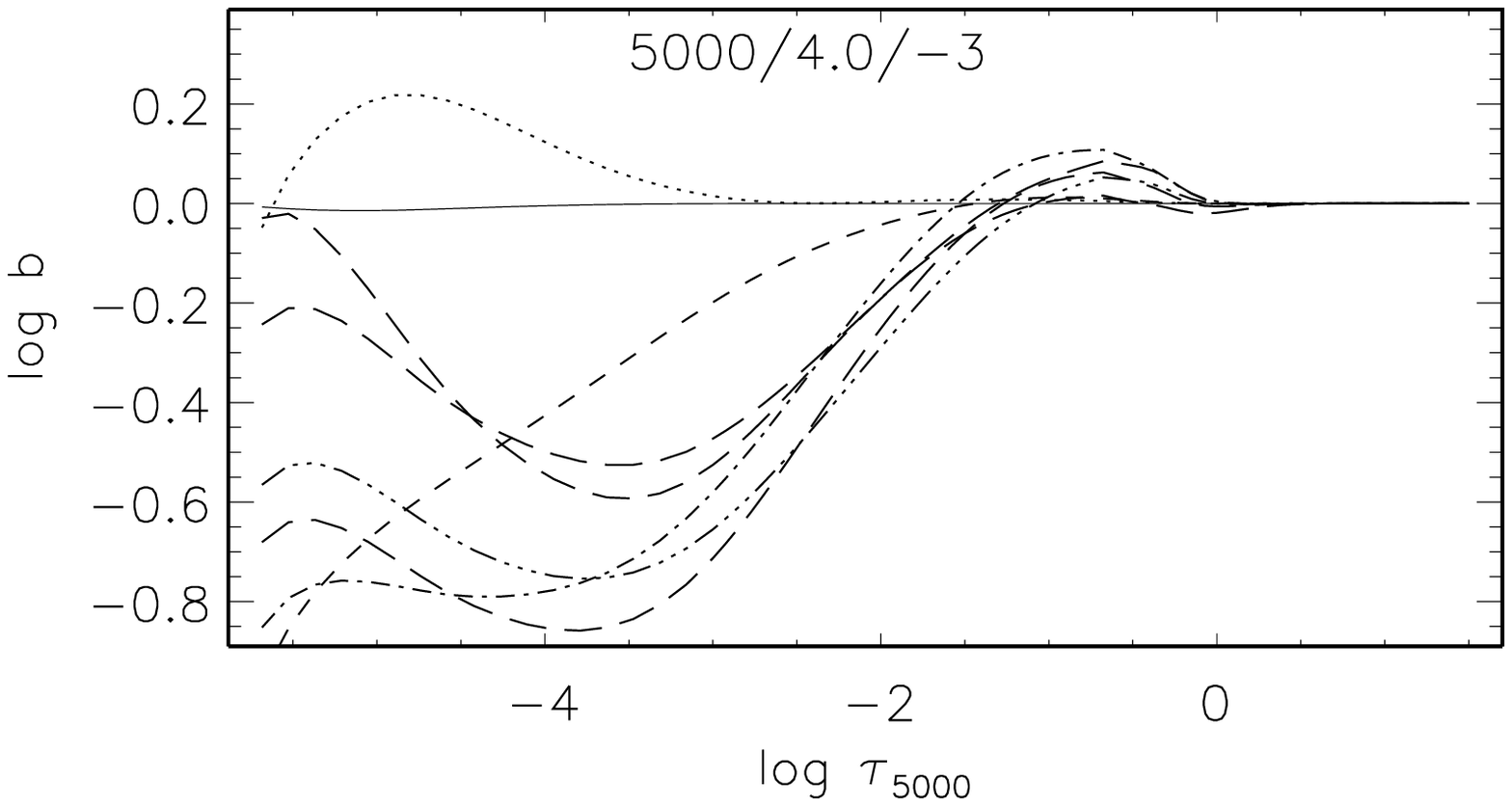}}
\hspace{-8mm} \rotatebox{0}{\includegraphics{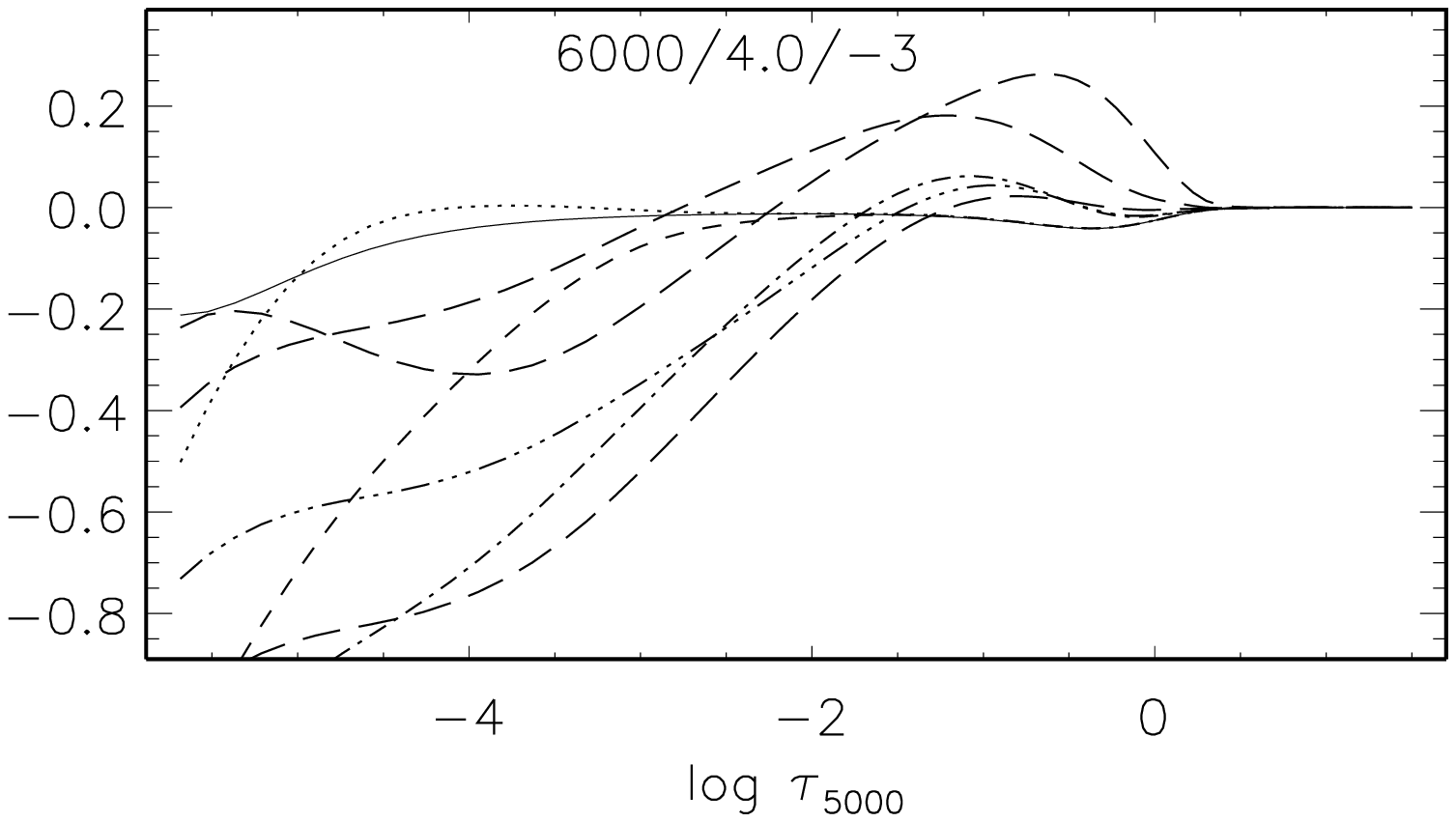}}}}

\vspace{3mm} \caption[]{The same as in Fig. \ref{bfca1} for
\ion{Ca}{ii}} \label{bfca2}
\end{figure}

Figure \ref{bfca2} shows departure coefficients for some important
levels of \ion{Ca}{ii} plotted for selected models of our grid. In
the temperature regime we are concerned with here ($\Teff$ =
5000\,K -- 6000\,K), \ion{Ca}{ii} dominates the element number
density over atmospheric depths. Thus, no process seems to affect
the \ion{Ca}{ii} ground state population, and $4s$ keeps its
thermodynamic equilibrium value. An exception is the uppermost layers above
$\log\tau_{5000}$ = $-$4 in the very metal-poor ([Fe/H] $\le -2$)
models with $\Teff$ = 6000\,K. \ion{Ca}{ii} competes there
with \ion{Ca}{iii} in contribution to element population,
and enhanced ionization of the low excitation levels of
\ion{Ca}{ii} leads to underpopulation of total \ion{Ca}{ii}. This
effect is amplified with decreasing metallicity. The levels $3d$
and $4p$ follow the ground state in deep layers, and their
coupling is lost at the depths where, for each model, photon
losses in the weakest line $\lambda\,8498$ of the multiplet $3d -
4p$ start to become important. These are the uppermost layers
above $\log\tau_{5000}$ = $-$5 in the solar metallicity models and
above $\log\tau_{5000}$ = $-$3 in the models with [Fe/H] = --2. At
[Fe/H] = --3, detailed balance
in the transition $3d - 4p$ is destroyed in the deeper layers
around $\log\tau_{5000}$ = $-$2. The departure coefficients of
\eu{4d}{2}{D}{}{} and the higher excitation levels begin to
deviate from 1 far inside the atmosphere even in the  solar
metallicity models due to photon losses in the transitions to low
excitation levels.

NLTE leads to strengthened \ion{Ca}{ii} lines and negative NLTE abundance corrections.
In the following, we consider a behaviour of individual lines.

In the stellar parameter range we are concerned with in this
paper, departures from LTE occur only in the very core of
\ion{Ca}{ii} K $\lambda\,3933$. The NLTE abundance correction does
not exceed 0.02\,dex in absolute value. This can be understood
because the line remains to be strong even at [Ca/H] = --3, and its
line strength is dominated by the line wings.

Similar to the resonance line, the IR lines of multiplet $3d -
4p$, $\lambda\,8498$, $\lambda\,8542$, and $\lambda\,8662$, reveal
NLTE effects only in the Doppler core.
The line core is strengthened because the line source function drops
below the Planck function due to $b_u/b_l < 1$ and the lower
level of the transition is overpopulated ($b_l > 1$) at the line
core formation depths. The NLTE correction is larger in absolute value
for the weakest line, $\lambda\,8498$, compared to the other two due
to smaller contribution of the line wings to the overall line strength.
However, even for $\lambda\,8498$, $\Delta_{\rm NLTE}$ is small
for the models with [Fe/H] = 0 and $-$1, $\vert\Delta_{\rm
NLTE}\vert \le 0.05$ (Table\,\ref{corr_full}). The van der Waals
broadened wings are weakened with decreasing Ca abundance and, at
fixed metallicity, with decreasing surface gravity and increasing
temperature. NLTE effects
are amplified in the same directions such that $\Delta_{\rm NLTE}$
($\lambda\,8498$) reaches --0.41\,dex for the model 6000\,K / 3.0
/ $-$3. Significant NLTE effects can be expected to occur in the
low-density atmospheres of metal-poor red giants. Further investigations
will be conducted when we have proper observations at our disposal.

The IR high excitation lines of multiplets $5p - 5d$,
$\lambda\,8248$ and $\lambda\,8254$, and $4d - 4f$,
$\lambda\,8912$ and $\lambda\,8927$, can be good candidates for
determination of \ion{Ca}{ii} abundance in close to solar metallicity cool
stars. They are of intermediate strength and nearly free of
blends.
It can be seen from
Fig.\,\ref{bfca2} that the high excitation levels are all depleted
relative to their LTE populations at line formation depths in the
models with [Fe/H] $\ge -2$. However, the lower levels of the
considered transitions, $5p$ and $4d$, are underpopulated to a lesser
extent than are the upper levels, $5d$ and $4f$, resulting in
smaller line source functions compared to the Planck function and
enhanced line absorption. NLTE effects are significant even for
the solar metallicity models and become stronger with
increasing $\Teff$ and decreasing $\log g$. The lines of multiplet
$5p - 5d$ are weaker compared to $\lambda\,8912$ and
$\lambda\,8927$ and, in general, reveal smaller NLTE effects. In
Table\,\ref{corr_full}, we present $\Delta_{\rm NLTE}$ for
$\lambda\,8248$ and $\lambda\,8927$. NLTE corrections for the
second line of multiplet $4d - 4f$ are close to the corresponding
values for $\lambda\,8927$ and, in general, smaller by up to
0.03\,dex in absolute value. For $\lambda\,8254$, NLTE effects are
very small with $\Delta_{\rm NLTE}$ at the level of a few parts in
a hundred.

\subsection{Error estimates for the \ion{Ca}{i/ii} SE calculations}\label{error}

To assess the effects of crucial atomic data on the accuracy of
NLTE Ca abundances derived from the \ion{Ca}{i} and \ion{Ca}{ii}
lines, test calculations were performed for three models with
$\Teff$ = 5500\,K, $\log g$ = 4.0 and [Fe/H] = $-$1, $-$2, and $-$3.
For each parameter or set
of cross-sections varied, we computed for a given line a small
grid at different abundances to determine the systematic shift in
Ca abundance needed to fit the NLTE equivalent width evaluated using our
standard set of atomic data. The results of the tests are
summarized for selected lines in Table\,\ref{corr_diff}.

\begin{table*}
\caption{Uncertainties in the NLTE analysis of \ion{Ca}{i/ii}.
Blanks denote NLTE theoretical equivalent widths below 5 m\AA}
\label{corr_diff} \tabcolsep2.5mm
\begin{tabular}{llrrrrrrrrr}
\hline\noalign{\smallskip}
 & & \multicolumn{9}{c}{changes in NLTE abundance corrections (dex)} \\
\noalign{\smallskip} \cline{3-11} \noalign{\smallskip}
\multicolumn{2}{c}{input parameter} & \ion{Ca}{i} 4226 &  4425 &  5588 & 5857 & 6162 &  6169 &  6439 & \ion{Ca}{ii} 8498 &  8927  \\
\noalign{\smallskip} \hline \noalign{\smallskip}
 & & \multicolumn{9}{l}{ ~~~~~~5500\,K / 4.0 / $-$1} \\
Photoionization: & $\sigma_{ph}$/2  &$-$0.02 & $-$0.02 & $-$0.05 & $-$0.06 & $-$0.03 & $-$0.04 & $-$0.06 &  0.00 &  0.00 \\
$e$-collisions:& $C_{ij}$(Reg)    &$-$0.02 & $-$0.01 & $-$0.03 &  0.01 &  0.00 & $-$0.02 & $-$0.02 &  0.00 &  0.14 \\
H collisions:    & \kH\ = 1        &$-$0.12 & $-$0.08 & $-$0.05 & $-$0.03 & $-$0.04 & $-$0.08 & $-$0.04 &  0.01 &  0.16 \\
\multicolumn{2}{l}{$\Delta\Vmic = +0.5$\,\kms} & 0.00 & $-$0.01 & $-$0.02 &  0.00 & $-$0.02 &  0.00 & $-$0.03 &  0.00 &  0.01 \\
 & & \multicolumn{9}{l}{~~~~~~5500\,K / 4.0 / $-$2} \\
Photoionization: &$\sigma_{ph}$/2  &$-$0.01 & $-$0.02 & $-$0.02 & $-$0.03 & $-$0.01 & $-$0.01 & $-$0.03 &  0.00 &  0.00 \\
$e$-collisions:&$C_{ij}$(Reg)    &$-$0.04 & $-$0.01 & $-$0.03 &  0.04 &  0.00 & $-$0.01 & $-$0.02 &  0.02 &  0.08 \\
H collisions:    &\kH\ = 1        &$-$0.18 & $-$0.17 & $-$0.12 & $-$0.10 & $-$0.06 & $-$0.19 & $-$0.10 &  0.03 &  0.09 \\
\multicolumn{2}{l}{$\Delta\Vmic = +0.5$\,\kms} & 0.00 &  0.00 &  0.01 &  0.00 &  0.00 &  0.00 &  0.01 & $-$0.01 &  0.00 \\
 & & \multicolumn{9}{l}{~~~~~~5500\,K / 4.0 / $-$3} \\
Photoionization: &$\sigma_{ph}$/2  &$-$0.01 & $-$0.01 & $-$0.01 & $-$0.02 & $-$0.01 &      & $-$0.01 &  0.00 &      \\
$e$-collisions: &$C_{ij}$(Reg)    &$-$0.03 & $-$0.01 & $-$0.03 &  0.02 &  0.00 &      & $-$0.01 &  0.07 &      \\
H collisions:    &\kH\ = 1        &$-$0.19 & $-$0.20 & $-$0.21 & $-$0.12 & $-$0.10 &      & $-$0.18 &  0.10 &      \\
\multicolumn{2}{l}{$\Delta\Vmic = +0.5$\,\kms} &$-$0.01 &  0.00 &  0.00 &  0.00 & 0.00 &   &  0.00 & $-$0.03 &      \\
\noalign{\smallskip}\hline
\end{tabular}
\end{table*} %

 OP photoionization data based on the R-matrix calculations typically are accurate to 10\%. In test computations, 
we assume a factor of two uncertainty of photoionization cross-sections
as a worst case. As expected, a variation in the
photoionization rates affects the lines of the minority
species (\ion{Ca}{i}) by way of a displaced ionization balance
and does not affect the lines of the majority species
(\ion{Ca}{ii}). Corrections for the \ion{Ca}{i} lines decrease with
decreasing metallicity. The explanation lies with an increasing
fraction of \ion{Ca}{iii} at the same optical depth in the models
with decreasing metallicity. When cross-sections of both the
\ion{Ca}{i} and \ion{Ca}{ii} levels change, the
\ion{Ca}{i}/\ion{Ca}{ii}/\ion{Ca}{iii} ionization balance is
established in such way that
the \ion{Ca}{i}/\ion{Ca}{ii} ratio is influenced to a lesser
extent in the more metal-poor models.

In the present work, detailed {\it electronic collision
excitation data} are used for a considerable number of transitions
in \ion{Ca}{ii} and for all important transitions from the ground
state in \ion{Ca}{i}. In test calculations, we vary collisional
rates for the remaining transitions applying the van Regemorter's
formula (\cite{Reg}) instead of IPM data. Van Regemorter's
collisional rates $C_{ij}$(Reg) are, in general, larger than the
corresponding IPM-based values, by up to 2 orders of
magnitude for the transitions with energy separation $\Delta E <$
2~eV (see Fig.~2 in Mashonkina (\cite{mash96})
for the transitions in \ion{Mg}{i}). As a result, NLTE effects are
weakened compared to the
standard collisional recipe and a NLTE Ca abundance is obtained to
be closer to the LTE one (lower from the \ion{Ca}{i} lines and
higher from the \ion{Ca}{ii} lines, Table\,\ref{corr_diff}). The
only exception is \ion{Ca}{i} 5857 for which NLTE effects are
strengthened. In any case, a variation of electronic collision
rates only weakly affects the \ion{Ca}{i} lines, such that the
derived Ca abundance changes by 0.04\,dex, at maximum. In
contrast, the effect is significant for the \ion{Ca}{ii} lines, and a
change in Ca abundance may consist of 0.07\,dex to 0.14\,dex for
different lines. A different reaction of the \ion{Ca}{i} and
\ion{Ca}{ii} lines can be understood because the main mechanism of
departures from LTE is connected with $b-f$ transitions for
\ion{Ca}{i}, while with $b-b$ transitions for \ion{Ca}{ii}.

We inspect also the effect of including {\it hydrogenic collisions}
in our SE calculations. Table\,\ref{corr_diff} shows the
difference in Ca abundance derived assuming \kH\ = 1 and ignoring
H collisions (\kH\ = 0). As expected, departures from LTE are weakened
for the \ion{Ca}{ii} lines. Changes in \ion{Ca}{ii} abundance are
comparable to those obtained when varying electronic collision
rates. A somewhat unexpected behaviour is seen for most \ion{Ca}{i} lines in
the model with [Fe/H] = $-$1 and for $\lambda\,6439$ in the model
with [Fe/H] = $-$2: NLTE effects seem to be amplified when the total
collisional rates increase.
This can be understood because departures from LTE go down in the line
wings, as expected, but are hardly changed in the cores of strong lines
due to the inefficiency of collisions in the uppermost atmospheric
layers. Thus, the line wings do not act anymore to reduce the
combined NLTE effect on the overall line strength, and it becomes larger
compared to the case of electronic collisions only.

In addition, we examine the effect of a shift of the line
formation depths due to increasing the microturbulence value in the
model. An increase of $\Vmic$ by 0.5\,\kms\ has a negligible effect
on $\Delta_{\rm NLTE}$ for every Ca line in the models with [Fe/H]\,=\,$-$2 and $-$3 and leads to strengthening NLTE effects for the
model with [Fe/H]\,=\,$-$1. The maximum correction amounts to $-$0.03\,dex.

Thus, the largest uncertainty of NLTE results for \ion{Ca}{i/ii}
is caused by poor knowledge of collision processes. Below we
empirically constrain collisional data
by way of analysis of Ca lines in solar (Sect.\,\ref{sun})
and stellar (Sect.\,\ref{stars1}) spectra.

\section{Solar Ca abundance and \ion{Ca}{i} /\ion{Ca}{ii} ionization equilibrium }\label{sun}

In this section, we derive solar Ca abundance from the \ion{Ca}{i}
subordinate lines and the \ion{Ca}{ii} high excitation lines and
examine atomic data used in SE calculations and element abundance
determinations. Taking advantage of the applied NLTE approach,
both weak and strong Ca lines are included in analysis. An
exception is the resonance lines in \ion{Ca}{i} and \ion{Ca}{ii} and
the \ion{Ca}{ii} lines of multiplet $3d - 4p$, because their cores
and inner wings are influenced by the chromospheric temperature
rise and non-thermal and depth-dependent chromospheric velocity
field that is not part of the MAFAGS model of the solar atmosphere.
However, we check here the wings of \ion{Ca}{i} $\lambda\,4226$ and
\ion{Ca}{ii} $\lambda\,8498$.

We use solar flux observations taken from the Kitt Peak Solar
Atlas (Kurucz et al. \cite{Atlas}) and select the Ca lines free of
blends, or with only one line wing distorted  (e.g., \ion{Ca}{i}
$\lambda\,4425$), or with the blending lines which can
be correctly taken into account (e.g., \ion{Ca}{i} $\lambda$6572 in
the wing of Balmer line H$\alpha$). The investigated lines are
listed in Table\,\ref{line_list}.

Four different sets of oscillator strengths are applied and
compared in this study. (i) $f_{ij}$ obtained from laboratory
measurements. They are available for all selected \ion{Ca}{i}
lines and shown in Table\,\ref{line_list} (column LAB) together with
their sources,
(ii) $f_{ij}$ based on OP calculations. They are available for all
optically permitted transitions in \ion{Ca}{i} and \ion{Ca}{ii},
(iii) and (iv) the data from NIST and VALD databases, respectively.

We use the radiative widths obtained by Kurucz (\cite{cdrom18}) from
radiative lifetimes and accessible via the VALD database.

\begin{figure}
\resizebox{88mm}{!}{\includegraphics{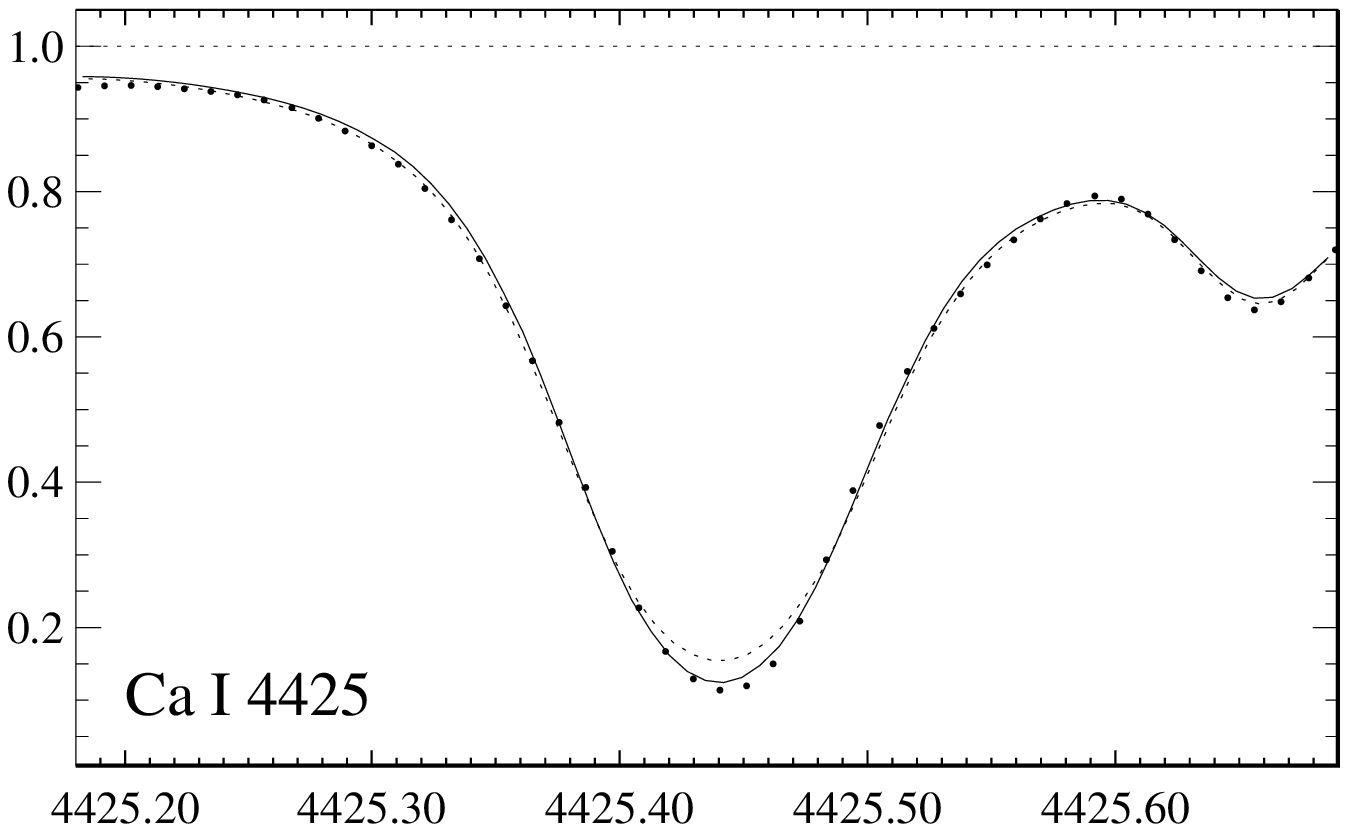}}

\vspace{-5mm} \resizebox{88mm}{!}{\includegraphics{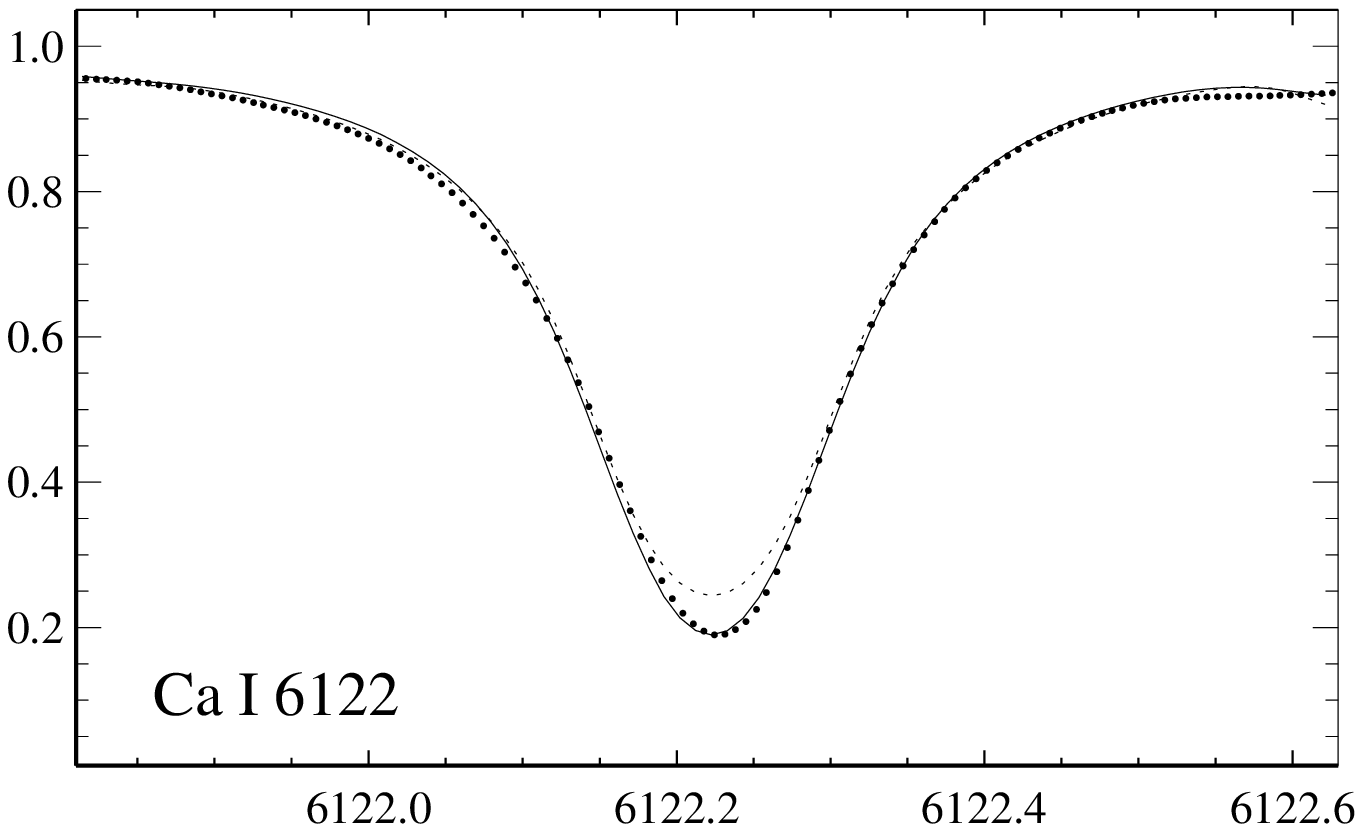}}

\vspace{-5mm} \resizebox{88mm}{!}{\includegraphics{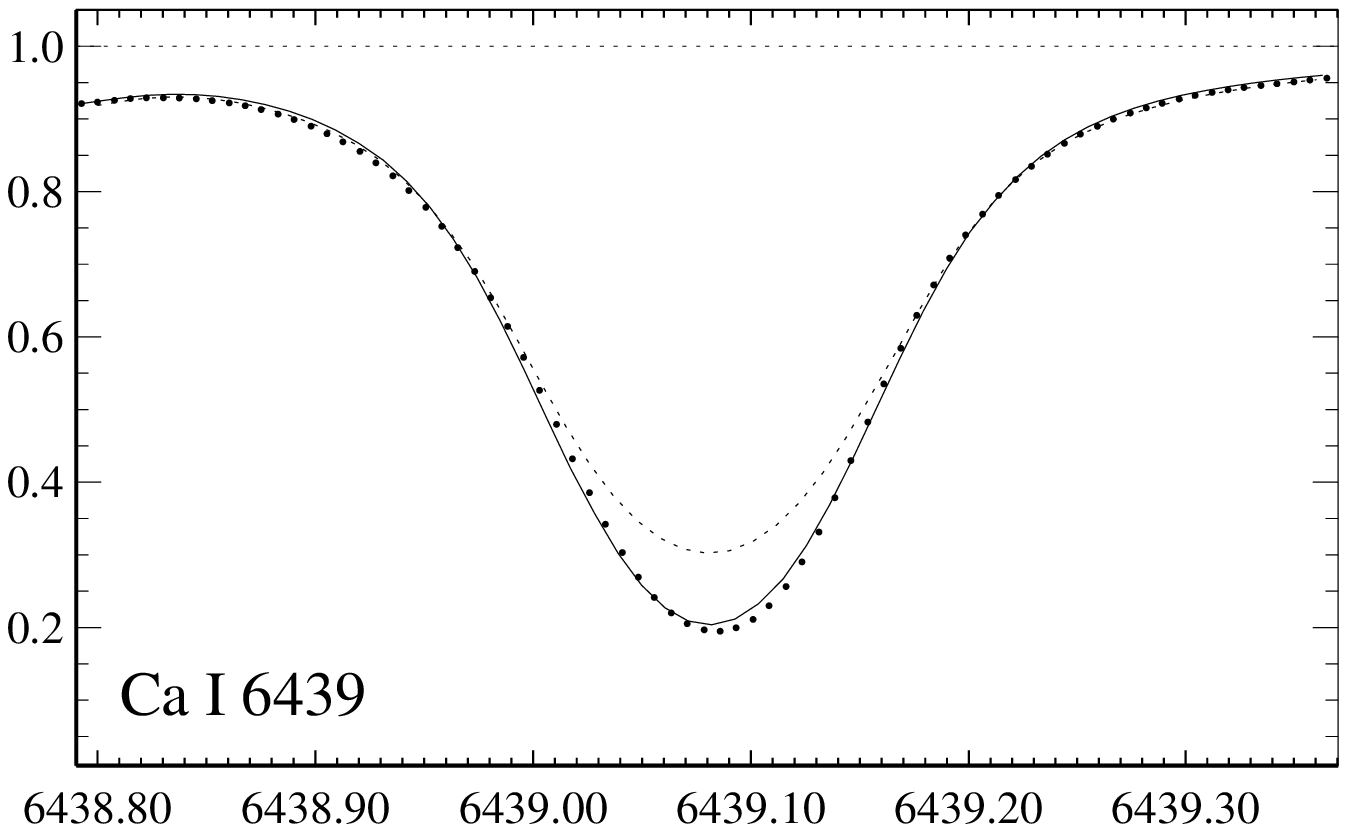}}

\vspace{-3mm} \resizebox{88mm}{!}{\includegraphics{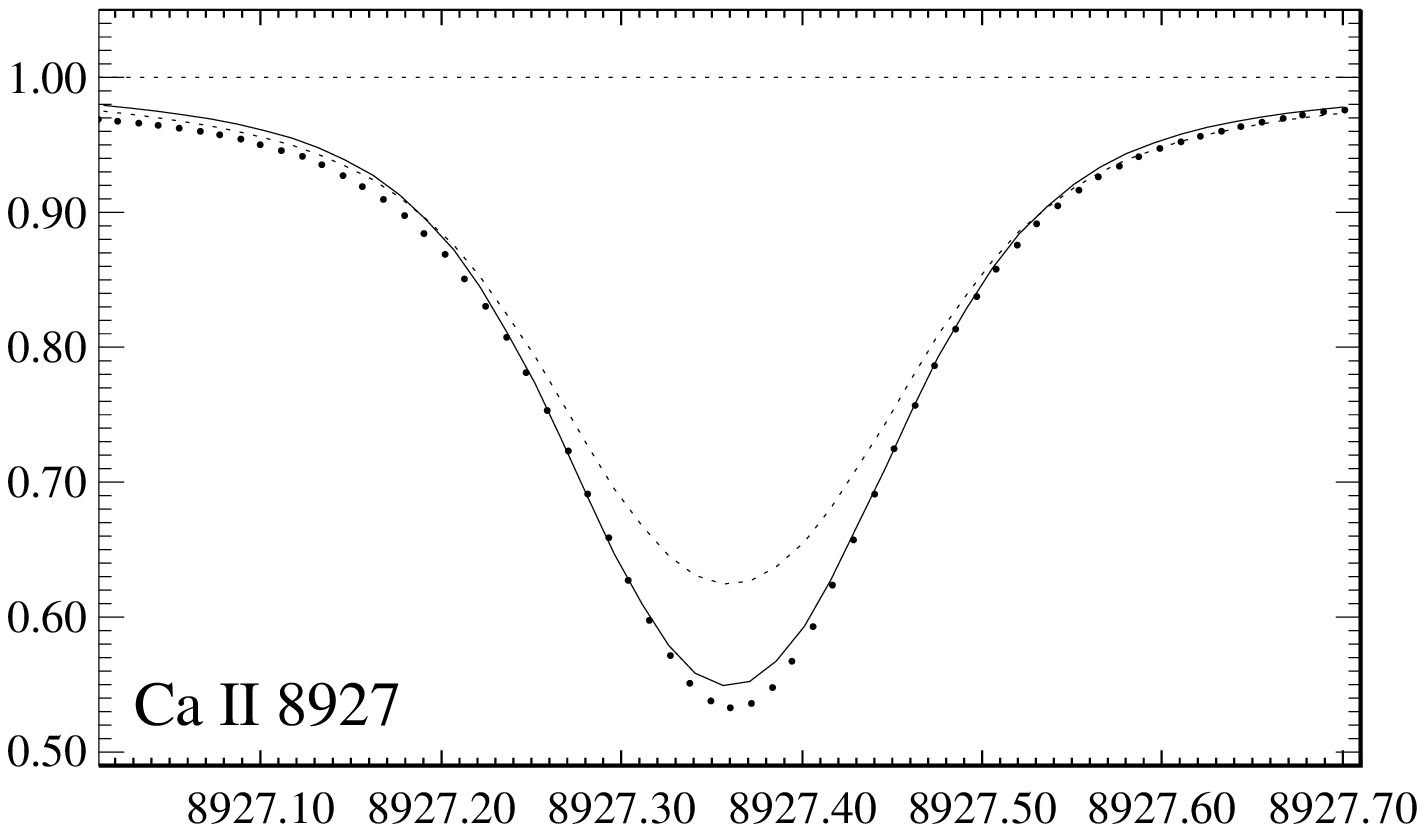}}

\vspace{-3mm} \caption[]{Synthetic NLTE (\kH\ = 0, continuous
line) and LTE (dotted line) flux profiles of the \ion{Ca}{i}
$\lambda\,4425$, $\lambda\,6122$, and $\lambda\,6439$ and
\ion{Ca}{ii} $\lambda\,8927$ lines compared with the observed
spectrum of the Kurucz et al. (1984) solar flux atlas (bold dots).
The fitting parameters are presented in Table\,\ref{line_list}. It
should be stressed that different Ca abundances are required to
achieve the best fits of the observed profile in the NLTE (\kH\ =
0) and LTE cases. Everywhere $\Vmic = 0.9$\,\kms}
\label{solar_line}
\end{figure}

For 17 \ion{Ca}{i} lines, the $C_6$ values are computed from
damping parameters given by Smith (\cite{ca_81}) and based on
measured parameters for broadening by helium. For the remaining
lines, we adopt $C_6$ values based on either $A\&O'M$ or Kurucz
calculations, giving a preference to the first source. The
necessary data are accessible via the VALD database. The van
der Waals damping parameters based on the perturbation theory of
$A\&O'M$ agree within 0.1\,dex of $\log C_6$ with quantum mechanic
computations by Spielfiedel et al. (\cite{ca6122_c6}) for the
\ion{Ca}{i} multiplet 3 (\eu{4p}{3}{P}{\circ}{} -
\eu{5s}{3}{S}{}{}) and by Kerkeni et al. (\cite{c6_4226}) for the
\ion{Ca}{i} resonance line $\lambda\,4226$. For four among five
common \ion{Ca}{i} multiplets, the predicted $A\&O'M$ parameters
lead to stronger collisional broadening compared to that from the
experimental data of Smith (\cite{ca_81}). The difference in $\log
C_6$ ranges from 0.19\,dex to 0.32\,dex. The opposite is the case
for \ion{Ca}{i} $\lambda\,5261$ line with the experimental value
of $\log C_6$ larger by 0.25\,dex compared to the predicted one.
For \ion{Ca}{i} $\lambda\,4425$ and the lines of \ion{Ca}{ii}
multiplets $5p - 5d$ and $4d - 4f$, the $C_6$ values were obtained
empirically from the fitting of solar line profiles. We note that
not only $\lambda\,4425$ but also two other lines of the
\ion{Ca}{i} multiplet \eu{4p}{3}{P}{\circ}{} - \eu{4d}{3}{D}{}{},
$\lambda\,4454$ and $\lambda\,4455$, though being blended,
certainly cannot be fitted with $\log C_6 = -30.23$ computed from
$A\&O'M$ data. We find $\log C_6 = -30.9$ for these lines. The
best fit of $\lambda\,4425$ is shown in Fig.\ref{solar_line}. For
the \ion{Ca}{ii} multiplet $5p - 5d$, $\log C_6 = -30.85$ was
found from the requirement that element abundances derived from
the weaker line, $\lambda\,8254$ (W$_\lambda =$ 18 m\AA), and the
stronger line, $\lambda\,8248$ (W$_\lambda =$ 65 m\AA), must be
equal. Contrary to multiplet $5p - 5d$, the lines of the
\ion{Ca}{ii} multiplet $4d - 4f$, $\lambda\,8912$ and
$\lambda\,8927$, are both sensitive to a variation of the van der
Waals damping constant. We {\it assume} that the broadening
parameter calculated by Kurucz is underestimated to the same
extent as the corresponding value for multiplet $5p - 5d$ and,
thus, obtain $\log C_6 = -31.1$ for multiplet $4d - 4f$. The best
fit of $\lambda\,8927$ is shown in Fig.\ref{solar_line}. The
atomic data are presented in Table\,\ref{line_list}.

A depth-independent microturbulence of 0.9\,\kms\, is adopted. Our
synthetic flux profiles are convolved with a profile that combines
a rotational broadening of 1.8\,\kms\ and broadening by
macroturbulence with a radial-tangential profile of $\Vmac$ =
3\,\kms\ to $\Vmac$ = 4\,\kms\ for different lines.

For each investigated line, the product $\log gf\varepsilon_{\rm
Ca}^\odot$ was obtained from solar line-profile fitting under various
line-formation assumptions: NLTE \kH\ = 0, \kH\ = 0.1, and
\kH\ = 1 and LTE. Table\,\ref{line_list} presents the NLTE
values $\log gf\varepsilon_{\rm Ca}^\odot$ derived assuming \kH\ =
0 and abundance corrections $\Delta_{\rm X} = \eps{X} - \eps{S_H =
0}$ where X takes the meaning of 0.1, 1, and LTE for \kH\ = 0.1,
\kH\ = 1, and the LTE assumption, respectively.
By definition, $\Delta_{\rm LTE}$ (Table 4) = $-\Delta_{\rm NLTE}$ (Table 2). It is
interesting to note that, for every investigated line,
$\Delta_{\rm NLTE}$ given in the first string of
Table\,\ref{corr_full} does not coincide in absolute value with
$\Delta_{\rm LTE}$ in Table\,\ref{line_list}. This can be
understood because NLTE corrections in  Table\,\ref{corr_full}
were calculated from comparison of NLTE and LTE {\it equivalent
widths}, while abundance corrections in Table\,\ref{line_list} are
based on the analysis of {\it line profiles}. For
each spectral line, the best NLTE fit, independent of \kH\ value,
reproduces the line core better than the LTE one. However, for
many lines, even the NLTE fit is not perfect. As an example, we
show in Fig.\,\ref{solar_line} the best NLTE (\kH\ = 0) and LTE
fits for two lines of \ion{Ca}{i} and $\lambda\,8927$ of
\ion{Ca}{ii}. It is clearly seen that the observed line core of
\ion{Ca}{i} $\lambda\,6439$ is asymmetric. This is, most probably,
due to atmospheric inhomogeneity and, therefore, cannot be
reproduced in the framework of a 1D analysis.

\begin{table*}
\caption{Atomic data for the \ion{Ca}{i} and \ion{Ca}{ii} lines:
column 2 contains the multiplet numbers accordingly to Moore
(1972); columns 4 - 7 give $\log gf$ from various sources
including laboratory measurements (column LAB); $\log C_6$ of
individual lines are from the sources cited in the bottom part of
this table except for the unmarked values which are based on the
Smith (1981) data. {\it Solar} $\log gf\varepsilon_{\rm Ca}^\odot$
values were determined from NLTE analysis of the Ca line profiles
in the Kitt Peak Solar Atlas (Kurucz et al. 1984) neglecting
hydrogenic collisions (\kH\,=\,0). Columns 10 -- 12 present the
abundance corrections $\Delta_{\rm X} = \eps{X} - \eps{S_H = 0}$
where X takes the meaning 0.1, 1, and LTE for \kH\,=\,0.1, \kH\,=\,1
and the LTE assumption, respectively. The last column contains
solar equivalent widths} \label{line_list} \tabcolsep1.2mm
\begin{center}
\begin{tabular}{lrcrrrrlcrrrr}
\hline\noalign{\smallskip}
 \multicolumn{1}{c}{$\lambda$} &    mult & E$_{low}$ &  \multicolumn{4}{c}{$\log gf$}   &\multicolumn{1}{c}{log C$_6$} & $\log gf\varepsilon_{\rm Ca}^\odot$ & $\Delta_{0.1}$
& $\Delta_1$ & $\Delta_{\rm LTE}$ & W$_\lambda$\ \ \ \\
\cline{4-7}
 \multicolumn{1}{c}{[\AA]} & & [eV] &  LAB &              NIST &  VALD  &  OP & & & & & & [m\AA] \\
\hline
\multicolumn{1}{c}{1} & \multicolumn{1}{c}{2} & 3 & \multicolumn{1}{c}{4} & \multicolumn{1}{c}{5} & \multicolumn{1}{c}{6}
 & \multicolumn{1}{c}{7} & \multicolumn{1}{c}{8} & 9 & \multicolumn{1}{c}{10} & \multicolumn{1}{c}{11} &
 \multicolumn{1}{c}{12} & \multicolumn{1}{c}{13} \\
\hline \noalign{\smallskip}
\multicolumn{2}{c}{ \ion{Ca}{i}} &  & & & & & & & & & & \\
  4226.73  &   2 &0.00& ~0.244$^a$& ~0.244& ~0.265& ~0.276& $-$31.23$^g$& $-$5.47 &           $-$0.03   &      $-$0.06   &   $-$0.08    &   1500 \\
  4425.44  &   4 &1.87& $-$0.358$^b$& $-$0.358& $-$0.286& $-$0.566& $-$30.90$^*$&$-$5.95 &             0.00   &      $-$0.03   &   $-$0.03    &    174 \\
  4578.56  &  23 &2.51& $-$0.697$^c$& $-$0.558& $-$0.697& $-$0.804& $-$30.30 &$-$6.33 &             0.00   &      $-$0.04   &   $-$0.06    &        87\\
  5261.71  &  22 &2.51& $-$0.579$^c$& $-$0.73 & $-$0.579& $-$0.836& $-$30.86 &$-$6.14 &             0.00   &      $-$0.03   &   $-$0.05    &       102 \\
  5349.47  &  33 &2.70& $-$0.310$^c$& ~~$-$   & $-$0.310& $-$0.730& $-$31.45 &$-$5.92 &             0.02   &       0.02   &    0.00    &        97 \\
  5512.98$^1$&   &2.92& $-$0.464$^d$& $-$0.30 & $-$0.447& $-$0.396& $-$30.61 &$-$6.05 &             0.00   &       0.02   &    0.00    &        89 \\
  5588.76  &  21 &2.51& ~0.358$^c$& ~0.21 & ~0.358& ~0.339& $-$31.39 &$-$5.36 &             0.00   &       0.00   &    0.00    &       171 \\
  5590.12  &  21 &2.51& $-$0.571$^c$& $-$0.71 & $-$0.571& $-$0.590& $-$31.39 &$-$6.18 &             0.00   &      $-$0.03   &   $-$0.03    &        97 \\
  5857.45  &  47 &2.92& ~0.240$^c$& ~0.23 & ~0.240& ~0.025& $-$30.61 &$-$5.40 &             0.00   &      $-$0.02   &   $-$0.02    &       160 \\
  5867.57$^2$&   &2.92& $-$1.570$^d$& ~~$-$   & $-$0.801& $-$2.402& $-$30.97$^h$&$-$7.12 &         $-$0.02   &      $-$0.06   &   $-$0.09    &        26 \\
  6122.22  &   3 &1.88& $-$0.315$^b$& $-$0.315& $-$0.386& $-$0.356& $-$30.30$^g$&$-$6.06 &          0.01   &      $-$0.01   &   $-$0.02    &       228 \\
  6162.17  &   3 &1.89& $-$0.089$^b$& $-$0.089& $-$0.167& $-$0.137& $-$30.30$^g$& $-$5.83 &         0.02   &       0.00   &   $-$0.02    &       289 \\
  6161.29  &  20 &2.51& $-$1.266$^c$& $-$1.03 & $-$1.266& $-$1.290& $-$30.48 &$-$6.85 &            $-$0.01   &      $-$0.06   &   $-$0.07     &       67 \\
  6166.44  &  20 &2.51& $-$1.142$^c$& $-$0.90 & $-$1.142& $-$1.166& $-$30.48 &$-$6.70 &            $-$0.02   &      $-$0.06   &   $-$0.07     &       77 \\
  6169.06  &  20 &2.51& $-$0.797$^c$& $-$0.54 & $-$0.797& $-$0.821& $-$30.48 &$-$6.40 &            $-$0.02   &      $-$0.03   &   $-$0.03     &      100 \\
  6169.56  &  20 &2.51& $-$0.478$^c$& $-$0.27 & $-$0.478& $-$0.502& $-$30.48 &$-$6.09 &            $-$0.02   &      $-$0.03   &   $-$0.03     &      129 \\
  6439.07  &  18 &2.51& ~0.390$^c$& ~0.47 & ~0.390& ~0.301& $-$31.58 &$-$5.32 &            $-$0.02   &      $-$0.01   &   $-$0.01     &      188 \\
  6471.66  &  18 &2.51& $-$0.686$^c$& $-$0.59 & $-$0.686& $-$0.775& $-$31.58 &$-$6.33 &            $-$0.02   &       0.01  &     0.01     &       97 \\
  6493.78  &  18 &2.51& $-$0.109$^c$& ~0.14 & $-$0.109& $-$0.198& $-$31.58 &$-$5.77 &            $-$0.03   &      $-$0.01   &    0.00    &       139\\
  6499.65  &  18 &2.51& $-$0.818$^c$& $-$0.59 & $-$0.818& $-$0.907& $-$31.58 &$-$6.41 &            $-$0.03   &      $-$0.06  &     0.00     &       89 \\
  6449.81  &  19 &2.51& $-$0.502$^c$& $-$0.55 & $-$0.502& ~~$-$   & $-$31.45 &$-$6.14 &            $-$0.02   &       0.00  &    0.00      &      106 \\
  6455.60  &  19 &2.51& $-$1.34$^d$ & $-$1.36 & $-$1.29 & ~~$-$   & $-$31.45 &$-$6.95 &            $-$0.02   &      $-$0.04  &   $-$0.05      &       59 \\
  6572.78  &   1 &0.00& $-$4.24$^e$ & $-$4.30 & $-$4.104& ~~$-$   & $-$31.54$^g$& $-$9.90  &       $-$0.03   &      $-$0.08   &   $-$0.10     &       31 \\
  7326.15  &  44 &2.92& $-$0.208$^d$&~~$-$    & ~0.073& ~0.021& $-$31.42$^h$& $-$5.74  &        0.02   &       0.00   &    0.02    &       116 \\
\multicolumn{2}{c}{ \ion{Ca}{ii}} &  & & & & & & & & & & \\
  5339.19  &  20 &8.40& ~~$-$   & ~~$-$   & $-$1.099& $-$0.079& $-$30.31$^h$& $-$5.77  &            0.00   &       0.00   &   0.00     &         6 \\
  5001.48  &  15 &7.47& ~~$-$   & $-$0.52 & $-$0.755& $-$0.505& $-$30.66$^h$& $-$6.18  &            0.00   &       0.01   &   0.03     &        11 \\
  6456.91  &  19 &8.40& ~~$-$   & ~~$-$   & $-$0.171& ~0.412& $-$30.77$^h$& $-$5.20  &            0.01   &       0.02   &    0.02     &       16\\
  8248.80  &  13 &7.48& ~~$-$   & ~0.57 & ~0.621& ~0.556& $-$30.85$^*$&$-$4.97  &             0.02   &       0.08   &    0.14     &       65\\
  8254.70  &  13 &7.48& ~~$-$   & $-$0.39 & $-$0.333& $-$0.398& $-$30.85$^*$&$-$5.92  &             0.00   &       0.01   &   0.02      &       18  \\
  8498.02  &   2 &1.69& $-$1.416$^f$&$-$1.318& $-$1.416& $-$1.465& $-$31.51$^g$ &$-$7.14 &          0.00   &       0.00   &    0.00     &     1132\\
  8912.07$^3$&     &7.05& ~~$-$   &~~$-$   & ~0.575& ~0.636& $-$31.10$^*$&$-$4.98  &            0.03   &       0.10   &    0.17     &      111\\
  8927.36$^4$&     &7.05& ~~$-$   & ~~$-$   & ~0.750& ~0.811& $-$31.10$^*$&$-$4.81  &           0.03   &       0.09   &    0.12     &      121\\
  9890.70$^5$&     &8.40& ~~$-$   &~~$-$   &  ~1.200 & ~1.270 & $-$31.34$^h$& $-$4.42  &          0.06   &       0.12   &    0.16     &       70 \\
\noalign{\smallskip}\hline
\multicolumn{3}{l}{$^1$ \eu{4p}{1}{P}{\circ}{} - \eu{4p^2}{1}{S}{}{}} & \multicolumn{4}{l}{$^a$ Smith \& Gallagher (1966) } &
\multicolumn{6}{l}{$^g$ $A\&O'M$ }\\
\multicolumn{3}{l}{$^2$ \eu{4p}{1}{P}{\circ}{} - \eu{6s}{1}{S}{}{}}  & \multicolumn{4}{l}{$^b$ Smith \& O'Neil (1975)} &
\multicolumn{6}{l}{$^h$ Kurucz's calculations} \\
\multicolumn{3}{l}{$^3$ \eu{4d}{2}{D}{}{3/2} - \eu{4f}{2}{F}{\circ}{5/2}}  & \multicolumn{4}{l}{$^c$ Smith \& Raggett (1981)} &
\multicolumn{6}{l}{$^*$ from solar line profile fitting}\\
\multicolumn{3}{l}{$^4$ \eu{4d}{2}{D}{}{5/2}  - \eu{4f}{2}{F}{\circ}{5/2,7/2}} & \multicolumn{10}{l}{$^d$ Smith (1988)} \\
\multicolumn{3}{l}{$^5$ \eu{4f}{2}{F}{\circ}{}  - \eu{5g}{2}{G}{}{}} & \multicolumn{10}{l}{$^e$ Drozdowski et al. (1997)} \\
\multicolumn{3}{l}{} & \multicolumn{10}{l}{$^f$ Theodosiou (1989) }\\
\end{tabular}
\end{center}
\end{table*}

\begin{figure}
\resizebox{88mm}{!}{\includegraphics{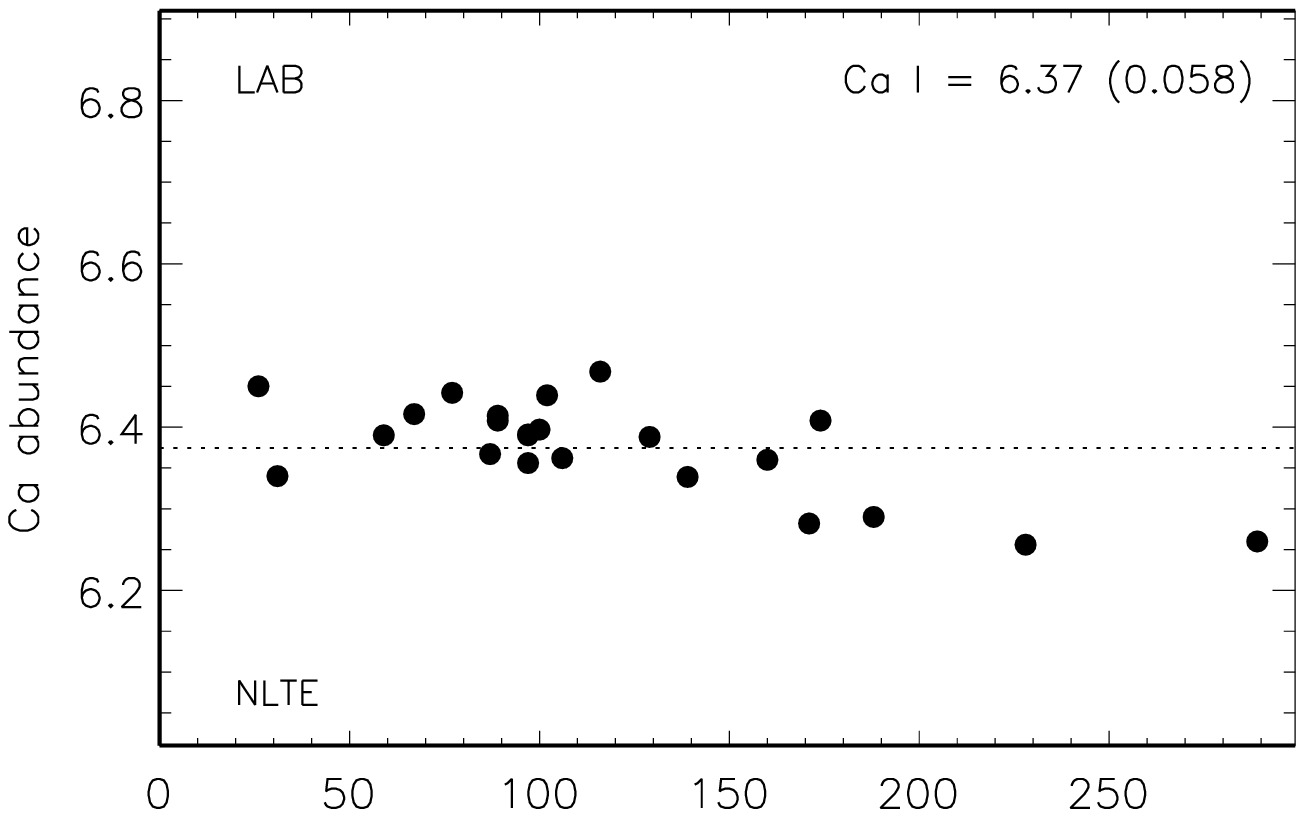}}

\vspace{-10mm}
\resizebox{88mm}{!}{\includegraphics{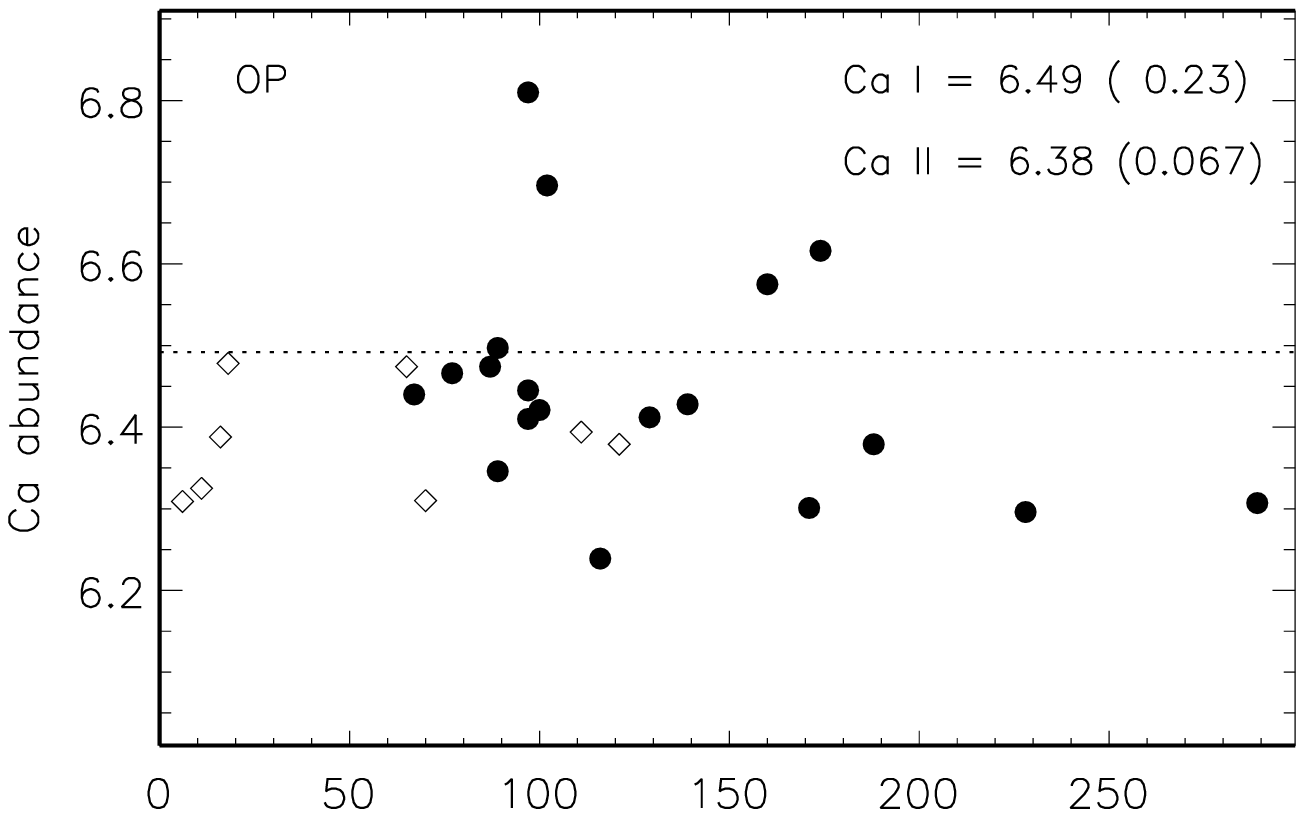}}

\vspace{-10mm}
\resizebox{88mm}{!}{\includegraphics{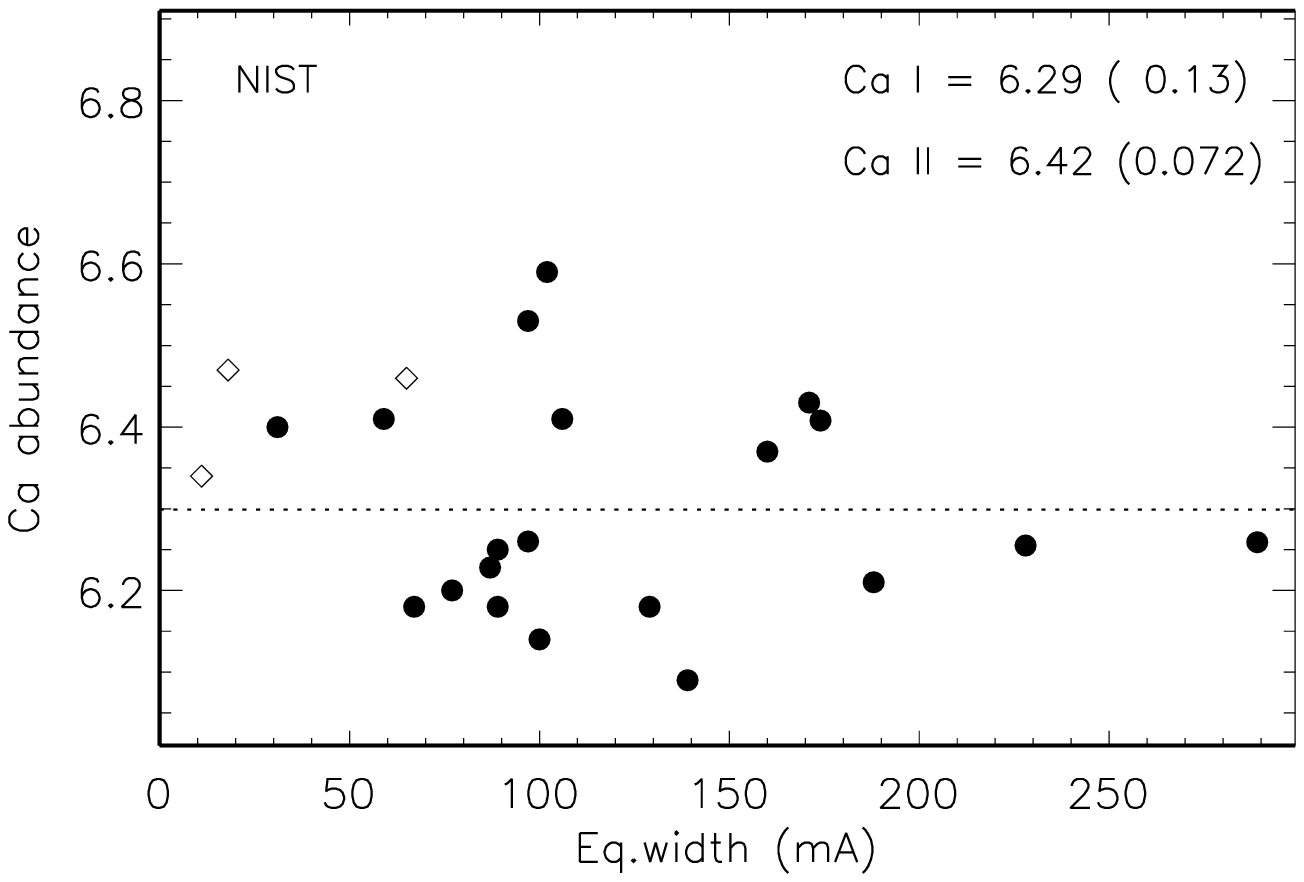}}

\vspace{-3mm}
\caption[]{Solar NLTE (\kH\ = 0) Ca abundances from the
\ion{Ca}{i} (filled circles) and \ion{Ca}{ii} (open diamonds)
lines plotted as function of the line equivalent width. Top,
middle, and bottom panels show results based on the measured
$f_{ij}$, OP, and NIST data, respectively. In each panel, the mean abundance derived from the \ion{Ca}{i} lines is shown by the dotted line and the mean values and their standard deviations are quoted for \ion{Ca}{i} and
\ion{Ca}{ii}}  \label{ca_solar}
\end{figure}

Using the obtained values $\log gf\varepsilon_{\rm Ca}^\odot$ and
 $f_{ij}$ from different sets, we compute Ca abundances
from individual lines. Fig.\,\ref{ca_solar} illustrates results
for NLTE (\kH\ = 0) calculations. It can be seen that the laboratory
oscillator strengths provide the highest accuracy in the {\it
absolute solar abundance} derived from the lines of \ion{Ca}{i}. The NLTE
and LTE averages from 23 lines are $\eps{CaI}$(LAB, \kH\ = 0) =
6.37 $\pm$ 0.06, $\eps{CaI}$(LAB, \kH\ = 0.1) = 6.36 $\pm$ 0.06,
$\eps{CaI}$(LAB, \kH\ = 1) = 6.35 $\pm$ 0.06, and $\eps{CaI}$(LAB,
LTE) = 6.34 $\pm$ 0.06. Throughout this paper, standard
deviations are quoted. NLTE effects for \ion{Ca}{i} in the Sun
are very small, and further we present the results only for \kH\ =
0. Ca abundances based on OP $f_{ij}$ for 20 \ion{Ca}{i} lines reveal
a large spread of data with the mean values $\eps{CaI}$(OP, \kH\ =
0) = 6.49 $\pm$ 0.23 and $\eps{CaI}$(OP, LTE) = 6.46 $\pm$ 0.21.
Excluding the one line, $\lambda\,5867$, with the largest contribution to
standard deviation, we obtain $\eps{CaI}$(OP, \kH\ = 0) = 6.45
$\pm$ 0.14 and $\eps{CaI}$(OP, LTE) = 6.43 $\pm$ 0.14. From 20
\ion{Ca}{i} lines with the NIST data available, the averages
equal $\eps{CaI}$(NIST, \kH\ = 0) = 6.29 $\pm$ 0.13 and
$\eps{CaI}$(NIST, LTE) = 6.26 $\pm$ 0.13. The results based on the
VALD data are close to that for laboratory measurements because,
for most lines investigated,
VALD's oscillator strengths have been taken from Smith \& Raggett
(\cite{ca_fij}). Tests show that
the NLTE abundance from \ion{Ca}{i} lines is not sensitive to a variation of
electronic collisional rates in SE calculations. Using the
formula of van Regemorter (\cite{Reg}) and IPM (Seaton
\cite{Seaton}) data leads to the mean Ca abundances, consistent
within 0.01\,dex. In each case, either NLTE or LTE, for any set of
$f_{ij}$, no significant correlation of individual Ca abundance is found with
the line strength.

Only predicted oscillator strengths are available for the
\ion{Ca}{ii} lines which are used to determine the {\it absolute solar abundance}
from \ion{Ca}{ii} lines.
OP data provide high accuracy for the desired value
provided that an SE approach is applied. NLTE corrections remove a trend of the
\ion{Ca}{ii} abundance with line strength obtained under the LTE
assumption. Averages from eight lines are $\eps{CaII}$(OP,
\kH\ = 0) = 6.38 $\pm$
0.07, $\eps{CaII}$(OP, \kH\ = 0.1) = 6.40 $\pm$ 0.06, and
$\eps{CaII}$(OP, \kH\ = 1) = 6.43 $\pm$ 0.08.
It should be
noted that a variation in $\log C_6$ values within 0.6\,dex for the
lines of multiplets $5p - 5d$ and $4d - 4f$ sensitive to van der
Waals broadening changes the mean \ion{Ca}{ii} abundance by only
0.01\,dex.
The NIST database contains only three \ion{Ca}{ii} lines of our interest and their
$f_{ij}$ are very close to the corresponding values from OP
calculations. For the VALD data, the NLTE and LTE mean
\ion{Ca}{ii} abundances equal $\eps{CaII}$(VALD, \kH\ = 0) = 6.65
$\pm$ 0.32 and $\eps{CaII}$(VALD, LTE) = 6.74 $\pm$ 0.29. The large
standard deviation is mainly caused by $\lambda\,6456$ and
$\lambda\,5339$. In contrast to \ion{Ca}{i}, NLTE \ion{Ca}{ii}
abundance is sensitive to a variation of electronic collisional
rates in SE calculations. The value determined using the formula
of van Regemorter (\cite{Reg}) and assuming \kH\ = 0 is larger by
0.05\,dex compared to the corresponding value obtained for the
standard recipe of electronic collisional rates.

Taking into account the highest accuracy of the mean values, we prefer
to use as final values the \ion{Ca}{ii} abundance based on OP
oscillator strengths and \ion{Ca}{i} abundance based on the
laboratory $f_{ij}$. They agree within $\le 0.04$\,dex
provided that \kH\ $\le 0.1$ and the standard recipe of electronic
collisional rates is applied. In any other case, the difference
equals 0.08\,dex (\kH\ = 1 or van Regemorter's electronic
collisional rates) to 0.12\,dex (LTE). Thus, using the theoretical
MAFAGS model atmosphere, we find solar Ca abundance
to lie between 6.36 and 6.40.


We obtain here also the fitting parameters of solar \ion{Ca}{i}
$\lambda\,4226$ and \ion{Ca}{ii} $\lambda\,8498$ lines necessary for
further analysis of metal-poor stars.
For both lines, oscillator
strengths are taken from laboratory measurements (Smith \&
Gallagher \cite{ca4226} for $\lambda\,4226$; Theodosiou
\cite{ca3933} for $\lambda\,8498$), and the van der Waals damping
constants are based on $A\&O'M$'s data (Table\,\ref{line_list}).
For $\lambda\,8498$, isotope structure is taken into account with
atomic data from Table\,\ref{iso8498}.
A good fit of the observed solar flux profile of $\lambda\,4226$
in the wings is achieved for $\eps{Ca}$ = 6.29 when
hydrogenic collisions are ignored and for $\eps{Ca}$ = 6.21 in the
LTE case.
NLTE effects are negligible for the $\lambda\,8498$ line wings. The
best fit is achieved at a Ca abundance of $\eps{Ca}$ = 6.28.

Recent determinations of solar photospheric Ca abundance based on a
3D LTE analysis (Asplund et al.~\cite{met05}) give "excellent
agreement between the two ionization stages" and the average of
the two: $\eps{Ca}$ = 6.31\,$\pm$\,0.04. The average obtained from our
1D NLTE analysis is $\eps{Ca}$ = 6.38\,$\pm$\,0.06. We showed above
that the absolute Ca abundance depends on the adopted values of
oscillator strengths and van der Waals damping constants. The
cited paper of Asplund et al.\ does not give enough
information for a discussion of possible sources of the found
discrepancy.

\section{Stellar sample, observations and stellar
parameters}\label{obs}

Our sample consists of eight stars. HD\,61421 (Procyon) is selected as
a fundamental star with nearly the same list of detected Ca lines as in the
Sun. It is especially important for checking the \ion{Ca}{ii} high
excitation lines. The other objects are metal-poor stars with
metallicity ranging between [Fe/H] = $-$1.35 and [Fe/H] = $-$2.43.
They give an opportunity to study a formation of some of the
strongest Ca lines, \ion{Ca}{i} $\lambda\,4226$ and \ion{Ca}{ii}
$\lambda\,8498$ which presumably become of purely photospheric
origin in this metallicity range.

\begin{table*}[htbp]
\caption{Stellar parameters and obtained NLTE elemental abundance
ratios  of the selected stars. $\Vmic$ is given in \kms.  }
\label{startab}
\setlength{\tabcolsep}{5mm} 
\begin{tabular}{rccccccc}
\noalign{\smallskip} \hline \noalign{\smallskip}
 HD/BD & $\Teff$ & $\log g$ & [Fe/H]$_{\rm II}$ & $\Vmic$ & [Mg/Fe]$_{\rm I}$ & [Ca/Fe]$_{\rm I}$ & [Ca/Fe]$_{\rm II}$ \\
\noalign{\smallskip} \hline \noalign{\smallskip}
 19445 & 6030 & 4.40\scriptsize{$\pm$0.06} & $-$2.08\scriptsize{$\pm$0.05} & 1.6\scriptsize{$\pm$0.2} & 0.49\scriptsize{$\pm$0.03} & ~~0.36\scriptsize{$\pm$0.04} & 0.40~~~~~~ \\ %
 29907 & 5500 & 4.64\scriptsize{$\pm$0.06} & $-$1.60\scriptsize{$\pm$0.02} & 0.7\scriptsize{$\pm$0.1} & 0.35\scriptsize{$\pm$0.01} & ~~0.33\scriptsize{$\pm$0.05} &   \\
 59392 & 6010 & 4.02\scriptsize{$\pm$0.15} & $-$1.60\scriptsize{$\pm$0.03} & 1.4\scriptsize{$\pm$0.2} & 0.33\scriptsize{$\pm$0.05} & ~~0.26\scriptsize{$\pm$0.04} &   \\
 61421 & 6510 & 3.96\scriptsize{$\pm$0.02} & $-$0.03\scriptsize{$\pm$0.04} & 1.8\scriptsize{$\pm$0.1} & 0.00\scriptsize{$\pm$0.06} & $-$0.11\scriptsize{$\pm$0.03} &   0.10\scriptsize{$\pm$0.05}  \\
 84937 & 6350 & 4.00\scriptsize{$\pm$0.12} & $-$2.16\scriptsize{$\pm$0.05} & 1.8\scriptsize{$\pm$0.2} & 0.39\scriptsize{$\pm$0.04} & ~~0.43\scriptsize{$\pm$0.04} & 0.38~~~~~~ \\
103095 & 5070 & 4.69\scriptsize{$\pm$0.06} & $-$1.35\scriptsize{$\pm$0.03} & 0.8\scriptsize{$\pm$0.1} & 0.25\scriptsize{$\pm$0.03} & ~~0.30\scriptsize{$\pm$0.04} & 0.29~~~~~~ \\
140283 & 5810 & 3.68\scriptsize{$\pm$0.06} & $-$2.43\scriptsize{$\pm$0.05} & 1.6\scriptsize{$\pm$0.2} & 0.32\scriptsize{$\pm$0.02} & ~~0.29\scriptsize{$\pm$0.06} & 0.24~~~~~~ \\ %
$-4^\circ$3208 & 6280 & 4.08\scriptsize{$\pm$0.24}& $-$2.30\scriptsize{$\pm$0.06} & 1.8\scriptsize{$\pm$0.2} & 0.41\scriptsize{$\pm$0.01} & ~~0.43\scriptsize{$\pm$0.04} & 0.50~~~~~~ \\
\noalign{\smallskip} \hline \noalign{\smallskip}
\end{tabular}
\end{table*}

Four metal-poor stars, HD\,29907, HD\,59392, HD\,140283 and
BD\,$-4^\circ$3208 were observed using the Ultraviolet and Visual
Echelle Spectrograph UVES at the 8m ESO VLT UT2 telescope on Cerro
Paranal. At least two exposures were obtained for each star.
Spectral resolving power is about 60\,000. The data cover an
approximate spectral range of 4000\,--\,7000\,\AA. The signal-to-noise
ratio is $200$ or higher over the whole spectral range. No near-IR spectra for HD\,29907 and HD\,59392 are available to us.

Observational data for all other stars (and for Ca\,{\sc ii} 8498 in BD\,$-4^\circ$3208) are taken from Korn
et al. (\cite{Korn03}) with overall similar data-quality specifications as for the UVES data. For these data, the wavelength coverage is 4200\,--\,9000\,\AA. High-quality observations for Ca\,{\sc ii} 8498 in HD\,140283 were taken from the ESO UVESPOP survey (Bagnulo et al. \cite{POP03}).

We use stellar parameters determined in our earlier studies (Korn
et al. \cite{Korn03}; Mashonkina et al. \cite{mash03}). In short,
effective temperatures were determined from Balmer line profile
fitting, H$\alpha$ and H$\beta$, with a statistical error
estimated at the level of 70\,K,  surface gravities from the {\sc
Hipparcos} parallaxes with masses determined from the tracks of
VandenBerg et al. (\cite{isohrone}). The errors obtained from
adding the squared errors of parallax and mass are quoted in
Table\,\ref{startab}. The iron abundance and microturbulence
velocity were obtained requiring the derived [Fe/H]$_{\rm II}$
\footnote{The subscript in [X/Y]$_{\rm II}$ indicates that the
abundance for element X is determined from the singly ionized
species. Likewise, a subscript I refers to neutral lines used in
the abundance determination.} abundances not to depend on line
strength.  For metal-poor stars, $\alpha$-enhanced models are
calculated with abundances of $\alpha$-elements O, Mg, Si and Ca
scaled by the stellar Mg/Fe ratio. The latter is determined in
this study from NLTE analysis of four \ion{Mg}{i} lines,
$\lambda\,4571$, $\lambda\,4703$, $\lambda\,5528$, and
$\lambda\,5711$. We use the model atom and atomic data for
\ion{Mg}{i} described by Gehren et al. (\cite{gehr_mg}). Stellar
parameters are given in Table\,\ref{startab}.

The microturbulence values are somewhat lower than values
previously published by us as we now utilize the broadening
parameters of Barklem \& Aspelund-Johansson (\cite{baj05}) which
play a role even in the metal-poor stars as the analysis is
differential to the Sun. In all metal-poor stars, the
discrimination of the microturbulence values relies heavily on the
three strong \ion{Fe}{ii} lines of multiplet 42 ($\lambda\lambda$
4923, 5018 and 5169\,\AA). All three lines seem to require
substantially higher microturbulence values in the Sun. Enforcing
a fixed solar microturbulence of $\Vmic$ = 0.9\,\kms\ thus means
that they cannot be well fitted. We estimate that, depending on
the choices made in the fitting of solar lines and the line
selection for the metal-poor stars, microturbulence values can
easily vary by 0.2\,\kms. In particular, lower values seem
possible (e.g. disregarding \ion{Fe}{ii} 5169\,\AA). This is
important to bear in mind for the discussion of the strong lines
in Sections \ref{stars1} and \ref{stars2} .

Our results are based on line profile analysis. In order to compare
with observations, computed synthetic profiles are convolved with
a profile that combines instrumental broadening with a Gaussian
profile and broadening by macroturbulence with a radial-tangential
profile. Only slow rotators are included in our sample, $V \sin i
\le 1.5$\,\kms\ except for Procyon with $V \sin i = 2.6$\,\kms\
(Fuhrmann \cite{Fuhr3}) and therefore the rotational velocity and
macroturbulence value cannot be separated at the spectral
resolving power of our spectra. We thus treat their overal effect as
radial-tangential macroturbulence. Only for Procyon,
rotational broadening and broadening by macroturbulence are
treated separately.
The macroturbulence value was determined for each star in our previous
studies from the analysis of an extended list of lines of \ion{Fe}{i/ii},
\ion{Mg}{i}, etc. Here, $\Vmac$ was allowed to vary by
$\pm$0.3\,\kms\ (1$\sigma$).

\begin{figure}
\hspace{3mm}
\resizebox{88mm}{!}{\includegraphics{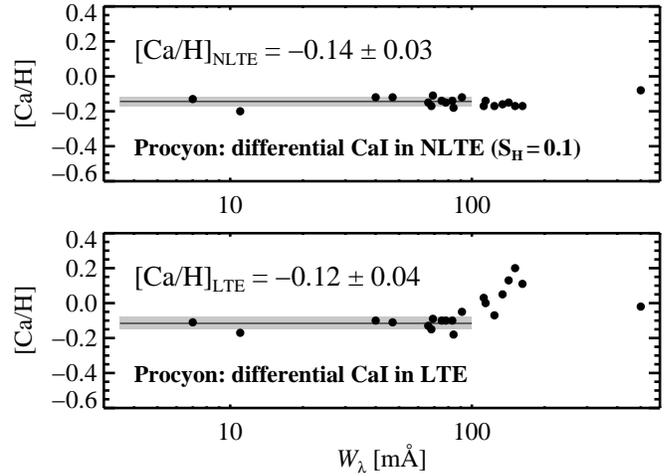}}

\vspace{1mm}
\caption[]{ Trends of abundance with line strength determined from the \ion{Ca}{i} lines in Procyon for our best NLTE model ({\em top}) and under the assumption of LTE ({\em bottom}). Note the steep trend of the LTE [Ca/H] values with line strength above $W_\lambda$ = 100\,m\AA. The mean values given are computed from weak lines only ($W_\lambda \leq$ 100\,m\AA).} \label{procyon}
\end{figure}

\section{The \ion{Ca}{i} versus \ion{Ca}{ii} in the selected stars}
\label{stars1}

In this section, we test NLTE formation of the \ion{Ca}{ii} lines
in two steps. In the first one, the \ion{Ca}{ii} high excitation
lines are examined in the solar metallicity star Procyon. These
lines are more sensitive to details of NLTE calculations than
$\lambda\,8498$ (see Table\,\ref{corr_diff}). We then derive element abundances from
the \ion{Ca}{i} lines and from \ion{Ca}{ii} $\lambda\,8498$ for five metal-poor stars and
inspect the difference $\Delta\eps{}$(\ion{Ca}{i} -- \ion{Ca}{ii}).
In the temperature regime we are concerned with and
at metallicities [Fe/H] between $-$2 and $-$3, the equivalent width
ratio $W$(\ion{Ca}{i})/$W$(\ion{Ca}{ii} 8498) is weakly sensitive
to a variation of $\Teff$ and $\log g$, independent of what
\ion{Ca}{i} line is taken. For example, $\log W$(\ion{Ca}{i} 6439)
/ $W$(\ion{Ca}{ii} 8498) = $-$0.88, $-$0.81, $-$0.77, and $-$0.78 for
the models with $\Teff$ / $\log g$ / [Fe/H] = 5000\,K / 3.0 / $-$2,
5000\,K / 4.0 / $-$2, 6000\,K / 3.0 / $-$2, and 6000\,K / 4.0 / $-$2,
respectively. This can be understood because the van der Waals
broadened line wings give a significant contribution to
$W$(\ion{Ca}{ii} 8498) even at [Ca/H] = $-$2.6 (for the models with
[Fe/H] = $-$3). When $\log g$ increases, strengthening of the
$\lambda\,8498$ line wings is compensated by weakening NLTE
effects. In such conditions, a difference between Ca abundances
determined from two ionization stages, if present, will point to
shortcomings of the NLTE treatment of Ca lines rather than to the
uncertainty of stellar parameters.

Analysis of stellar spectra is made line-by-line differentially
with respect to the Sun. At the LTE assumption, the cores of many lines cannot be fitted. In such cases, Ca abundance is
derived from the line wing fitting.

\noindent
\underline{Procyon}\\
Irrespective of the assumed efficiency of hydrogen collisions, average NLTE effects are very small among weak \ion{Ca}{i} lines ($W_\lambda \leq$ 100\,m\AA) and vary between $-$0.02 dex (\kH\ = 1) and $-$0.03 dex (\kH\ = 0). However, the SE approach is able to remove a steep trend with line strength among strong \ion{Ca}{i} lines seen in LTE (see Fig. \ref{procyon}). For individual strong lines (like \ion{Ca}{i} 6439\,\AA) NLTE corrections exceed $-$0.3 dex.
Among weak lines ($W_\lambda$ $<$ 100\,m\AA), the line-to-line scatter is reduced by all three NLTE models and attains its minimal value ($\sigma$ = 0.027 dex) at \kH\ = 0.1. Surprisingly, the mean abundance derived from \ion{Ca}{i} lines is clearly subsolar,
[Ca/H]$_{\rm I}$ = $-$0.12\,$\pm$\,0.04 (LTE) to $-0.15\pm0.04$ (\kH\ = 0).

\begin{figure}
\resizebox{88mm}{!}{\includegraphics{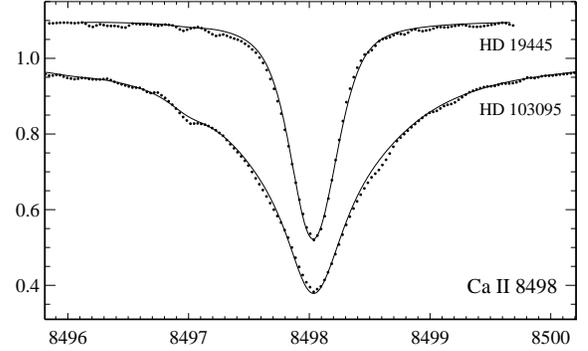}}

\vspace{-2mm}
\caption[]{The best NLTE (\kH\ = 0.1) fits (continuous line) of
the observed profiles of \ion{Ca}{ii} $\lambda\,8498$ in spectra
of HD\,19445 and HD\,103095 (bold dots). The obtained NLTE Ca
abundances are presented in Table \ref{startab}. For HD\,19445,
line profiles are shifted along Y axis} \label{star8498}
\end{figure}

Abundances for \ion{Ca}{ii} are derived from six high-excitation
lines. The mean NLTE value is [Ca/H]$_{\rm II}$(NLTE) = 0.07\,$\pm$\,0.05
independent of the assumed strength of hydrogenic
collisions. LTE abundances show a mean value of [Ca/H]$_{\rm II}$(LTE) = 0.14\,$\pm$\,0.05. An application of the formula
of van Regemorter (\cite{Reg}) for calculation of electronic
collisional rates results in an even larger difference
$\Delta\eps{}$(\ion{Ca}{i} -- \ion{Ca}{ii}).

Can the stellar parameters be the cause of this discrepancy?
Allende Prieto et al. (\cite{procyon}) estimate an error of the
effective temperature of 49\,K. Ramirez \& Melendez
(\cite{irfm05}) revise the fundamental effective temperature of
Procyon to $T_{\rm eff}^{\rm fund} = 6532\pm39$\,K and derive
$T_{\rm eff}^{{\rm IRFM}} = 6591\pm73$\,K using their infrared
flux method. Assuming $\Teff = 6590$\,K (a value favoured by Korn
et al. \cite{Korn03} for the NLTE ionization equilibrium of
Fe\,{\sc i/ii}) and keeping the values of $\log g$, [Fe/H] and
$\Vmic$ unchanged we obtain mean values [Ca/H]$_{\rm I}$(NLTE) =
$-$0.10\,$\pm$\,0.04 and [Ca/H]$_{\rm II}$(NLTE) =
0.05\,$\pm$\,0.05. As the discrepancy lies with weak lines,
modifications to the microturbulence value cannot remove this
discrepancy, either. Aufdenberg et al. (\cite{aufdenberg})
conclude that the interferometric data provide evidence for
convective overshooting in Procyon's atmosphere. We checked a
model which includes an ``approximate overshooting'' prescription
for convective flux transport in a mixing-length formalism
according to Castelli et al. (\cite{castelli}). The investigated
lines of both \ion{Ca}{i} and \ion{Ca}{ii} become weaker in this
model compared to our standard MAFAGS model, and the derived
abundances of \ion{Ca}{i} and \ion{Ca}{ii} increase by 0.08 -
0.10\,dex for different lines. However, the difference
$\Delta\eps{}$(\ion{Ca}{i} -- \ion{Ca}{ii}) remains at the same
level.

The effect of atmospheric temperature inhomogeneities on \ion{Ca}{i/ii} is expected to be of the same order of magnitude as that for neutral and singly ionized iron lines calculated by Allende Prieto et al. (\cite{procyon}). They show that weak lines ($W_\lambda \le$ 50\,m\AA) of both ionization stages, \ion{Fe}{i} and \ion{Fe}{ii}, are weakened compared to a classical 1D analysis, such that the derived Fe abundance increases by 0.05\,dex and 0.04\,dex, respectively. Three-dimensional simulations might lead to the higher Ca abundance for Procyon compared to our results, however, the discrepancy between \ion{Ca}{i} and \ion{Ca}{ii} is unlikely to be removed.
Thus, while our NLTE analysis leads to a better agreement of the Ca abundance from the two ionization stages, the discrepancy exceeds 3$\sigma$ under standard assumptions about Procyon's stellar parameters.

\noindent
\underline{HD\,19445}\\
The abundance is determined from 15 \ion{Ca}{i} lines with
equivalent widths between 4~m\AA\, and 63~m\AA. The accuracy of
the obtained differential abundances is at the level of $\sigma =
0.04$\,dex for \kH\ = 0, 0.1 and the LTE case and slightly worse,
$\sigma = 0.06$\,dex for \kH\ = 1. The difference between NLTE and
LTE \ion{Ca}{i} abundances depends strongly on the assumed value
for \kH. It equals +0.13\,dex for \kH\ = 0 and reduces down to
+0.03\,dex for \kH\ = 1. In contrast, NLTE abundances from
\ion{Ca}{ii} $\lambda\,8498$ are smaller than in LTE. The best fit
of $\lambda\,8498$ achieved at \kH\ = 0.1 and [Ca/Fe] = 0.40 is
shown in Fig.\,\ref{star8498}.
The difference between \ion{Ca}{i} and \ion{Ca}{ii} abundances is
found to be 0.04\,dex, $-$0.04\,dex, $-$0.11\,dex and $-$0.19\,dex
for \kH = 0, \kH\ = 0.1, \kH\ = 1 and LTE, respectively. An
uncertainty in microturbulence of 0.2\,\kms\ hardly affects the
\ion{Ca}{i} abundance and leads to a change in \ion{Ca}{ii} of
0.03\,dex. This does not destroy the agreement of the NLTE
\ion{Ca}{i} and \ion{Ca}{ii} abundances if \kH\ = 0 or 0.1 is
assumed.

\noindent
\underline{HD\,84937}\\
The abundance obtained from \ion{Ca}{i} lines vary between
[Ca/Fe]$_{\rm I}$ = 0.47\,$\pm$\,0.04 and [Ca/Fe]$_{\rm I}$ =
0.39\,$\pm$\,0.04, when changing \kH\ between 0 and $\infty$
(LTE). Similarly to HD\,19445, the abundances from the two
ionization stages agree much better in the NLTE case than in LTE
one. $\Delta\eps{}$(\ion{Ca}{i} -- \ion{Ca}{ii}) = +0.11\,dex and
$-$0.02\,dex for NLTE abundances at \kH\ = 0 and 1, respectively,
while the difference of LTE abundances equals --0.23\,dex. The
best agreement is found for an \kH\ value between 0.1 and 1.

\noindent
\underline{HD\,103095}\\
The determined NLTE abundances range between [Ca/Fe]$_{\rm I}$ =
0.37\,$\pm$\,0.04 and [Ca/Fe]$_{\rm I}$ = 0.29\,$\pm$\,0.04,
depending on the assumed \kH\ value. NLTE effects for \ion{Ca}{ii}
$\lambda\,8498$ occur only in the very core, and the same Ca
abundance is obtained with [Ca/Fe]$_{\rm II}$ = 0.29, independent
of the used theory of line formation. The best fit of this line is
achieved for \kH\ = 0.1 (Fig.\,\ref{star8498}). The difference
between \ion{Ca}{i} and \ion{Ca}{ii} is within the mutual error
bars in all cases: $\Delta\eps{}$(\ion{Ca}{i} -- \ion{Ca}{ii}) =
+0.08\,dex, +0.01\,dex, 0.00\,dex, and +0.03\,dex for \kH\ = 0,
0.1, 1, and LTE, respectively. It should be noted that the
discrepancy is larger when hydrogenic collisions are neglected
(cf.\ HD\,84937 above).

\noindent
\underline{HD\,140283}\\
The [Ca/Fe]$_{\rm I}$ ratios vary between +0.16\,$\pm$\,0.03 (LTE)
and +0.33\,$\pm$\,0.03 (\kH\ = 0) and are determined from six
lines with equivalent widths between 6 and 40 m\AA. At a
microturbulence value of $\Vmic$\,=\,1.65\,\kms\ the \ion{Ca}{i}
resonance line ($W_\lambda$ = 146 m\AA) is in good agreement with
the weak lines under the assumption of LTE, but yields too low
abundances for all NLTE models. We comment on this behavior in the
next section. Like in the case of HD\,84937,
$\Delta\eps{}$(\ion{Ca}{i} -- \ion{Ca}{ii}) vanishes between
\kH\,=\,0.1 and 1. In LTE, however, its value is --0.3\,dex.

\noindent
\underline{BD\,$-4^\circ$3208}\\
This star is found to be very similar to HD\,84937, both with
respect to its stellar parameters and its behavior in terms of
$\Delta\eps{}$(\ion{Ca}{i} -- \ion{Ca}{ii}). The abundances
obtained from 14 \ion{Ca}{i} lines with equivalent widths between
4\,m\AA\ and 40\,m\AA\ give mean values of [Ca/Fe]$_{\rm
I}$(\kH\,=\,0.1) = 0.43\,$\pm$\,0.04 and [Ca/Fe]$_{\rm I}$(LTE) =
0.34\,$\pm$\,0.04. Good agreement between \ion{Ca}{i} and
\ion{Ca}{ii} is again found for an \kH\ range of 0 to 0.1, while
the difference increases above \kH\,=\,0.1 and reaches --0.35\,dex
in the case of LTE.

Summarizing our results for the five metal-poor stars, we conclude
that, within the modelling uncertainties, NLTE leads to consistent
Ca abundances derived from the two ionization stages while LTE
fails to give consistent results. We find that ignoring hydrogenic
collisions results in too strong NLTE effects for \ion{Ca}{i/ii}.
Based on the results obtained for the Sun and the stars presented
in this section, our best choice for a scaling factor applied to
Steenbock \& Holweger's (\cite{hyd}) formula is \kH\ = 0.1. Final
\ion{Ca}{i} and \ion{Ca}{ii} abundances corresponding to \kH\ =
0.1 are presented in Table\,\ref{startab}.

\section{The \ion{Ca}{i} resonance line versus subordinate lines
in metal-poor stars} \label{stars2}

\ion{Ca}{i} $\lambda\,4226$ is the only neutral calcium line which
can be detected in extremely metal-poor stars. In this section, we
test NLTE formation of this line using the metal-poor stars where,
on one hand, the resonance line becomes, probably, of purely
photospheric origin and, on other hand, \ion{Ca}{i} subordinate
lines are still measurable and provide a reliable value of the Ca
abundance. In addition to the five metal-poor stars discussed
above, another two stars are studied here. We start by determining
the [Ca/Fe]$_{\rm I}$ ratio for them.

\noindent
\underline{HD\,29907}\\
The \ion{Ca}{i} abundance is determined from 16
lines with equivalent widths between 18\,m\AA\ and 174\,m\AA. NLTE
removes a trend with line strength displayed by LTE abundances
and results in the smaller standard deviation, $\sigma =
0.05$\,dex, compared to $\sigma = 0.08$\,dex in the LTE case. Mean
NLTE (\kH\,=\,0.1) and LTE abundances agree within 0.01 dex.

\begin{figure}
\resizebox{88mm}{!}{\includegraphics{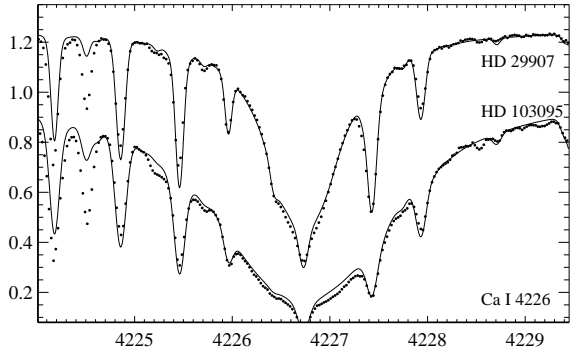}}


\vspace{-3mm}
\caption[]{Theoretical NLTE (\kH\ = 0.1, continuous line) profiles of the \ion{Ca}{i} $\lambda\,4226$ line
compared with the observed spectra of HD\,29907 and HD\,103095
(bold dots).
For each star, the Ca abundance adopted in the NLTE calculations
is taken from the analysis of
the \ion{Ca}{i} subordinate lines (see Table \ref{startab}), and
this abundances allows to fit the resonance line as well.
Offsets in y are applied for better illustration} \label{star4226}
\end{figure}

\noindent
\underline{HD\,59392}\\
The NLTE and LTE mean abundances derived from 17 \ion{Ca}{i} lines
turn out quite similar with [Ca/Fe](\kH\,=\,0.1) =
0.26\,$\pm$\,0.04 and [Ca/Fe](LTE) = 0.24\,$\pm$\,0.05.

Our calculations show that, for every star, overionization in the
atmospheric layers below $\log \tau_{5000} = -1$
(Fig.\,\ref{bfca1}) leads to weakening the line wings of
$\lambda\,4226$ compared to LTE case, while a steep decrease of
the departure coefficient ratio $b(4p)/b(4s)$ above that depth
point results in opposite effect for the line core. In the two
coolest stars of our small sample, HD\,29907 and HD\,103095, the
\ion{Ca}{i} resonance line is very strong with the core formed in
the uppermost atmospheric layers near $\log \tau_{5000} = -4.4$
and $-$4.9, respectively (we indicate here a location of line
center optical depth unity). However, total line absorption is
dominated by the van der Waals broadened wings and not by the
core. For both stars, we find that the NLTE theoretical profile
computed with Ca abundance determined from the \ion{Ca}{i}
subordinate lines describes the observed profile well except for
the very core which can be influenced by the star's chromosphere.
Results are illustrated in Fig.\,\ref{star4226}.

For HD\,84937 and  HD\,19445, the agreement between subordinate
lines of \ion{Ca}{i}, \ion{Ca}{ii} 8498 and \ion{Ca}{i} 4226 is excellent
when \kH = 0.1 is assuming. \ion{Ca}{i} 4226 requires a [Ca/Fe] ratio of 0.41
for HD\,84937 and of 0.36 for HD\,19445.

A less consistent picture emerges for the remaining three stars.
When Ca abundance is fixed at the value derived from the analysis of the \ion{Ca}{i}
subordinate lines, the half-width of the theoretical NLTE profile of
$\lambda\,4226$ is obtained to be larger than the observed one.
To fit the observed line width, a smaller Ca abundance is required: by 0.08\,dex for HD\, 59292,
0.09\,dex for HD\,140283, and 0.22\,dex for BD$-4^\circ 3208$. Discrepancies are
smaller in LTE for these stars, but larger for e.g. HD\,84937.

In each star investigated,
\ion{Ca}{i} 4226 lies on the saturated part of the curve of
growth, and such discrepancy could well be related to
uncertainties in the microturbulence value: $\Delta\Vmic =
-0.1$\,\kms\ translates to $\Delta\eps{} (\lambda\,4226)$ =
+0.03\,dex to +0.04\,dex for different stars. However, for a star
like BD$-4^\circ 3208$, a substantial reduction of the
microturbulence would be required, which in turn would affect the
good agreement between \ion{Ca}{i} and \ion{Ca}{ii} reported
above. We note in passing that a similar discrepancy is found for
subordinate lines of \ion{Mg}{i} and the \ion{Mg}{i}\,$b$ triplet
lines in these stars.

Are these then shortcomings of our SE calculations?
Collisions are inefficient in statistical equilibrium of atoms in
the layers, where the Doppler core of $\lambda\,4226$ is formed,
between $\log \tau_{5000} = -1$ and $-$3, for the most metal-poor
stars of our sample, BD\,$-4^\circ 3208$ and HD\,140283. The
related non-LTE effect is therefore entirely due to
photoionization which should be modelled accurately (given the
correctness of the OP photoionization cross-sections.)
The explanation can lie with the adopted one-dimensional atmospheric models.
Based on the recent results of Shchukina et al. (\cite{3Dfe}) for
weak and moderately strong ($W_\lambda \le$ 80\,m\AA) lines of
\ion{Fe}{i} in HD\,140283, we expect an overall small effect of
atmospheric-temperature and velocity inhomogeneities on the
\ion{Ca}{i} subordinate lines in our sample of metal-poor stars.
The \ion{Ca}{i} resonance line may be a different case because it
is formed over the more extended range of atmospheric depths. For
a fully quantitative understanding of its formation in cool stars,
further studies are required on the basis of advanced model
atmospheres. One clear advantage of such studies will be the
removal of adjustable parameters like microturbulence from which
the current modelling potentially suffers.

\section{The \ion{Ca}{i}/\ion{Ca}{ii} ionization equilibrium as indicator
of surface gravity for extremely metal-poor stars} \label{extrim}

\begin{figure}
\resizebox{88mm}{!}{\includegraphics{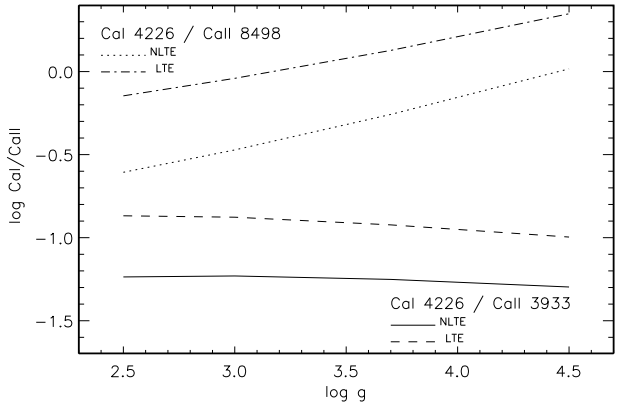}}

\vspace{-3mm}
\caption[]{The \ion{Ca}{i} 4226 / \ion{Ca}{ii} 3933 and
\ion{Ca}{i} 4226 / \ion{Ca}{ii} 8498 equivalent width ratios as a
function of the surface gravity for the models with $\Teff = 5500$\,K
and [Fe/H] = $-$4.34. A Ca abundance of [Ca/H] = $-$4.9 is adopted
everywhere} \label{g5500}
\end{figure}

In this section, we consider the possibility of using the
$W$(\ion{Ca}{i} 4226) / $W$(\ion{Ca}{ii} 3933) and $W$(\ion{Ca}{i}
4226) / $W$(\ion{Ca}{ii} 8498) equivalent width ratios as
indicators of surface gravity for extremely metal-poor stars. At
extremely low Ca abundance, the line wings do not contribute
anymore to $W$(\ion{Ca}{ii} 8498). The line is expected to be
strengthened with decreasing $\log g$ due to decreasing H$^-$
continuous absorption. NLTE effects for \ion{Ca}{ii}
$\lambda\,8498$ are amplified in the same direction. Thus, the
$W$(\ion{Ca}{i} 4226) / $W$(\ion{Ca}{ii} 8498) ratio should
increase when $\log g$ goes up. The \ion{Ca}{ii} resonance lines
remain strong even at [Ca/H] = $-$5, and their van der Waals
broadened wings are weakened with decreasing $\log g$. The
\ion{Ca}{i} resonance line is weakened in the same direction due
to amplified overionization. Thus, the $W$(\ion{Ca}{i} 4226) /
$W$(\ion{Ca}{ii} 3933) ratio is rather insensitive to variation of
$\log g$. Both ratios were calculated for a small grid of models
with [Fe/H] = $-$4.34; the Ca abundance was adopted to be [Ca/H] =
$-$4.9. The results corresponding to $\Teff = 5500 K$ are plotted
in Fig.\,\ref{g5500} as a function of $\log g$. It is clear that the
$W$(\ion{Ca}{i} 4226) / $W$(\ion{Ca}{ii} 8498) ratio can be used
to determine the surface gravity for extremely metal-poor stars, contrary to
the ratio involving the \ion{Ca}{ii} resonance line(s) which is
nearly constant over the $\log g$ range between 2.5 and 4.5.
An application of this technique to the two known ultra-metal-poor stars
(HE~0107$-$5240, Christlieb et al. \cite{chri02}; HE~1327$-$2326, Frebel et al. \cite{frebel05})
will be presented in a forthcoming paper.

\section{Concluding remarks} \label{conclusion}

In this study, NLTE line formation of an extended list of
\ion{Ca}{i} and \ion{Ca}{ii} lines was considered for the
temperatures ranging between $\Teff = 5000 K$ and $\Teff = 6000
K$, surface gravities $\log g =$ 3.0 and 4.0, and metallicities
from [Fe/H] = 0 down to [Fe/H] = $-$4.34. For every \ion{Ca}{i}
line, departures from LTE affect significantly its profile over
the whole range of stellar parameters. If a line is strong
(multiplets 2, 3, 4, 18, 21, and 47 at [Fe/H] $\ge -2$ and the
remaining multiplets at [Fe/H] $\ge -1$), its wings are weakened
but the core is strengthened compared to the LTE case. The value
and sign of NLTE abundance correction are defined by a relative
contribution of the core and the wings to the overall line
strength. When the line becomes weak due to decreasing Ca
abundance, NLTE leads to depleted total absorption in the line and
positive abundance correction. NLTE effects are very large at
extremely low Ca abundance. For example, at [Ca/H] = $-$4.9 NLTE
abundance correction for \ion{Ca}{i} $\lambda\,4226$ can exceed
0.5\,dex. Thus, for a given line, $\Delta_{\rm NLTE}$ depends on
$\Teff$, $\log g$, [Ca/H], and microturbulence value, and, for a
given model, $\Delta_{\rm NLTE}$ is different in value and sign
for a variety of \ion{Ca}{i} lines. Any interpolation of NLTE
results for \ion{Ca}{i} has to be performed with caution. A
different situation is found for \ion{Ca}{ii}. For every
\ion{Ca}{ii} line, NLTE leads to enhanced absorption in the line
core and negative abundance correction over the whole range of
stellar parameters. The absolute value of $\Delta_{\rm NLTE}$ is
defined by the relative contribution of the core and the wings to
equivalent width. For example, NLTE corrections remain very small,
$\le 0.02$\,dex, for the resonance lines and grow in absolute
value with decreasing Ca abundance for the IR lines of multiplet
$3d$ - $4p$ exceeding 0.4\,dex in a metal-poor model with $T_{\rm
eff}$\,=\,6000\,K, $\log g$\,=\,3.0 and [Fe/H]\,=\,$-$3. Such
corrections are important to be considered for the correct
interpretation of the near-IR spectroscopic data collected in the
RAVE survey and the future ESA Gaia satellite mission.

Empirical evidence is found from the analysis of stellar spectra
that inelastic collisions with hydrogen atoms serve as an
additional source of thermalization in cool (metal-poor) stars.
 Our best choice for a
scaling factor to the formula of Steenbock \& Holweger
(\cite{hyd}) for calculation of hydrogenic collisions is \kH\ =
0.1. Disregarding hydrogenic collisions completely will lead to
overestimated NLTE effect at low metallicity.

Taking advantage of our SE approach and accurate atomic data for the
investigated lines, we obtain good agreement, within 0.04\,dex, between
absolute solar \ion{Ca}{i} (from 23 lines) and \ion{Ca}{ii} (from 8 lines)
abundances, and the average of the two is $\eps{Ca}^\odot$ = 6.38\,$\pm$\,0.06.

Calcium abundances from two ionization stages are examined for the first time
for five metal-poor stars. We show that NLTE largely removes obvious discrepancies between
\ion{Ca}{i} and \ion{Ca}{ii} obtained under the LTE assumption.

Solving the restricted NLTE problem for calcium in ``classical''
one-dimensional LTE model atmospheres, we qualitatively understand
the formation of the \ion{Ca}{i} resonance line in metal-poor
stars, where it is of purely photospheric origin. In agreement
with observations, NLTE predicts weakening of the line wings and
strengthening of the line core compared to the LTE case. Residual
discrepancies may be related to the use of classical model
atmospheres with adjustable parameters like microturbulence.

\begin{acknowledgements}
We are grateful to Thomas Gehren for providing a Windows version
of the code DETAIL and Tatyana Ryabchikova for help with
collecting atomic data. LM acknowledges with gratitude the
Institute of Astronomy and Astrophysics of Munich University for
warm hospitality during a productive stay in May -- August 2005.
This research was supported by the Deutsche Forschungsgemeinschaft
with grant 436 RUS 17, the Russian Foundation for Basic Research
with grant 05-02-39005-GFEN-a, the Royal Swedish Academy of
Sciences with grant 11630102, and the Presidium RAS Programme
``Origin and evolution of stars and the Galaxy''. AJK acknowledges
support from the Leopoldina foundation/Germany under grant
BMBF-LPD 9901/8-87 and the Swedish Research Council. AJK also
thanks Nikolai Piskunov for travel support for a visit to Moscow
in April 2005. We thank the anonymous referee for valuable
suggestions and comments. We made ample use of data collected in
the NIST and VALD databases.
\end{acknowledgements}


\begin{thebibliography}{99}
\bibitem[1973]{Allen}
Allen, C.W. 1973, Astrophysical Quantities. Athlone Press
\bibitem[2002]{procyon}
Allende Prieto, C., Asplund, M., Garsia Lopez, R.J. \& Lambert,
D. 2002, ApJ, 567, 544
\bibitem[1989]{AG}
Anders, E. \& Grevesse, N. 1989, Geoch. \& Cosmochim Acta, 53,
197
\bibitem[2005]{andretta05}
Andretta, V. Busa, I., Gomez, M.T. \& Terranegra, L. 2005, A\&A,
430, 669
\bibitem[1995]{omara_sp}
Anstee, S.D. \& O'Mara, B.J. 1995, MNRAS, 276, 859 ($A\&O'M$)
\bibitem[2006]{frebel06}
Aoki, W., Frebel, A., Christlieb, N., Norris, J.E., Beers, T.C.,
et al. 2006, ApJ, 639, 897
\bibitem[1991]{ardaco91}
Armandroff, T.E., \& da Costa, G.S.~1991, \aj, 101, 1329
\bibitem[2005]{met05}
Asplund, M., Grevesse, N. \& Sauval, A.J. 2005, ASP Conf.Ser.,
336, 25
\bibitem[2005]{aufdenberg}
Aufdenberg, J.P., Ludwig, H.-G., \& Kervella, P. 2005, ApJ, 633, 424
\bibitem[2005]{POP03}
Bagnulo, S., Jehin, E., Ledoux, C. et al. 2003, ESO Messenger, 114,
10
\bibitem[2005]{baj05}
Barklem, P.S. \& Aspelund-Johansson, J. 2005, A\&A, 435, 373
\bibitem[1997]{omara_pd}
Barklem, P.S. \& O'Mara, B.J. 1997, MNRAS, 290, 102 ($A\&O'M$)
\bibitem[1998]{omara_df}
Barklem, P.S., O'Mara, B.J. \& Ross, J.E. 1998, MNRAS, 296, 1057
($A\&O'M$)
\bibitem[1998]{omara_ion}
Barklem, P.S. \& O'Mara, B.J. 1998, MNRAS, 300, 863 ($A\&O'M$)
\bibitem[1999]{belyaev99}
Belyaev, A.K., Grosser, J., Hahne, J. \& Menzel, T. 1999, Phys.
Rev., A60, 2151
\bibitem[2003]{belyaev03}
Belyaev, A.K. \& Barklem, P. 2003, Phys. Rev., A68, 062703
\bibitem[1995]{burgess}
Burgess, A., Chidichimo, M.C. \& Tully, J.A. 1995, A\&A, 300, 627
\bibitem[1985]{detail}
Butler, K. \& Giddings, J. 1985, Newsletter on the analysis of
   astronomical spectra No. 9, University of London
\bibitem[1997]{castelli}
Castelli, F., Gratton, R.G., \& Kurucz, R.L. 1997,  A\&A, 318, 841
\bibitem[2004]{cayrel04}
Cayrel, R., Depagne, E., Spite, M., Hill, V., Spite, F., et al.
2004, A\&A, 416, 1117
\bibitem[2001]{cenarroetal01}
Cenarro, A.J., Cardiel, N., Gorgas, J., Peletier, R.F., Vazdekis, A., \&
Prada, F.~2001, \mnras, 326, 959
\bibitem[2002]{chri02}
Christlieb, N., Bessel, M., Beers, T., Gustafsson, B., Korn, A.,
et al. 2002, Nature, 419, 904
\bibitem[1992]{topbase}
Cunto, W. \& Mendoza, C. 1992, Rev. Mex. Astrofis., 23, 107
\bibitem[1968]{Doyle}
Doyle, R.O. 1968, ApJ, 153, 987
\bibitem[1991]{Drake}
Drake, J. 1991, MNRAS, 251, 369
\bibitem[1968]{D68}
Drawin, H.W. 1968, Z.Physik, 211, 404
\bibitem[1969]{D69}
Drawin, H.W. 1969, Z. Physik, 225, 483
\bibitem[1997]{ca6572}
Drozdowski, R., Ignaciuk, M., Kwela ,J. \& Heldt, J. 1997,
Z.Phys., D41, 125
\bibitem[2005]{frebel05}
Frebel, A., Aoki, W., Christlieb, N., Ando, H., Asplund, M., et
al. 2005, Nature, 434, 871
\bibitem[1998]{Fuhr3}
Fuhrmann, K. 1998, A\&A, 338, 161
\bibitem[1997]{Fuhr1}
Fuhrmann, K., Pfeiffer, M., Frank, C., Reetz, J. \& Gehren, T.
  1997, A\&A, 323, 909
\bibitem[1975]{gehren75}
Gehren, T. 1975, A\&A, 38, 289
\bibitem[2004]{gehr_mg}
Gehren, T., Liang, Y.C., Shi, J.R., Zhang, H.W. \& Zhao, G.
2004 A\&A, 413, 1045
\bibitem[2005]{ibataetal05}
Ibata, R., Chapman, S., Ferguson, A.M.N., Lewis, G., Irwin, M., \& Tanvir,
N.~2005, \apj, 634, 287
\bibitem[2000]{th-mg}
Idiart, T.P. \& Thevenin, F. 2000, ApJ, 541, 207
\bibitem[1992]{ca92}
J\o rgensen, U.G., Carlsson, M. \& Johnson, R. 1992 A\&A, 254,
258
\bibitem[2004]{c6_4226}
Kerkeni, B., Barklem, P.S., Spielfiedel, A. \& Feautrier N. 2004, J. Phys., B37, 677
\bibitem[2003]{Korn03}
Korn, A., Shi, J. \& Gehren, T. 2003 A\&A, 407, 691
\bibitem[1999]{vald}
Kupka, F., Piskunov, N., Ryabchikova, T.A., Stempels, H.C. \&
Weiss, W.W. 1999, A\&AS, 138, 119
\bibitem[1992]{cdrom18}
Kurucz, R.L. 1992, CD-ROM No. 18; CD-ROM No. 23
\bibitem[1984]{Atlas}
Kurucz, R.L., Furenlid, I., Brault, J. \& Testerman, L. 1984,
Solar Flux Atlas from 296 to 1300 nm. Nat. Solar Obs., Sunspot,
New Mexico
\bibitem[2004]{ca_is_res}
Lucas, D.M., Ramos, A., Home, J.P., McDonnell, M.J., Nakayama, S.,
et al. 2004, Phys. Rev., A69, 012711
\bibitem[1997]{mallik97}
Mallik, S.V. 1997, A\&AS, 124, 359
\bibitem[1996]{mash96}
Mashonkina, L.I. 1996, Model Atmospheres and Spectrum Synthesis,
ASP Conf. Ser., 108, 140
\bibitem[2003]{mash03}
Mashonkina, L., Gehren, T., Travaglio, C. \& Borkova, T. 2003,
A\&A, 397, 275
\bibitem[1972]{moore72}
Moore, C.E. 1972, A Multiplet Table of Astrophysical Interest.
 NSRDS-NBS 40
\bibitem[2002]{nis02}
Nissen, P.E., Primas, F., Asplund, M. \& Lambert, D.L. 2002,
A\&A, 390, 235
\bibitem[1998]{ca_is}
N\"ortersh\"auser, W., Blaum, K., Icker, K., M\"uller, P.,
Schmitt, A., et al. 1998, Eur. Phys. J., D2, 33
\bibitem[2001]{perrymanetal01}
Perryman, M.A.C., de Boer, K.S., Gilmore, G., H{\o}g, E., Lattanzi, M.G., et
al.~2001, \aap, 369, 339
\bibitem[2005]{irfm05}
Ramirez, I. \& Melendez, J. 2005, ApJ, 626, 446
\bibitem[1997]{rutledgeetal97}
Rutledge, G.A., Hesser, J.E., \& Stetson, P.B.~1997, \pasp, 109, 907
\bibitem[1991]{rh91}
Rybicki, G.B. \& Hummer, D.G. 1991, A\&A, 245, 171
\bibitem[1992]{rh92}
Rybicki, G.B. \& Hummer, D.G. 1992, A\&A, 262, 209
\bibitem[2002]{Sagliaetal02}
Saglia, R.P., Maraston, C., Thomas, D., Bender, R., \& Colless, M.~2002,
\apj, 579, L13
\bibitem[2001]{samson}
Samson, A.M. \& Berrington, K.A. 2001, ADNDT, 77, 87
\bibitem[1962a]{Seaton}
Seaton, M.J. 1962, Proc. Phys. Soc. London, 79, 1105
\bibitem[1962b]{seat_i}
Seaton, M.J. 1962, in ``Atomic and Molecular Processes'', New York
Academic Press
\bibitem[1994]{OP}
Seaton, M.J., Mihalas, D. \& Pradhan, A.K. 1994, MNRAS, 266, 805
\bibitem[1963]{ca_np}
Shabanova, L.N. 1963, Opt. Spectros. (USSR), 15, 450
\bibitem[2005]{3Dfe}
Shchukina, N.G., Trujillo Bueno, J. \& Asplund, M. 2005,
ApJ, 618, 939
\bibitem[1974]{ca74}
Shine ,R.A. \& Linsky, J.L. 1974, Solar Phys., 39, 49
\bibitem[1981]{ca_81}
Smith, G. 1981, A\&A, 103, 351
\bibitem[1988]{ca_88}
Smith, G. 1988, J. Phys., B21, 2827
\bibitem[1966]{ca4226}
Smith, W.W. \& Gallagher, A. 1966, Phys. Rev., 145, 26
\bibitem[1975]{ca_75}
Smith, G. \& O'Neil, J.A. 1975, A\&A, 38, 1
\bibitem[1981]{ca_fij}
Smith, G. \& Raggett, D.St.J. 1981, J. Phys., B14, 4015
\bibitem[1991]{ca6122_c6}
Spielfiedel, A., Feautrier, N., Chambaud, G. \& Levy, B.
1991, J. Phys., B24, 4711
\bibitem[1984]{hyd}
Steenbock, W. \& Holweger, H. 1984, A\&A, 130, 319
\bibitem[2006]{steinmetzetal06}
Steinmetz, M., Zwitter, T., Siebert, A., Watson, F.G., Freeman, K.C., et
al.~2006, \aj, accepted (astro-ph/0606211)
\bibitem[1985]{ca_nist}
Sugar, J. \& Corliss, C. 1985, J. Phys. Chem. Ref. Data, 14,
Suppl. No. 2
\bibitem[1989]{ca3933}
Theodosiou, C.E. 1989, Phys. Rev., A39, 4880
\bibitem[2001]{tolstoyetal01}
Tolstoy, E., Irwin, M.J., Cole, A.A., Pasquini, L., Gilmozzi, R., \&
Gallagher, J.S.~2001, \mnras, 327, 918
\bibitem[2000]{isohrone}
VandenBerg, D.A., Swenson, F.J., Rogers, F.J., Iglesias, C.A. \&
Alexander, D.R. 2000, ApJ, 532, 430
\bibitem[1962]{Reg}
van Regemorter, H. 1962, ApJ, 136, 906
\bibitem[1985]{ca1}
Watanabe, T. \& Steenbock, W. 1985, A\&A, 149, 21
\end{thebibliography}
\end{document}